\documentclass{iopart}
\usepackage{graphicx}
\begin{document}
\review[Metal-insulator transition in 2D]{Metal-insulator transition in two-dimensional electron systems}
\author{S~V Kravchenko}
\address{Physics Department, Northeastern University, Boston, MA 02115, USA}
\author{M~P Sarachik}
\address{Physics Department, City College of the City University of New York, New York, NY 10031, USA}
\begin{abstract}
The interplay between strong Coulomb interactions and randomness has been a long-standing problem in condensed matter physics.  According to the scaling theory of localization, in two-dimensional systems of noninteracting or weakly interacting electrons, the ever-present randomness causes the resistance to rise as the temperature is decreased, leading to an insulating ground state. However, new evidence has emerged within the past decade indicating a transition from insulating to metallic phase in two-dimensional systems of {\em strongly} interacting electrons.  We review earlier experiments that demonstrate the unexpected presence of a metallic phase in two dimensions, and present an overview of recent experiments with emphasis on the anomalous magnetic properties that have been observed in the vicinity of the transition.
\end{abstract}

\ead{s.kravchenko@neu.edu; sarachik@sci.ccny.cuny.edu}\vspace{2cm}
Journal ref.: {\it Rep.\ Prog.\ Phys}.\ {\bf 67}, 1 (2004)\vspace{5mm}\\
\maketitle

\tableofcontents

\newpage

\section{INTRODUCTION}

In two-dimensional electron systems, the electrons are confined to move in a plane in the presence of a random potential.  According to the scaling theory of localization (Abrahams~\etal 1979), these systems lie on the boundary between high and low dimensions insofar as the metal-insulator transition is concerned.  The carriers are always strongly localized in one dimension, while in three dimensions, the electronic states can be either localized or extended.  In the case of two dimensions the electrons may conduct well at room temperature, but a weak logarithmic increase of the resistance is expected as the temperature is reduced.  This is due to the fact that, when scattered from impurities back to their starting point, electron waves interfere constructively with their time reversed paths.  While this effect is weak at high temperatures due to inelastic scattering events, quantum interference becomes increasingly important as the temperature is reduced and leads to localization of the electrons, albeit on a large length scale; this is generally referred to as ``weak localization''.  Indeed, thin metallic films and two-dimensional electron systems fabricated on semiconductor surfaces were found to display the predicted logarithmic increase of resistivity (Dolan and Osheroff 1979; Bishop~\etal 1980, 1982; Uren~\etal 1980), providing support for the weak localization theory.

The scaling theory does not explicitly consider the effect of the Coulomb interaction between electrons.  The strength of the interactions is usually characterized by the dimensionless Wigner-Seitz radius, $r_s=1/(\pi n_s)^{1/2}a_B$ (here $n_s$ is the electron density and $a_B$ is the Bohr radius in a semiconductor).  In the experiments mentioned above, the Coulomb interactions are relatively weak.  Indeed, these experiments are in agreement with theoretical predictions (Altshuler, Aronov and Lee 1980) that weak electron-electron interactions ($r_s\ll1$) increase the localization even further.  As the density of electrons is reduced, however, the Wigner-Seitz radius grows and the interactions provide the dominant energy of the system.  No analytical theory has been developed to date in the strongly-interacting limit ($r_s\gg1$).  Finkelstein (1983, 1984) and Castellani~\etal (1984) predicted that for weak disorder and sufficiently strong interactions, a 2D system should scale towards a {\em conducting} state as the temperature is lowered.  However, the scaling procedure leads to an increase in the effective strength of the interactions and to a divergent spin susceptibility, so that the perturbative approach breaks down as the temperature is reduced toward zero.  Therefore, the possibility of a 2D metallic ground state stabilized by strong electron-electron interactions was not seriously considered.

Recent progress in semiconductor technology has enabled the fabrication of high quality 2D samples with very low randomness in which measurements can be made at very low carrier densities.  The strongly-interacting regime ($r_s\gg1$) has thus become experimentally accessible.  Experiments on low-disordered 2D silicon samples (Kravchenko~\etal 1994, 1995, 1996) demonstrated that there are surprising and dramatic differences between the behaviour of strongly interacting systems at $r_s>10$ as compared with weakly-interacting systems: with increasing electron density, one can cross from the regime where the resistance diverges with decreasing temperature (insulating behaviour) to a regime where the resistance decreases strongly with decreasing $T$ (metallic behaviour).  These results were met with great scepticism and largely overlooked until 1997, when they were confirmed in silicon MOSFETs from a different source (Popovi\'{c}~\etal 1997) and in other strongly-interacting 2D systems (Coleridge~\etal 1997; Hanein~\etal 1998a; Papadakis and Shayegan 1998).  Moreover, it was found (Simonian~\etal 1997b; Pudalov~\etal 1997; Simmons~\etal 1998) that in the strongly-interacting regime, an external magnetic field strong enough to polarize the electrons' spins induces a giant positive in-plane magnetoresistance\footnote{The fact that the parallel magnetic field promotes insulating behaviour in strongly interacting 2D systems was first noticed by Dolgopolov~\etal (1992).} and completely suppresses the metallic behaviour, thus implying that the spin state is central to the high conductance of the metallic state. This finding was in qualitative agreement with the prediction of Finkelstein and Castellani~\etal that for spin-polarized electrons, only an insulating ground state is possible in a disordered 2D system even in the presence of strong interactions.   Subsequent experiments (Okamoto~\etal 1999; Kravchenko~\etal 2000b; Shashkin~\etal 2001, 2002; Vitkalov~\etal 2001b, 2002; Pudalov~\etal 2002b; Zhu~\etal 2003) have shown that there is a sharp enhancement of the spin susceptibility as the metal-insulator transition is approached; indications exist that in silicon MOSFETs, the spin susceptibility may actually diverge at some sample-independent electron density $n_\chi\approx8\cdot10^{10}$~cm$^{-2}$.

In silicon samples with very low disorder potential, the critical density for the metal-insulator transition was found to be at or very near $n_\chi$ (Shashkin~\etal 2001a, 2002; Vitkalov~\etal 2001b, 2002), indicating that the metal-insulator transition observed in these samples is a property of a clean disorder-free 2D system, rather than being a disorder-driven transition.  In such samples, the metallic and insulating regimes are divided by a temperature-independent separatrix with $\rho\approx3h/e^2$, in the vicinity of which the resistivity displays virtually universal critical behaviour. However, in samples with relatively strong disorder, the electrons become localized at densities significantly higher than $n_\chi$: from $1.44\cdot10^{11}$ to $6.6\cdot10^{11}$~cm$^{-2}$ (Prus~\etal 2002), and even as high as $1.6\cdot10^{12}$~cm$^{-2}$ (Pudalov~\etal 1999), indicating that the localization transition in these samples is driven by disorder.

We suggest that there has been a great deal of confusion and controversy caused by the fact that often no distinction has been made between results obtained in systems with relatively high disorder and those obtained for very clean samples, and also because in many experimental studies, a change in the sign of the derivative ${\rm d}R/{\rm d}T$ has always been assumed to signal a metal-insulator transition.  In this review, we focus our attention on results for very clean samples.

The experimental findings described above were quite unexpected.  Once accepted, they elicited strong and widespread interest among theorists, with proposed explanations that included unusual superconductivity (Phillips~\etal 1998), charging/discharging of contaminations in the oxide (Altshuler and Maslov 1999), the formation of a disordered Wigner solid (Chakravarty~\etal 1999), inter-subband scattering (Yaish~\etal 2000) and many more (for a review, see Abrahams, Kravchenko and Sarachik 2001). It is now well-documented that the metallic behaviour in zero magnetic field is caused by the delocalizing effect associated with strong electron-electron interactions which overpower quantum localization.  In the ``ballistic regime'' deep in the metallic state, the conductivity is linear with temperature and derives from coherent scattering of electrons by Friedel oscillations (Zala, Narozhny and Aleiner 2001).  Closer to the transition, in the ``diffusive'' regime, the temperature dependence of the resistance is well described by a renormalization group analysis of the interplay of strong interactions and disorder (Punnoose and Finkelstein 2002).  Within both theories (which consider essentially two limits of the same physical process) an external magnetic field quenches the delocalizing effect of interactions by aligning the spins, and causes a giant positive magnetoresistance. However, the metal-insulator transition itself, as well as the dramatic increase of the spin susceptibility and effective mass in its close vicinity, still lack adequate theoretical description; in this region the system appears to behave well beyond a weakly interacting Fermi liquid.

In the next three sections, we describe the main experimental results that demonstrate the unexpected presence of a metallic phase in 2D and present an overview of recent experiments with emphasis on the anomalous magnetic properties observed in the vicinity of the metal-insulator transition.

\section{EXPERIMENTAL RESULTS IN ZERO MAGNETIC FIELD}
\subsection{Resistance in zero magnetic field, experimental scaling,
reflection symmetry}

The first experiments that demonstrated the unusual temperature dependence of the resistivity (Kravchenko~\etal 1994, 1995, 1996) were performed on low-disordered silicon metal-oxide-semiconductor field-effect transistors (MOSFETs) with maximum electron mobilities reaching more than $4\cdot10^4$~cm$^2$/Vs, mobilities that were considerably higher than in samples used in earlier investigations. It was the very high quality of the samples that allowed access to the physics at electron densities below $10^{11}$~cm$^{-2}$.  At these low densities, the Coulomb energy, $E_C$, is the dominant parameter. Estimates for Si MOSFETs at $n_s=10^{11}$~cm$^{-2}$ yield $E_C\approx10$~meV, while the Fermi energy, $E_F$, is about $0.6$~meV (a valley degeneracy of two is taken into account when calculating the Fermi energy, and the effective mass is assumed to be equal to the band mass, $m_b$.) The ratio between the Coulomb and Fermi energies, $r^*\equiv E_C/E_F$, thus assumes values above 10 in these samples.

\begin{figure}
\begin{center}
\scalebox{.55}{\includegraphics{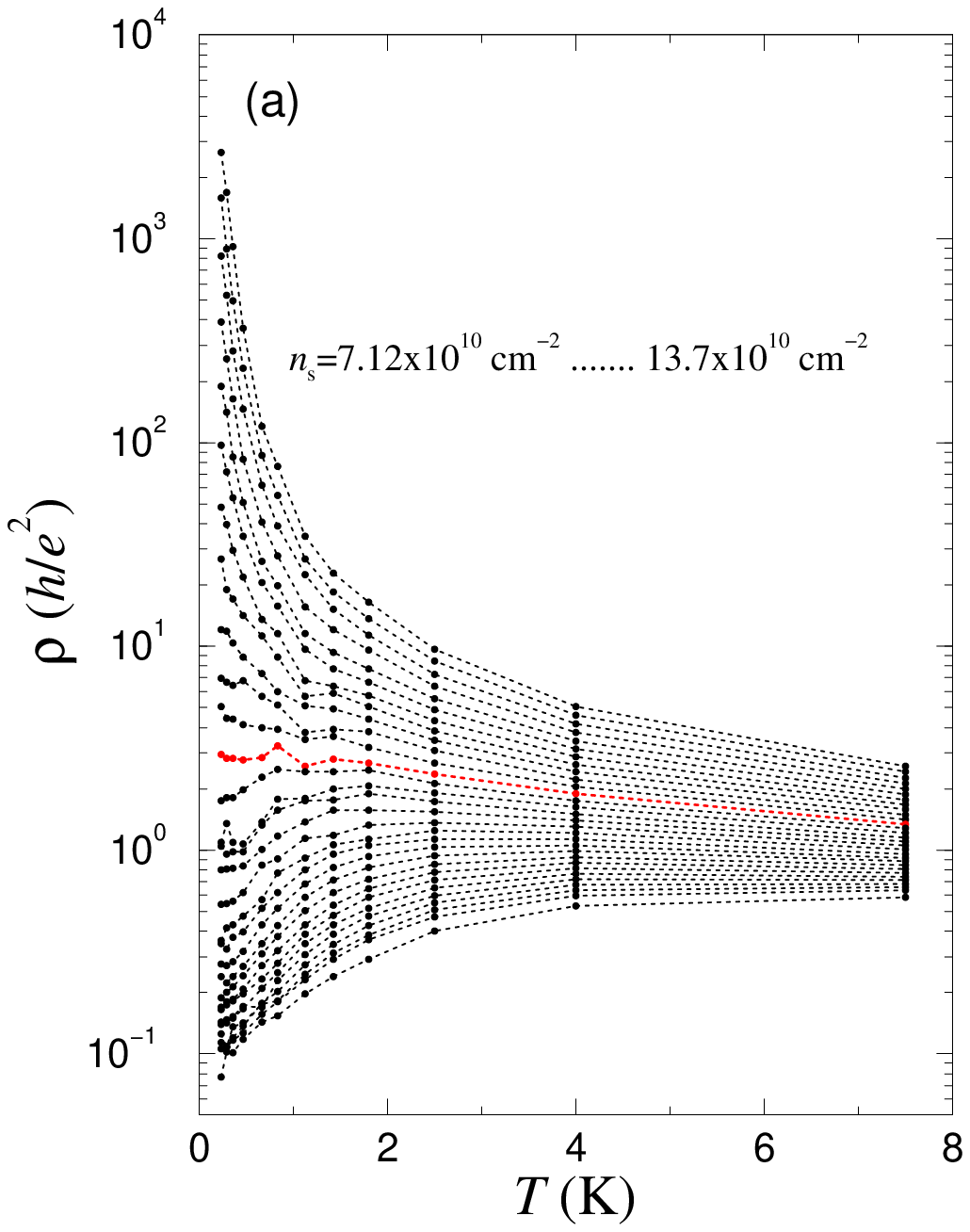}}
\hspace{2mm}
\scalebox{.55}{\includegraphics{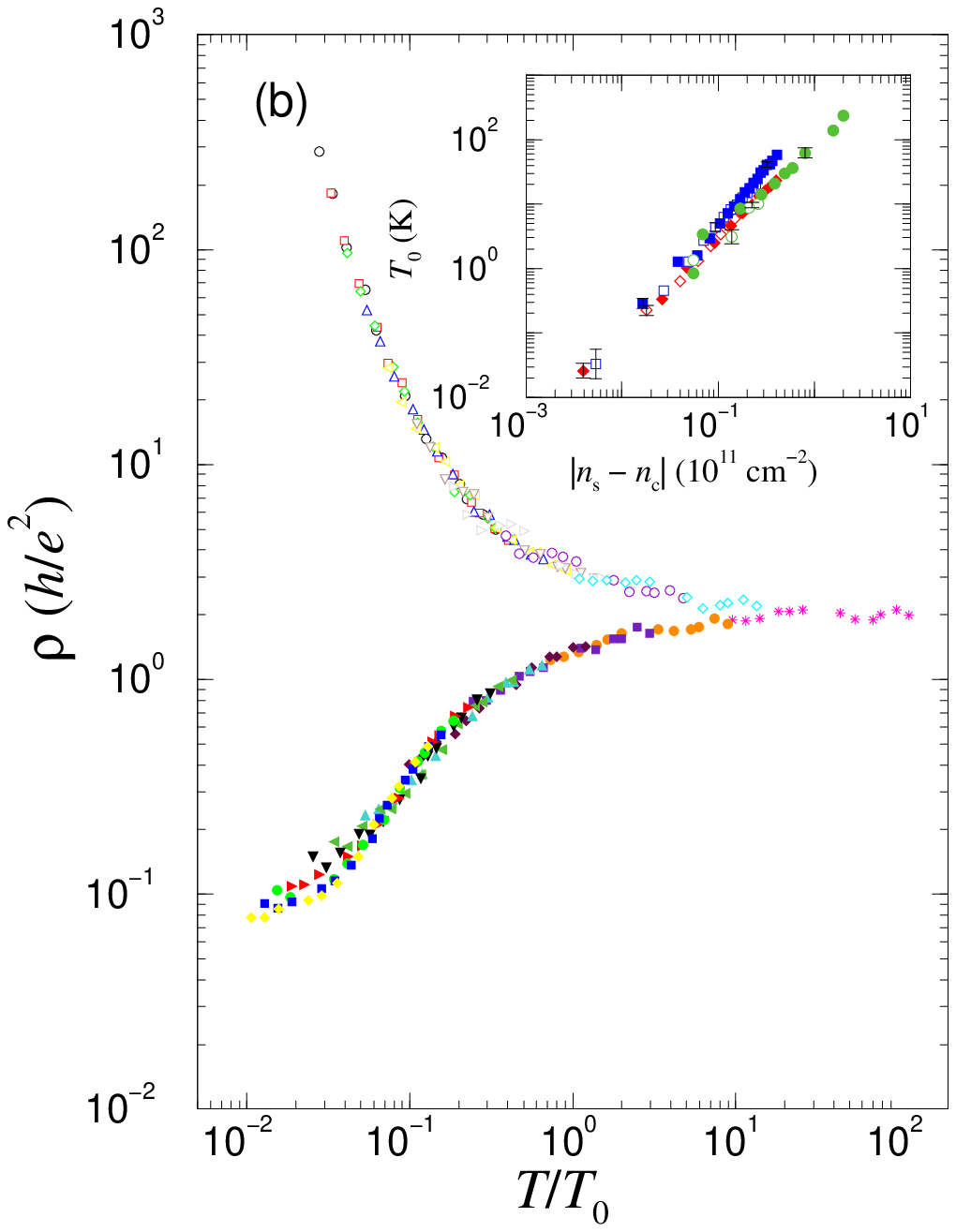}}
\end{center}
\caption{\label{r(t)first} (a):~temperature dependence of the $B=0$ resistivity in a dilute low-disordered Si MOSFET for 30 different electron densities $n_s$ ranging from $7.12$ to $13.7\cdot10^{10}$~cm$^{-2}$.  (b):~resistivity versus $T/T_0$, with $T_0(n_s)$ chosen to yield scaling with temperature.  The inset shows the scaling parameter, $T_0$, versus the deviation from the critical point, $|n_s-n_c|$; data are shown for silicon MOSFETs obtained from three different wafers.  Open symbols correspond to the insulating side and closed symbols to the metallic side of the transition.  From Kravchenko~\etal (1995).}
\end{figure}

Figure~\ref{r(t)first}~(a) shows the temperature dependence of the resistivity measured in units of $h/e^2$ of a high-mobility MOSFET for 30 different electron densities $n_s$ varying from $7.12\cdot10^{10}$ to $13.7\cdot10^{10}$~cm$^{-2}$.  If the resistivity at high temperatures exceeds the quantum resistance $h/e^2$ (the curves above the dashed red line), $\rho(T)$ increases monotonically as the temperature decreases, behaviour that is characteristic of an insulator.  However, for $n_s$ above a certain ``critical'' value, $n_c$ (the curves below the ``critical'' curve that extrapolates to $3h/e^2$ denoted in red), the temperature dependence of $\rho(T)$ becomes non-monotonic: with decreasing temperature, the resistivity first increases (at $T>2$~K) and then starts to decrease.  At yet higher density $n_s$, the resistivity is almost constant at $T>4$~K but drops by an order of magnitude at lower temperatures, showing strongly metallic behaviour as $T\rightarrow0$.

A striking feature of the $\rho(T)$ curves shown in Figure~\ref{r(t)first}~(a) for different $n_s$ is that they can be made to overlap by applying a (density-dependent) scale factor along the $T$-axis.  Thus, the resistivity can be expressed as a function of $T/T_0$ with $T_0$ depending only on $n_s$.  This was demonstrated for several samples over a rather wide range of electron densities (typically $(n_c-2.5\cdot10^{10})<n_s<(n_c+2.5\cdot10^{10}$~cm$^{-2}$)) and in the temperature interval 0.2 to 2~K.  The results of this scaling are shown in Fig.~\ref{r(t)first}~(b), where $\rho$ is plotted as a function of $T/T_0$.  One can see that the data collapse onto two separate branches, the upper one for the insulating side of the transition and the lower one for the metallic side.  The thickness of the lines is largely governed by the noise within a given data set, attesting to the high quality of the scaling.

The procedure used to bring about the collapse and determine $T_0$ for each $n_s$ was the following.  No power law dependence was assumed {\it a priori} for $T_0$ versus $(n_s-n_c)$; instead, the $\rho(T)$ curves were successively scaled along the $T$-axis to coincide: the second ``insulating'' curve from the top was scaled along the $T$-axis to coincide with the top-most curve and the corresponding scaling factor was recorded, then the third, and so on, yielding the upper curve in Fig.~\ref{r(t)first}~(b) (designated by open symbols).  The same procedure was applied on the metallic side of the transition starting with the highest-density curve,  giving the lower curve in Fig.~\ref{r(t)first}~(b) (shown as closed symbols).  A quantitative value was assigned to the scaling factor to obtain $T_0$ for the insulating curves by using the fact that the resistivity of the most insulating (lowest $n_s$) curve was shown (Mason~\etal, 1995) to exhibit the temperature dependence characteristic of hopping in the presence of a Coulomb gap: $\rho=\rho_0\,$exp$\left[\left(T_0/T\right)^{1/2}\right]$ (Efros and Shklovskii 1975).  $T_0$ was determined on the metallic side using the symmetry between metallic and insulating curves, as described in more detail in Kravchenko~\etal (1995).  For all samples studied, this scaling procedure yields a power law dependence of $T_0$ on $|\delta_n|\equiv|n_s-n_c|$ on both sides of the transition: $T_0\propto|\delta_n|^\beta$ with the average power $\beta=1.60\pm0.1$ for the insulating side and $1.62\pm0.1$ for the metallic side of the transition; this common power law can be clearly seen in the inset to Fig.~\ref{r(t)first}~(b), where for each sample the open (insulating side) and filled (metallic side) symbols form a single line.  The same power law was later observed by Popovi\'{c} (1997) in silicon samples from another source, thus establishing its universality and supporting the validity of the scaling analysis.

\begin{figure}
\begin{center}
\scalebox{.3}{\includegraphics[angle=-90]{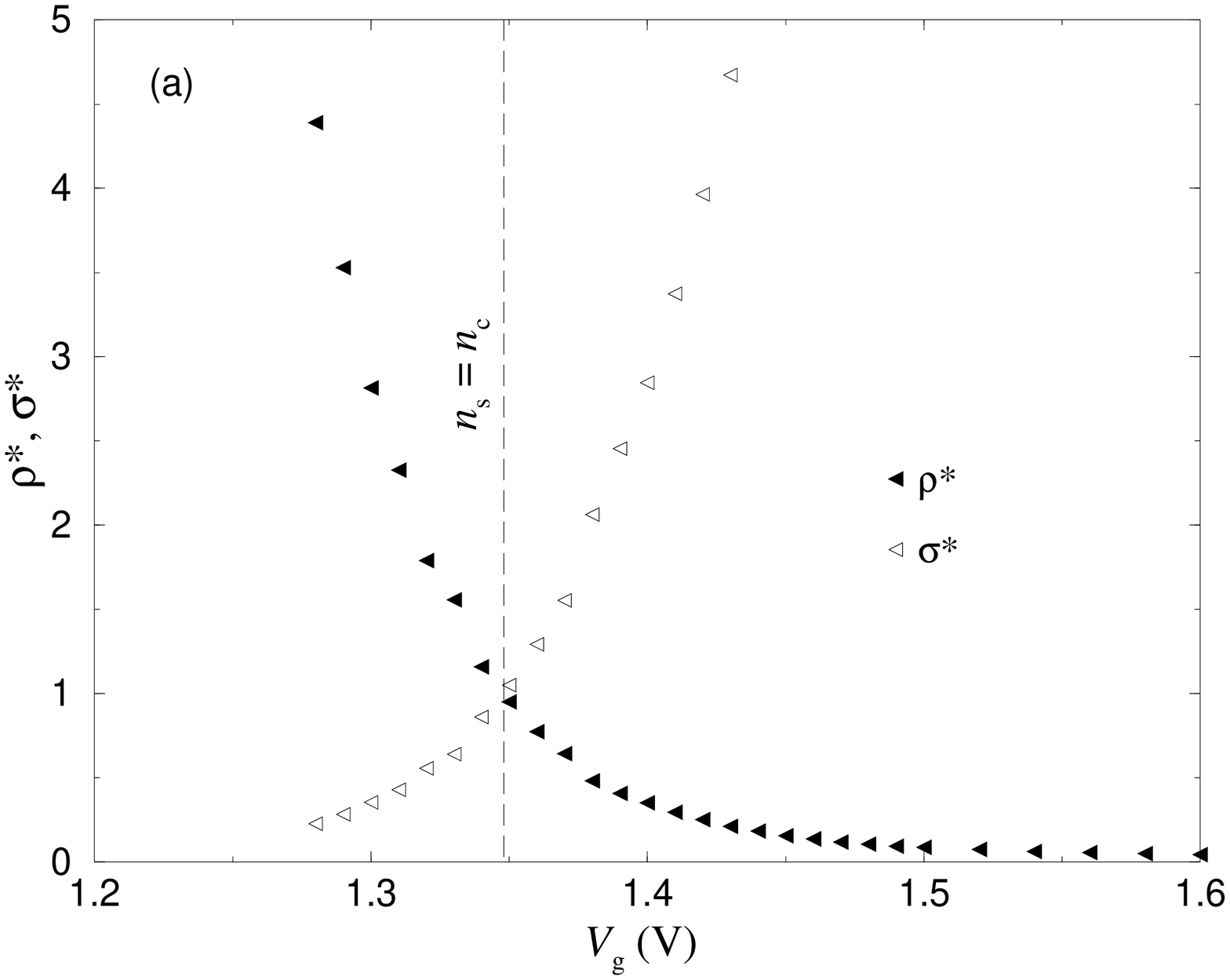}}\hspace{.2cm}
\scalebox{.3}{\includegraphics[angle=-90]{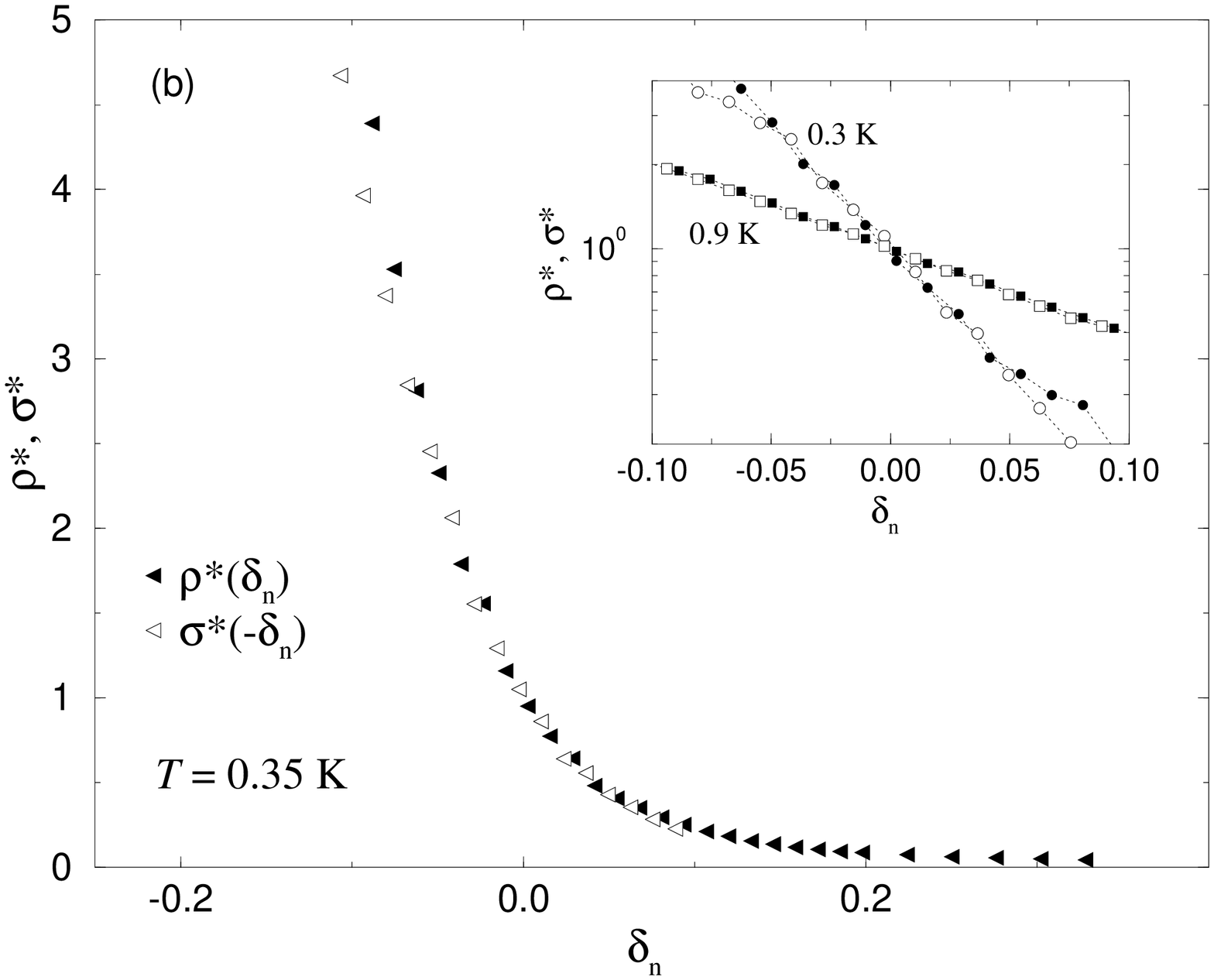}}
\end{center}
\caption{\label{duality} (a)~For a silicon MOSFET, the normalized resistivity, $\rho^*$, and normalized conductivity, $\sigma^*$, as a function of the gate voltage, $V_g$, at $T=0.35$~K.  Note the symmetry about the line $n_s = n_c$. The electron density is given by $n_s=(V_g-0.58$V$)\cdot1.1\cdot 10^{11}$~cm$^{-2}$.  (b)~To demonstrate this symmetry explicitly, $\rho^*(\delta_n)$ (closed symbols) and $\sigma^*(-\delta_n)$ (open symbols) are plotted as a function of $\delta_n\equiv(n_s-n_c)/n_c$. Inset: $\rho^*(\delta_n)$ (closed symbols) and $\sigma^*(-\delta_n)$ (open symbols) versus $\delta_n$ at $T=0.3$~K and $T=0.9$~K, the lowest and highest measured temperatures. From Simonian~\etal (1997a).}
\end{figure}

Simonian~\etal (1997a) noted that the metallic and insulating curves are reflection-symmetric in the temperature range between approximately 300~mK and 2~K. In Fig.~\ref{duality}~(a), the normalized resistivity, $\rho^*\equiv\rho/\rho(n_c)$, is shown as a function of the gate voltage, $V_g$, together with the normalized conductivity, $\sigma^*\equiv1/\rho^*$; the apparent symmetry about the vertical line corresponding to the critical electron density can be clearly seen. Fig.~\ref{duality}~(b) demonstrates that the curves can be mapped onto each other by reflection, {\it i.e.}, $\rho^*(\delta_n)$ is virtually identical to $\sigma^*(-\delta_n)$.  This mapping holds over a range of temperature from $0.3$~K to $0.9$~K; however, the range $|\delta_n|$ over which it holds decreases as the temperature is decreased: for example, at $T=0.9$~K, $\rho^*$ and $\sigma^*$ are symmetric for $|\delta_n|<0.1$, while at $T=0.3$~K, they are symmetric only for $|\delta_n|<0.05$ (see inset to Fig.~\ref{duality}~(b)).  Similar symmetry was later reported by Popovi\'{c}~\etal (1997) and Simmons~\etal (1998).  This implies that there is a simple relation between the mechanism for conduction on opposite sides in the vicinity of the transition; the data bear a strong resemblance to the behaviour found by Shahar~\etal (1996) for the resistivity near the transition between the quantum Hall liquid and insulator, where it has been attributed to charge-flux duality in the composite boson description.  It was argued by Dobrosavljevi{\'c}~\etal (1997) that both the scaling and reflection symmetry are consequences of a simple analysis assuming that a $T=0$ quantum critical point describes the metal-insulator transition.

\subsection{How universal is $\rho(T)$?}

\begin{figure}
\scalebox{.41}{\includegraphics{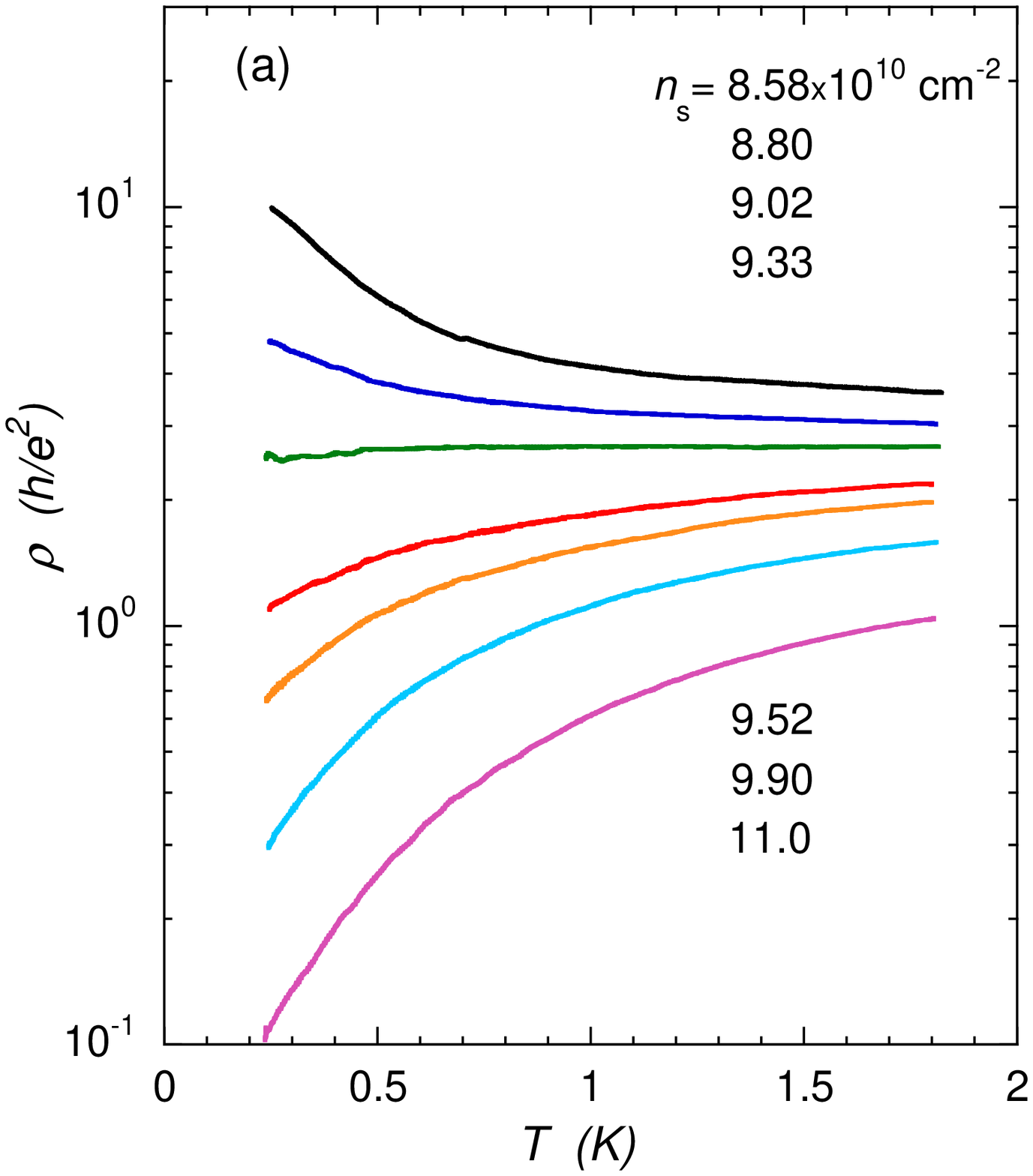}}\hspace{.5cm}
\scalebox{.41}{\includegraphics{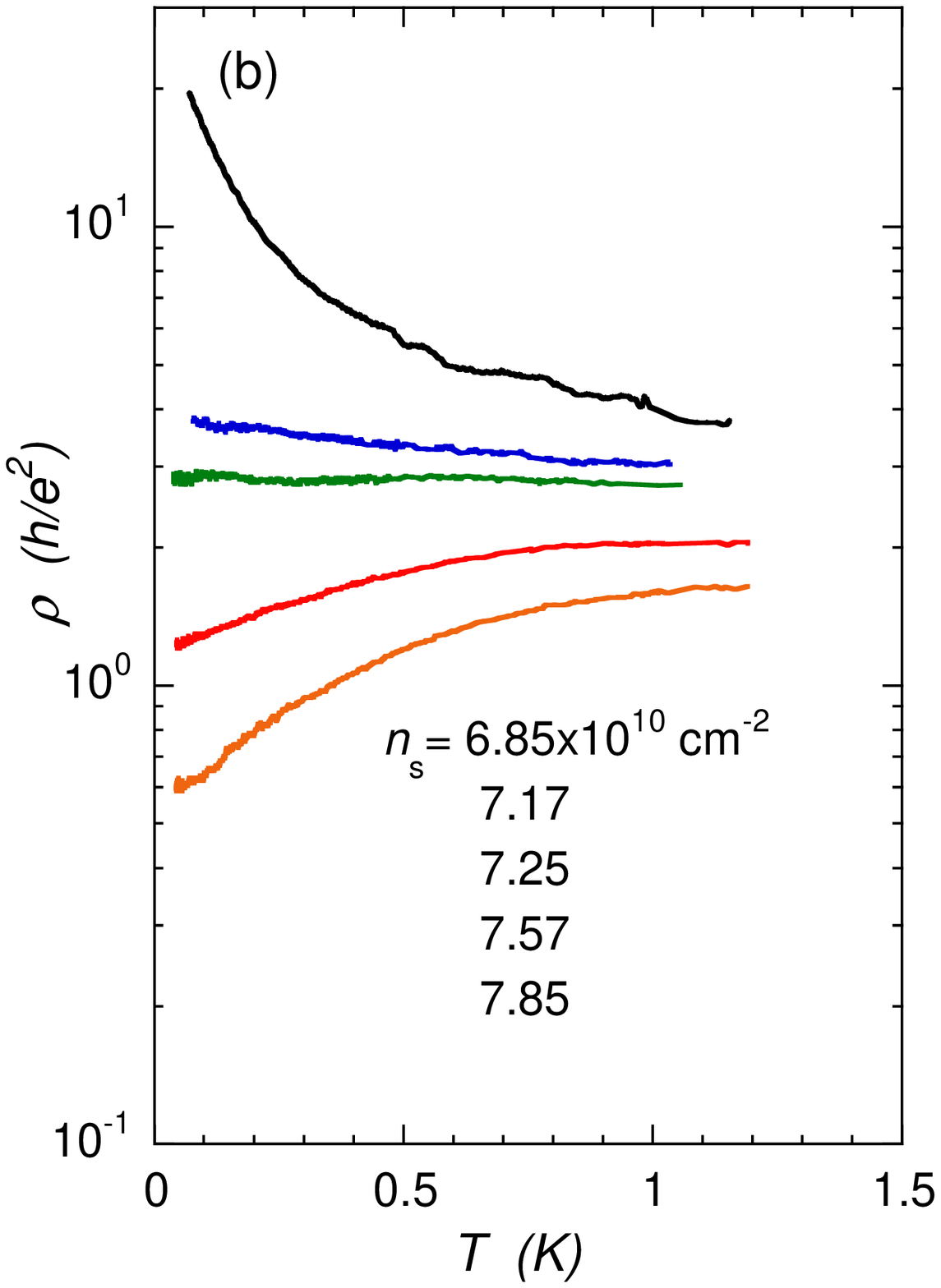}}
\begin{center}
\scalebox{.46}{\includegraphics[width=12cm,height=15cm]{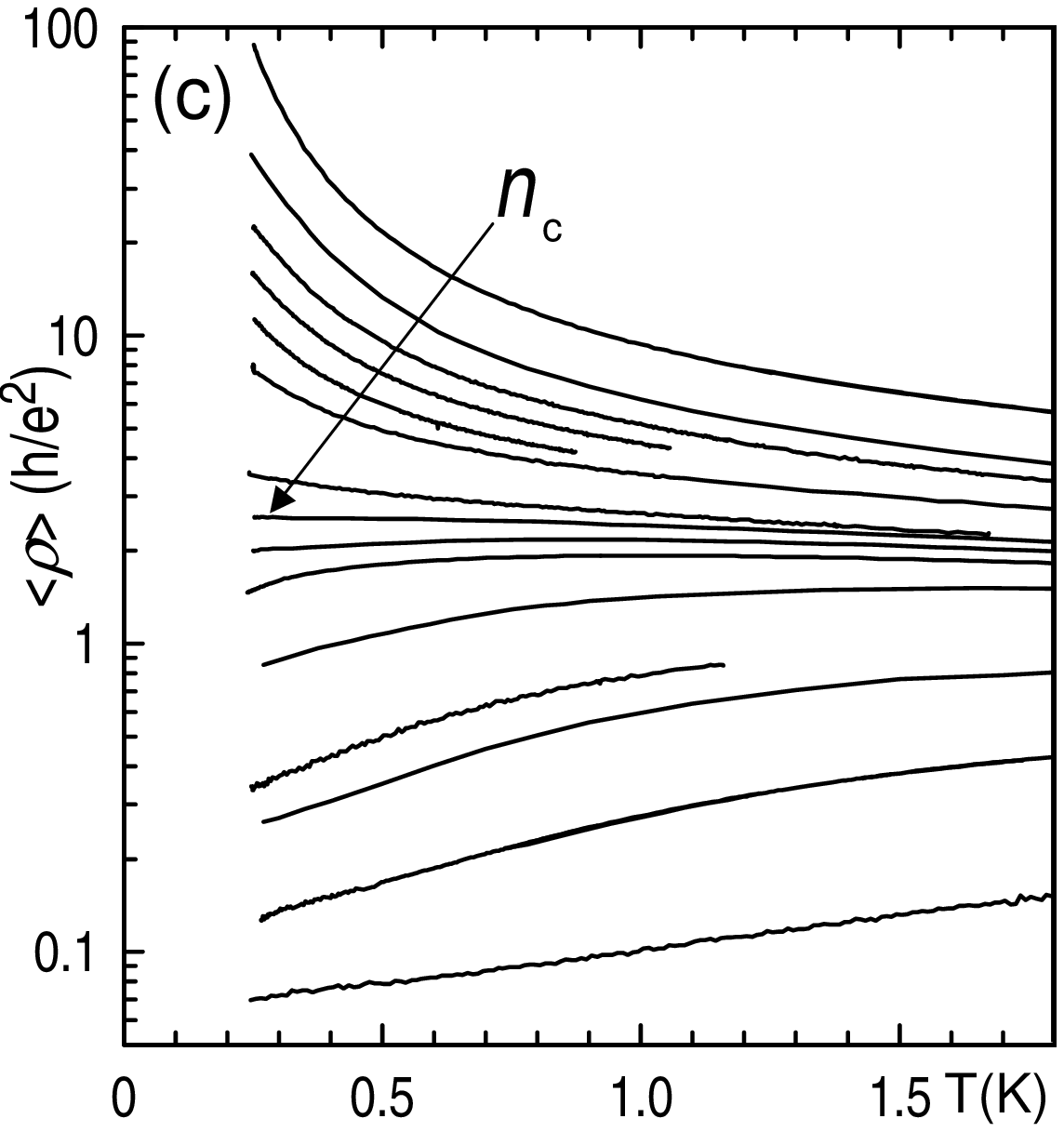}}
\end{center}
\caption{\label{r(t)universal} Universal behaviour in ultra-clean silicon MOSFETs: the resistivity versus temperature in three samples from different sources. Note that the resistivities are essentially the same at the separatrices for all samples even though their critical densities are different. (a)~high-mobility sample provided by V~M Pudalov (graph adopted from Sarachik and Kravchenko 1999), (b)~sample fabricated by Heemskerk and Klapwijk (1998) (adopted from Kravchenko and Klapwijk 2000a) and (c)~sample of Heemskerk and Klapwijk but from a different wafer (graph adopted from Jaroszy\'nski~\etal 2002).  Electron densities in (c) vary from $8.55$ to $14\cdot10^{10}$~cm$^{-2}$ (top to bottom).
}
\end{figure}
\begin{figure}
\begin{center}
\scalebox{.5}{\includegraphics{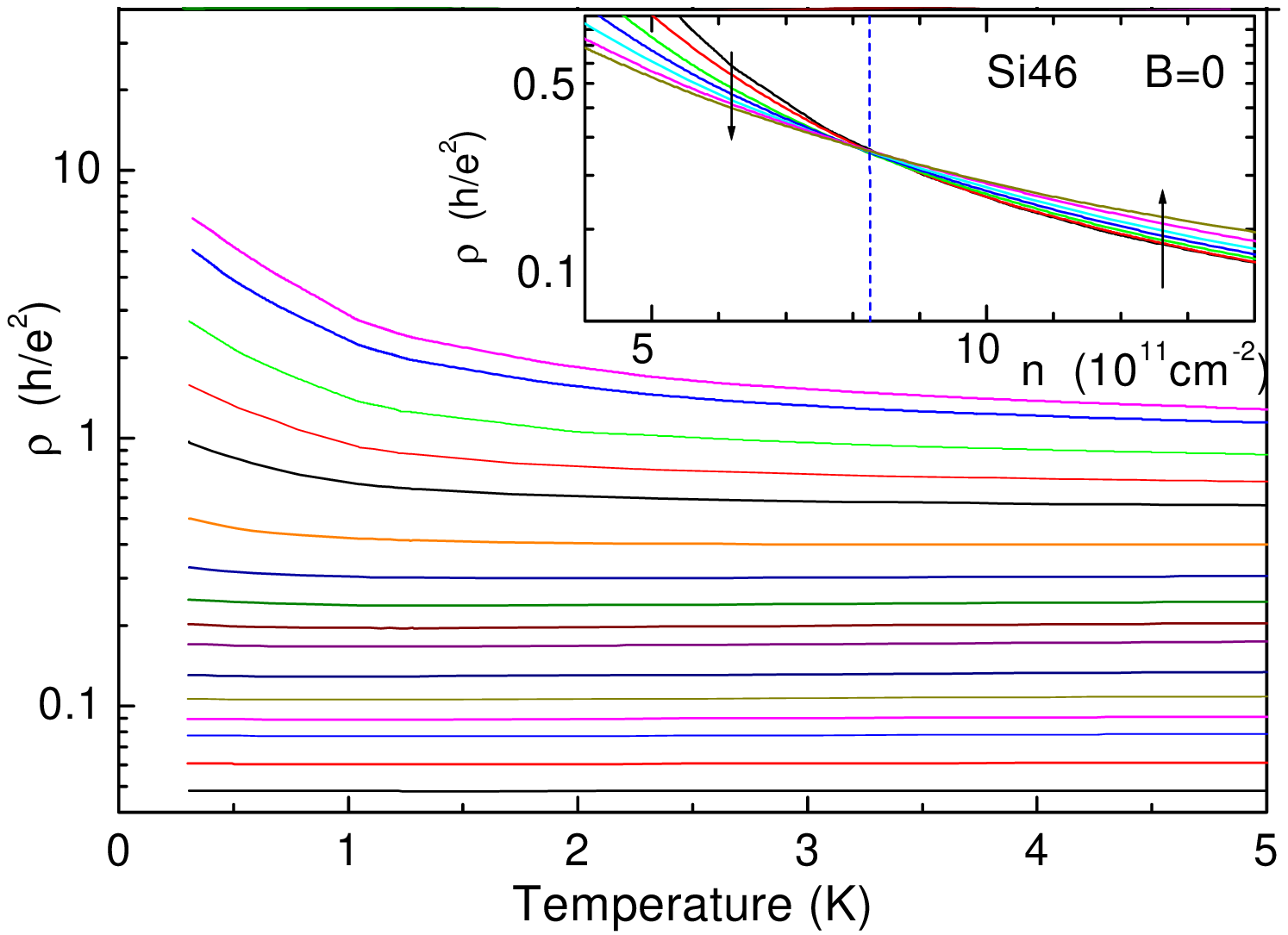}}
\end{center}
\caption{\label{pudalov2001dirty} Resistivity versus temperature of a very disordered silicon MOSFET.  Note that the vertical scale is similar to Fig.~\protect\ref{r(t)universal}.  The electron densities are (in units of $10^{11}$~cm$^{-2}$): 3.85, 4.13, 4.83, 5.53, 6.23, 7.63, 9.03, 10.4, 11.8, 13.2, 16.0, 18.8, 21.6, 24.4, 30.0 and 37.  Adopted from Pudalov~\etal (2001).  Even though there is an apparent crossing point on the $\rho(n_s)$ isotherms (see the inset), the temperature dependence of the resistivity does not resemble the critical behaviour seen in low-disordered samples.}
\end{figure}

The temperature dependence of $\rho(T)$ is very similar for clean silicon MOSFETs.  As an example, Fig.~\ref{r(t)universal} shows the resistivity as a function of temperature of three low-disordered samples obtained from different sources.  The behaviour is quantitatively similar in the critical region in the vicinity of the ``separatrix'', which is the horizontal curve for which the resistivity is independent of temperature. In all samples, the separatrix is remarkably flat at temperatures below 1~K\footnote{At $T>2$~K, the resistivity of the separatrix slowly decreases with increasing temperature, as can be seen in Fig.~\ref{r(t)first}, where $\rho(T)$ curves are shown in a much wider temperature interval.}, and the resistivity is essentially the same numerically at slightly less than $3h/e^2$.  The curves below the separatrix exhibit strongly metallic temperature dependence (${\rm d}\rho/{\rm d}T>0$) with no low-$T$ saturation down to the lowest temperature ($40$~mK or lower; see Fig.\ref{r(t)universal}~(b)); the drop in resistivity reaches as much as a factor of 10 for the bottom curve in Fig.\ref{r(t)universal}~(a).  Alternative methods used to determine the critical electron density in low-disordered samples yield the same value as the density for the separatrix (see sec.~\ref{n_c}).  It is important to note that the separatrix in ultra-clean samples represents the ``upper limit'' of the resistivity for which metallic behaviour (as characterized by ${\rm d}\rho/{\rm d}T>0$) can exist: metallic $\rho(T)$ has never been observed in any 2D samples at resistivities above $\approx3h/e^2$, in quantitative agreement with the predictions of the renormalization group theory (see sec.\ref{sec:rg}).

In more disordered samples, however, the behaviour of the resistivity is very different.  Even though the $\rho(n_s)$ isotherms apparently cross at some electron density (see the inset to Fig.\ref{pudalov2001dirty}), the temperature dependence of the resistivity does not resemble the critical behaviour seen in low-disordered samples. An example of $\rho(T)$ curves in a disordered sample is shown in Fig.\ref{pudalov2001dirty}.  The sample is insulating at electron densities up to $\sim8\cdot10^{11}$~cm$^{-2}$ which is an order of magnitude higher than the critical density in low-disordered samples.  The metallic temperature dependence of the resistance visible at higher $n_s$ does not exceed a few percent (compared to a factor of 10 in low-disordered samples).  Most importantly, the density corresponding to the crossing point shown in the inset to Fig.\ref{pudalov2001dirty} does not coincide with the critical density determined by other methods discussed in the next subsection.

\begin{figure}\hspace{-1mm}
\scalebox{.37}{\includegraphics{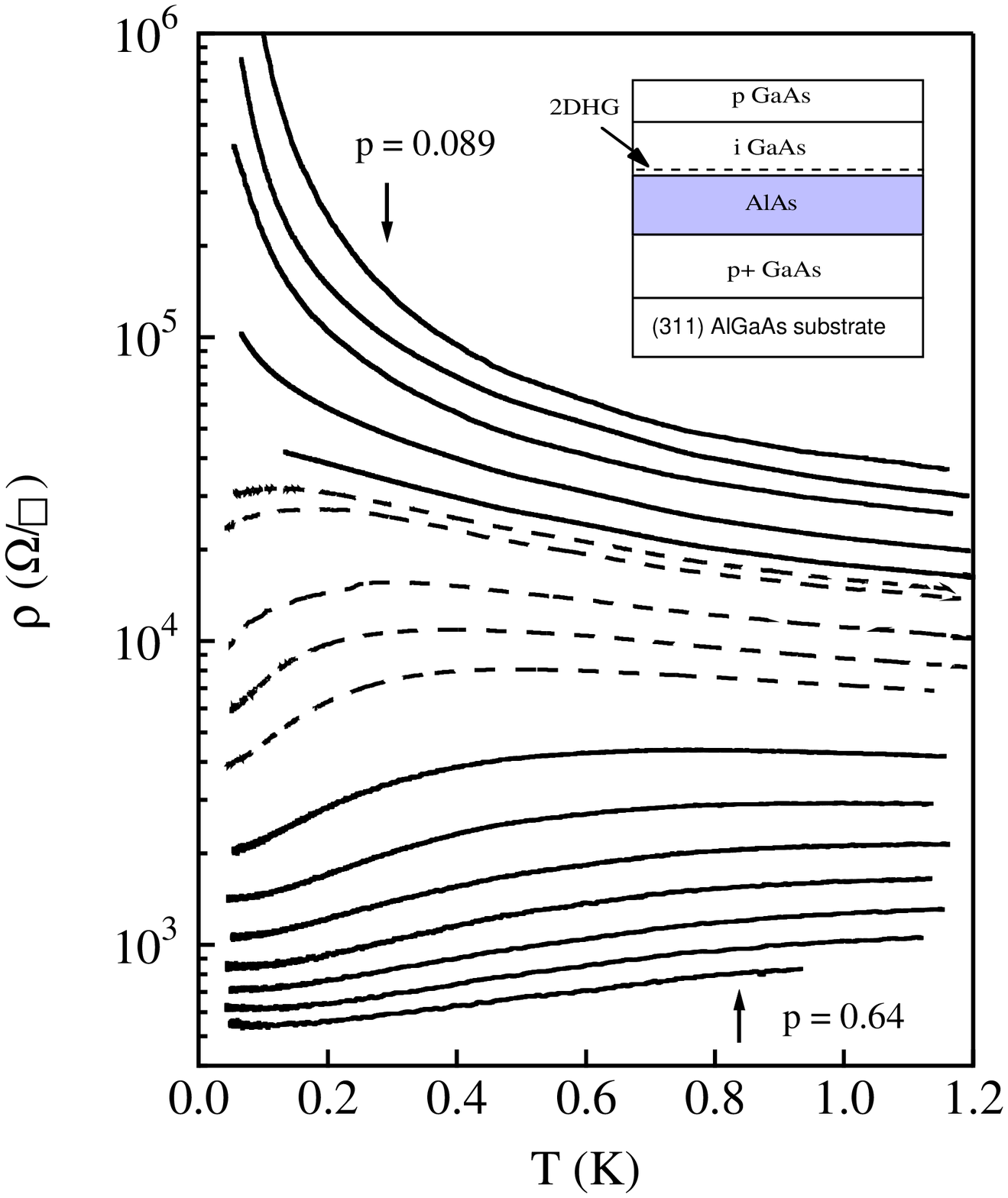}}
\scalebox{.412}{\includegraphics{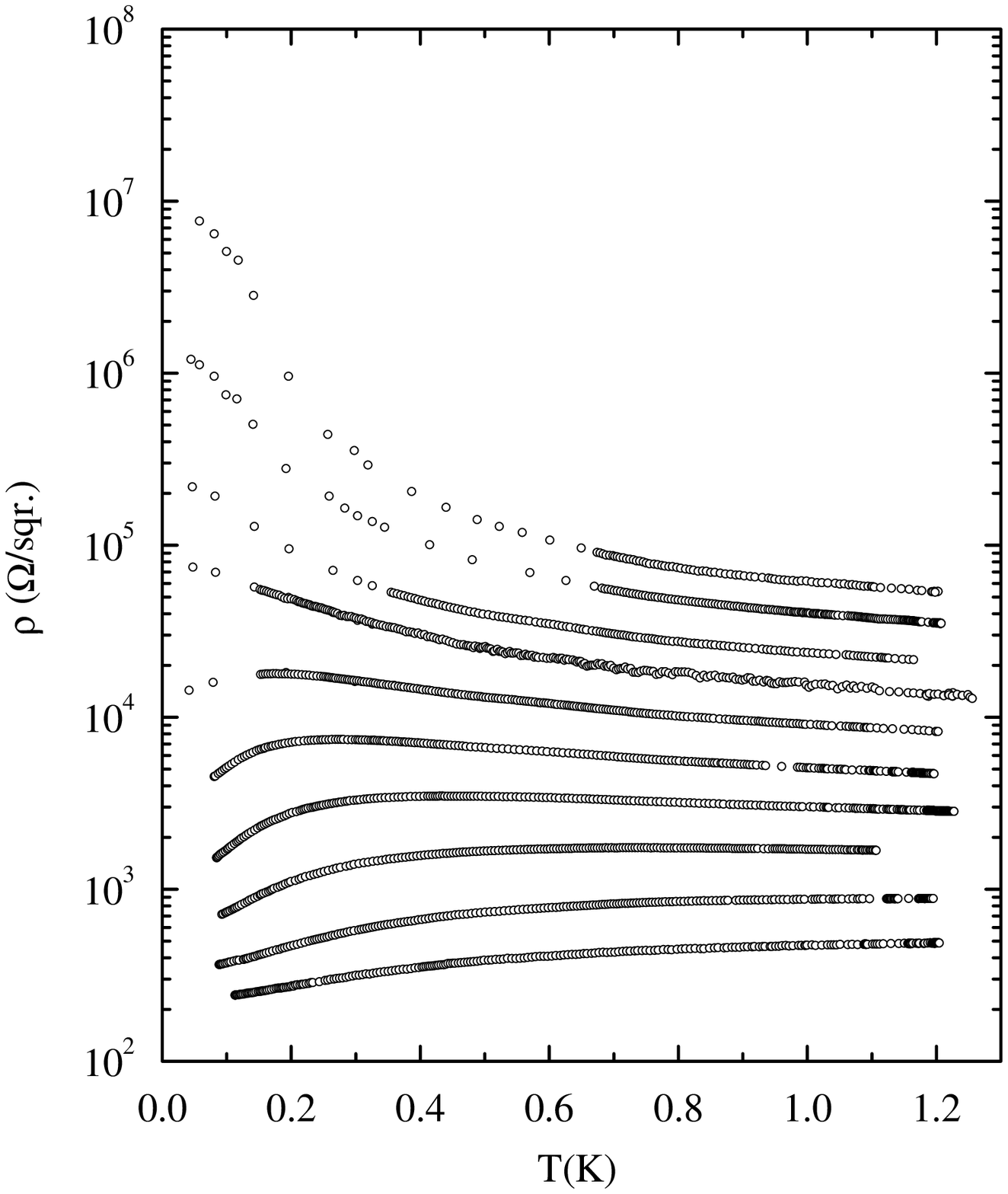}}
\caption{\label{r(t)gaas} For low-disordered 2D hole systems in $p$-GaAs/AlGaAs, the resistivity per square is shown as a function of temperature for $B=0$ at various fixed hole densities, $p$.  Left hand panel: ISIS (inverted semiconductor-insulator-semiconductor) structure with hole densities (from top to bottom) $p=$~0.89, 0.94, 0.99, 1.09, 1.19, 1.25, 1.30, 1.50, 1.70, 1.90, 2.50, 3.20, 3.80, 4.50, 5.10, 5.70 and 6.40$\cdot$10$^{10}$~cm$^{-2}$.  The inset shows a schematic diagram of the ISIS structure: The carriers are accumulated in an undoped GaAs layer situated on top of an undoped AlAs barrier, grown over a $p^+$ conducting layer which serves as a back-gate; the hole density, $p$, is varied by applying a voltage to the back gate.  From Hanein~\etal (1998a). Right hand panel: Temperature dependence of $\rho$ in an ultra high mobility $p$-type GaAs/AlGaAs heterostructure at $p=$~0.48, 0.55, 0.64, 0.72, 0.90, 1.02, 1.27, 1.98, 2.72 and 3.72$\cdot10^{10}$~cm$^{-2}$ (from top to bottom). From Yoon~\etal (1999).}
\end{figure}

A metal-insulator transition similar to that seen in clean silicon MOSFETs has also been observed in other low-disordered, strongly-interacting 2D systems: $p$-type SiGe heterostructures (Coleridge~\etal 1997), GaAs/AlGaAs heterostructures (Hanein~\etal 1998a; Yoon~\etal 1999; Mills~\etal 1999; Noh~\etal 2002 and others) and AlAs heterostructures (Papadakis and Shayegan 1998).  It is difficult to make a direct comparison of the resistivity observed in different material systems because the temperature scales are different, since the Coulomb and Fermi energies depend on the effective mass and carrier density.  For example, the characteristic temperature below which the metallic decrease in the resistivity occurs in $p$-type GaAs/AlGaAs samples is about ten times smaller than in silicon MOSFETs.  On the other hand, the values of the resistivity are quite similar in the two systems.  In Fig.~\ref{r(t)gaas}, the resistivity is plotted as a function of temperature for two $p$-type GaAs/AlGaAs samples produced using different technologies.  The data shown in the left hand panel were obtained by Hanein~\etal (1998a) for an inverted semiconductor-insulator-semiconductor (ISIS) structure with maximum mobility of $\mu_{\max}=1.5\cdot10^5$~cm$^{2}/$Vs, while the right-hand panel shows $\rho(T)$ measured by Yoon~\etal (1999) on a $p$-type GaAs/AlGaAs heterostructure with peak mobility by a factor of five higher ($7\cdot10^{5}$~cm$^{2}$/Vs).  The interaction parameter, $r_s$, changes between approximately 12 and 32 for the left hand plot and from 16 to 44 for the right hand plot\footnote{These $r_s$ values were calculated assuming that the effective mass is independent of density and equal to $0.37\,m_e$, where $m_e$ is the free-electron mass.}.  In spite of the difference in the sample quality and range of densities, the dependence of $\rho(T)$ on temperature is almost the same for the two samples.  The main features are very similar to those found in silicon MOSFETs: when the resistivity at ``high'' temperatures exceeds the quantum resistance, $h/e^2$ ({\it i.e.}, at hole densities below some critical value, $p_c$), the $\rho(T)$ curves are insulating-like in the entire temperature range; for densities just above $p_c$, the resistivity shows insulating-like behaviour at higher temperatures and then drops by a factor of 2 to 3 at temperatures below a few hundred mK; and at yet higher hole densities, the resistivity is metallic in the entire temperature range.  Note that the curves that separate metallic and insulating behaviour have resistivities that increase with decreasing temperature at the higher temperatures shown; this is quite similar to the behaviour of the separatrix in silicon MOSFETs when viewed over a broad temperature range (see Fig.~\ref{r(t)first}~(a)).  However, below approximately 150~mK, the separatrix in $p$-type GaAs/AlGaAs heterostructures is independent of temperature (Hanein~\etal, 1998b), as it is in Si MOSFETs below approximately 2~K.  The resistivity of the separatrix in both systems extrapolates to $2-3\,h/e^2$ as $T\rightarrow0$, even though the corresponding carrier densities are very different.

\begin{figure}\hspace{2cm}
\scalebox{.9}{\includegraphics{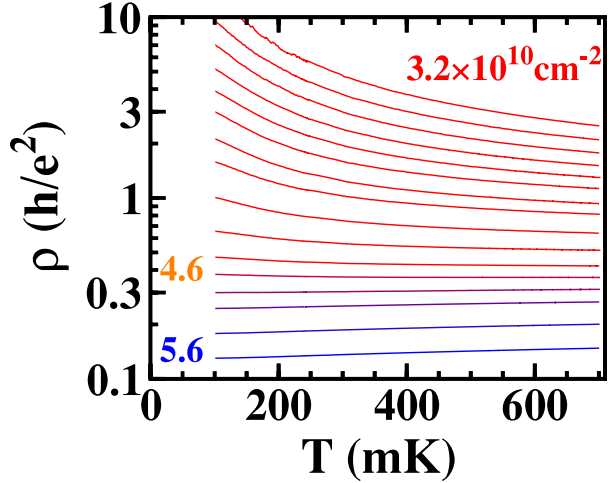}}
\caption{\label{simmons2000dirty} The resistivity as a function of temperature for a disordered $p$-type GaAs/AlGaAs heterostructure at hole densities $p=3.2-5.6\cdot10^{10}$~cm$^{-2}$. From Simmons~\etal (2000).}
\end{figure}

As in the case of highly disordered silicon MOSFETs, no critical behaviour of resistance is observed in disordered GaAs/AlGaAs heterostructures.  An example is shown in Fig.~\ref{simmons2000dirty} where the temperature dependence of the resistivity at $B=0$ is plotted for hole densities $p=3.2-5.6{\cdot}10^{10}$ cm$^{-2}$. Monotonic localized behaviour is observed even when the ``high-temperature'' resistivity lies well below $h/e^2$, at carrier densities up to $4.6{\cdot}10^{10}$~cm$^{-2}$; both samples shown in the previous figure would be in the deeply metallic regime at this hole density.  Above this density, the decrease in resistivity with decreasing temperature is very small (about 10\% in the temperature interval 0.7 to 0.1~K).

As indicated by the experimental results presented above, the $\rho(T)$ curves are nearly universal in the vicinity of the metal-insulator transition, but only in samples with very weak disorder potential. The strength of the disorder is usually characterized by the maximum carrier mobility, $\mu_{\max}$.  In general, the higher the maximum mobility ({\it i.e}.\ the lower the disorder), the lower the carrier density at which the localization transition occurs.  This was shown empirically by Sarachik (2002) to hold over a broad range (five decades in density) for all materials studied: the critical density follows a power law dependence on peak mobility (or scattering rate).  However, the data exhibit some scatter, and the correlation is not exact.  Thus, for example, the peak hole mobilities in samples used by Hanein~\etal (1998a) and Simmons~\etal (1998) are similar, while the localization transition occurs in the latter at a value of $p$ several times higher than the former.  This may be due to sample imperfections ({\it e.g.}, a slightly inhomogeneous density distribution), which are important at low carrier densities, while the maximum mobility is reached at relatively high carrier densities and may therefore be relatively insensitive to such effects.

A better indicator of the strength of the disorder potential near the MIT is how low the carrier density is at which the localization transition occurs.  In silicon MOSFETs, the experimental results obtained to date suggest that the resistivity near the transition approaches universal behaviour for samples in which the transition to a strongly localized state occurs at $n_s<1\cdot 10^{11}$~cm$^{-2}$. (In $p$-type GaAs/AlGaAs heterostructures, the corresponding density separating universal and non-universal behaviour appears to be about an order of magnitude lower, although the data are currently insufficient to determine the value reliably.)  Below, we will argue that the ``universal'' metal-insulator transition in very clean samples is not driven by disorder but by some other mechanism, possibly of magnetic origin.  In contrast, the transition is not universal in more disordered samples and is presumably due to Anderson localization, which is strong enough to overpower the metallic behaviour at low densities.

\subsection{Critical density}\label{n_c}

\begin{figure}
\begin{center}
\scalebox{.5}{\includegraphics{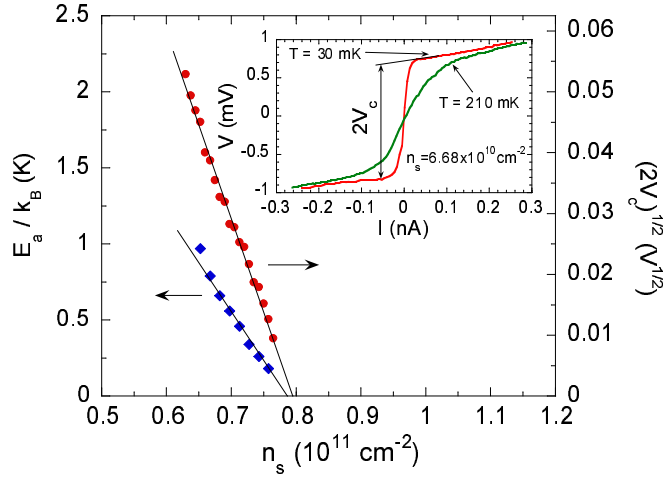}}
\end{center}
\caption{\label{mobility_edge} Activation energy (diamonds) and square root of the threshold voltage (circles) versus electron density in zero magnetic field in a low-disordered silicon MOSFET. The inset shows current-voltage characteristics recorded at $\approx30$ and $210$~mK, as labelled; note that the threshold voltage is essentially independent of temperature. From Shashkin~\etal (2001b).}
\end{figure}

To verify whether or not the separatrix corresponds to the critical density, an independent determination of the critical point is necessary: comparison of values obtained using different criteria provides an experimental test of whether or not a true MIT exists at $B=0$. One obvious criterion, hereafter referred to as the ``derivative criterion'', is a change in sign of the temperature derivative of the resistivity, ${\rm d}\rho/{\rm d}T$; this is the criterion often used to identify the MIT.  A positive (negative) sign of the derivative at the lowest achievable temperatures is empirically associated with a metallic (insulating) phase.  A weakness of this criterion is that it requires extrapolation to zero temperature.  A second criterion can be applied based on an analysis of a temperature-independent characteristic, namely, the localisation length $L$ extrapolated from the insulating phase.  These two methods have been applied to low-disordered silicon MOSFETs by Shashkin~\etal (2001b) and Jaroszy\'nski~\etal (2002).

\begin{figure}
\begin{center}
\scalebox{.45}{\includegraphics{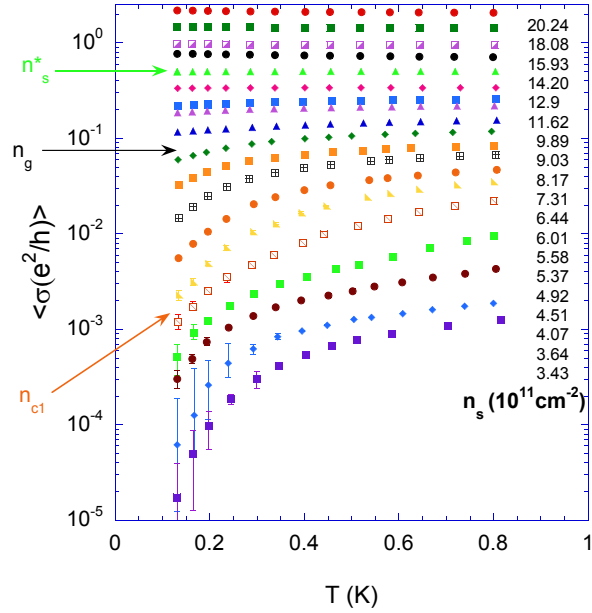}}
\end{center}
\caption{\label{popovic2002dirty} Conductivity versus temperature for different $n_s$ in a highly disordered silicon MOSFET.  (The error bars show the size of the fluctuations of the conductivity with time.)  The critical electron densities obtained by the derivative and localization length criteria ($n_{s}^{\ast}$ and $n_{c1}$, correspondingly) are marked by arrows.  $n_g$ is the density corresponding to the onset of glassy behaviour; see text.  The figure is adopted from Bogdanovich and Popovi\'{c} (2002).}
\end{figure}

As mentioned earlier, the temperature dependence of the resistance deep in the insulating phase obeys the Efros-Shklovskii variable-range hopping form (Mason~\etal 1995); on the other hand, closer to the critical electron density at temperatures that are not too low, the resistance has an activated form $\rho \propto e^{E_a/k_BT}$ (Pepper~\etal 1974; Pudalov~\etal 1993; Shashkin~\etal 1994) due to thermal activation to the mobility edge.  Fig.~\ref{mobility_edge} shows the activation energy $E_a$ as a function of the electron density (diamonds); the data can be approximated by a linear function which yields, within the experimental uncertainty, the same critical electron density as the ``derivative criterion''.

The critical density can also be determined by studying the nonlinear current-voltage $I-V$ characteristics on the insulating side of the transition.  A typical low-temperature $I-V$ curve is close to a step-like function: the voltage rises abruptly at low current and then saturates, as shown in the inset to Fig.~\ref{mobility_edge}; the magnitude of the step is $2\, V_c$.  The curve becomes less sharp at higher temperatures, yet the threshold voltage, $V_c$, remains essentially unchanged.  Closer to the MIT, the threshold voltage decreases, and at $n_s=n_{c1}=0.795\cdot10^{11}$~cm$^{-2}$, the $I-V$ curve is strictly linear (Shashkin~\etal 2001b).  According to Polyakov and Shklovskii (1993) and Shashkin~\etal (1994), the breakdown of the localized phase occurs when the localized electrons at the Fermi level gain enough energy to reach the mobility edge in an electric field, $V_c/d$, over a distance given by the localization length, $L$, which is temperature-independent:
$$eV_c(n_s)\; L(n_s)/d=E_a(n_s)$$
(here $d$ is the distance between the potential probes). The dependence of $V_c^{1/2}(n_s)$ near the MIT is linear, as shown in Fig.~\ref{mobility_edge} by closed circles, and its extrapolation to zero threshold value again yields approximately the same critical electron density as the two previous criteria.  The linear dependence $V_c^{1/2}(n_s)$, accompanied by linear $E_a(n_s)$, signals the localization length diverging near the critical density: $L(n_s)\propto1/(n_c-n_s)$.

These experiments indicate that in low-disordered samples, the two methods --- one based on extrapolation of $\rho(T)$ to zero temperature and a second based on the behaviour of the temperature-independent localization length --- give the same critical electron density: $n_c~\approx n_{c1}$. This implies that the separatrix remains ``flat'' (or extrapolates to a finite resistivity) at zero temperature.  Since one of the methods is independent of temperature, this equivalence supports the existence of a true $T=0$ MIT in low-disordered samples in zero magnetic field.

In contrast, we note that in highly-disordered samples, the localization length method yields a critical density noticeably lower than the derivative criterion.  Figure~\ref{popovic2002dirty} shows the (time-averaged) conductivity $\langle\sigma\rangle$ as a function of $T$ for different $n_s$.  $d\langle\sigma\rangle/dT$ changes sign at electron density $n_s^\ast=12.9\cdot10^{11}$~cm$^{-2}$.  The activation energy, however, vanishes at $n_{c1}\approx5.2\cdot10^{11}$~cm$^{-2}$, which is more than a factor of two lower than $n_s^\ast$: at densities between these two values, the resistivity does not diverge as $T\rightarrow0$, even though it exhibits an insulating-like temperature dependence. Thus, these two different methods yield different ``critical densities'' for a sample with strong disorder.  Moreover, from the study of low-frequency resistance noise in dilute silicon MOSFETs, Bogdanovich and Popovi\'{c} (2002) and Jaroszy\'nski~\etal (2002) have found that the behaviour of several spectral characteristics indicates a dramatic slowing down of the electron dynamics at a well-defined electron density $n_g$, which they have interpreted as an indication of a (glassy) freezing of the 2D electron system.  In low-disordered samples, $n_g$ nearly coincides with $n_c$, while in highly-disordered sample, $n_g$ lies somewhere between $n_{c1}$ and $n_s^*$, as indicated in Fig.~\ref{popovic2002dirty}.  The width of the glass phase ($n_{c1}<n_s<n_g$) thus strongly depends on disorder, becoming extremely narrow (or perhaps even vanishing) in low-disordered samples. The strong dependence on disorder of the width of the metallic glass phase is consistent with predictions of the model of interacting electrons near a disorder-driven metal-insulator transition (Pastor and Dobrosavljevi\'{c} 1999).  These observations all suggest that the origin of the metal-insulator transition is different in clean and strongly-disordered samples.

\subsection{Does weak localization survive in the presence of strong interactions?}

\begin{figure}\vspace{0.5cm}\hspace{2.4cm}
\scalebox{.5}{\includegraphics{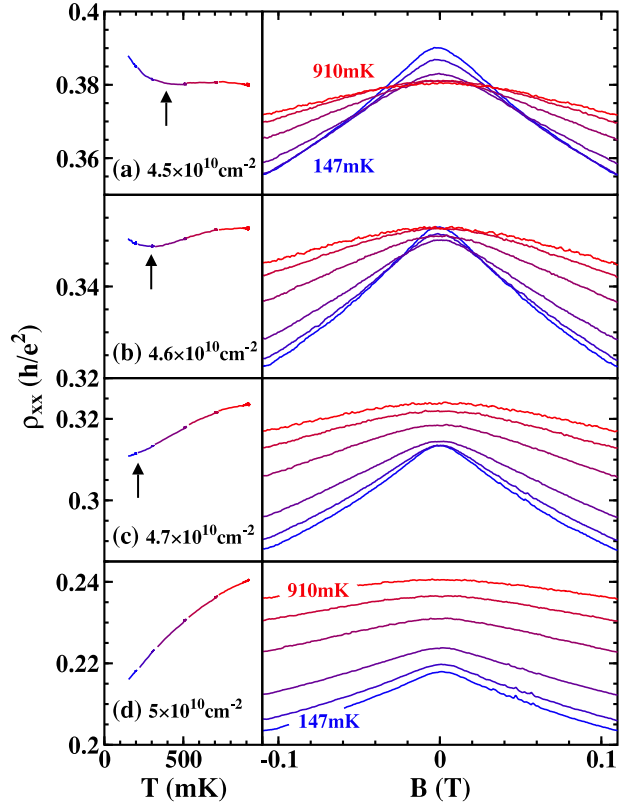}}
\caption{\label{simmons_R(T)} (a-d) The left hand panels show data for the resistivity versus temperature at $B=0$ in a disordered $p$-type GaAs/AlGaAs quantum well, illustrating the transition from insulating to metallic behaviour as the density increases. The right hand panels show the corresponding magnetoresistance traces for temperatures of 147, 200, 303, 510, 705, and 910~mK.  From Simmons~\etal (2000).}
\end{figure}

The theory of weak localization was developed for noninteracting systems, and it was not {\it a priori} clear whether it would work in the presence of strong interactions.  In 2000, Simmons~\etal studied transport properties of a dilute modulation-doped $p$-type GaAs/AlGaAs quantum well and observed a temperature-dependent negative magnetoresistance, consistent with the suppression of the coherent backscattering by the perpendicular magnetic field.  Magnetoresistance curves obtained by Simmons~\etal are plotted in the right hand panel of Fig.~\ref{simmons_R(T)}.  A characteristic peak develops in the resistivity at $B=0$ as the temperature is decreased, signalling that the weak localization is still present at $p$ as low as $4.5 \cdot 10^{10}$~cm$^{-2}$, corresponding to the interaction parameter $r_s\sim15$.  Simmons~\etal successfully fitted their magnetoresistance data by the Hikami-Larkin formula (Hikami~\etal 1980)
\begin{equation}
\Delta\sigma=-\frac{e^2}{\pi
h}\;\left[\Psi\left(\frac{1}{2}+\frac{\tau_B}{\tau}\right)
-\Psi\left(\frac{1}{2}+\frac{\tau_B}{\tau_\phi}\right)\right],\label{HL}
\end{equation}
and obtained reasonable values of the phase-breaking time, $\tau_\phi\sim10$ to $30$~ps, in the temperature interval 1 to 0.15~K (here $\tau$ is the elastic scattering time, $\Psi$ is the Digamma function, $\tau_B=\hbar/4eB_\perp D$, and $D$ is the diffusion coefficient).  Weak negative magnetoresistance was also observed by Brunthaler~\etal (2001) in silicon MOSFETs at electron densities down to $1.5 \cdot 10^{11}$~cm$^{-2}$, although they found values of $\tau_\phi$ that were about an order of magnitude shorter than those expected theoretically, $\tau_{\phi}\sim\sigma\hbar^2/e^2k_{B}T$.

\begin{figure}\vspace{2mm}\hspace{.5cm}
\scalebox{.52}{\includegraphics{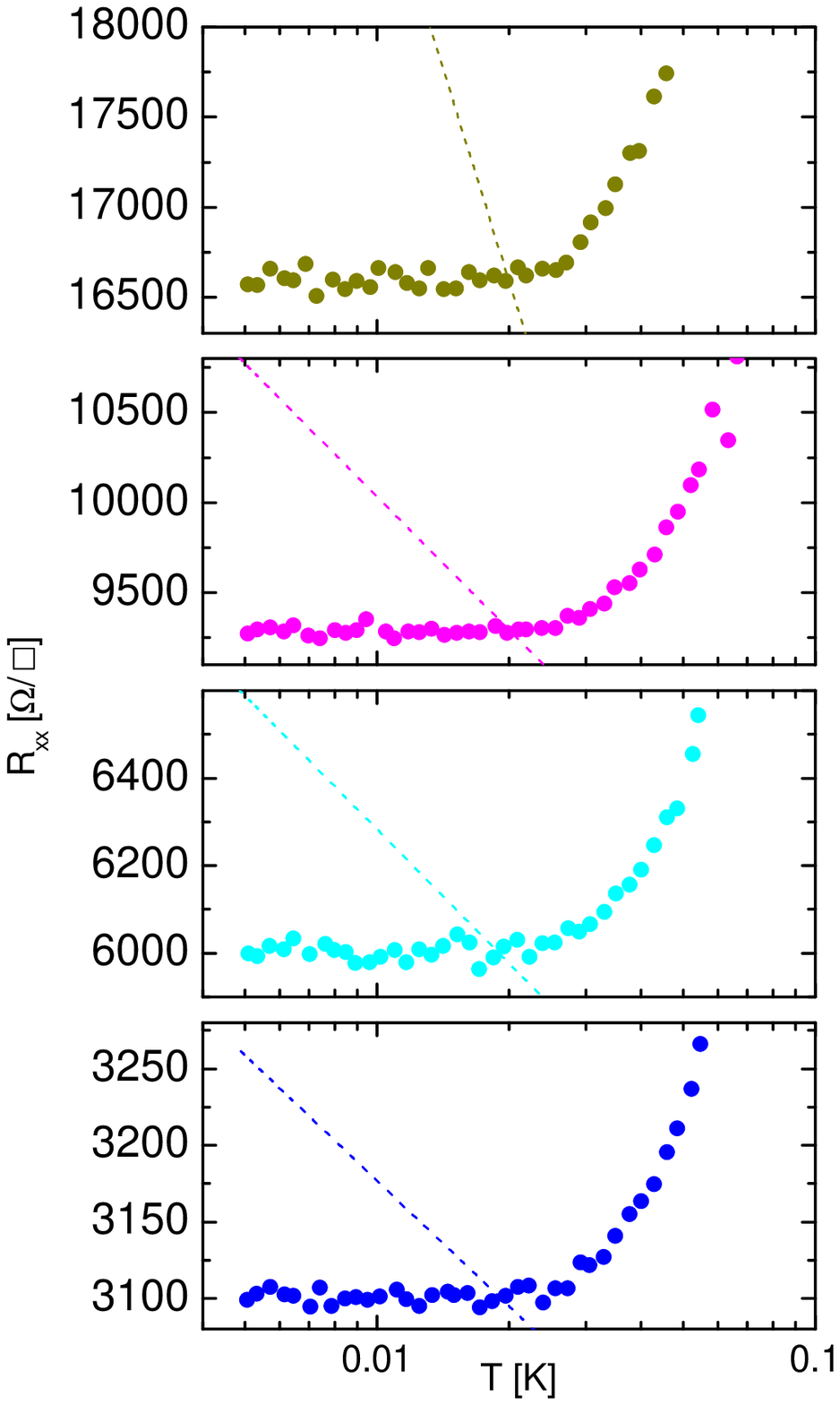}}
\scalebox{.627}{\includegraphics{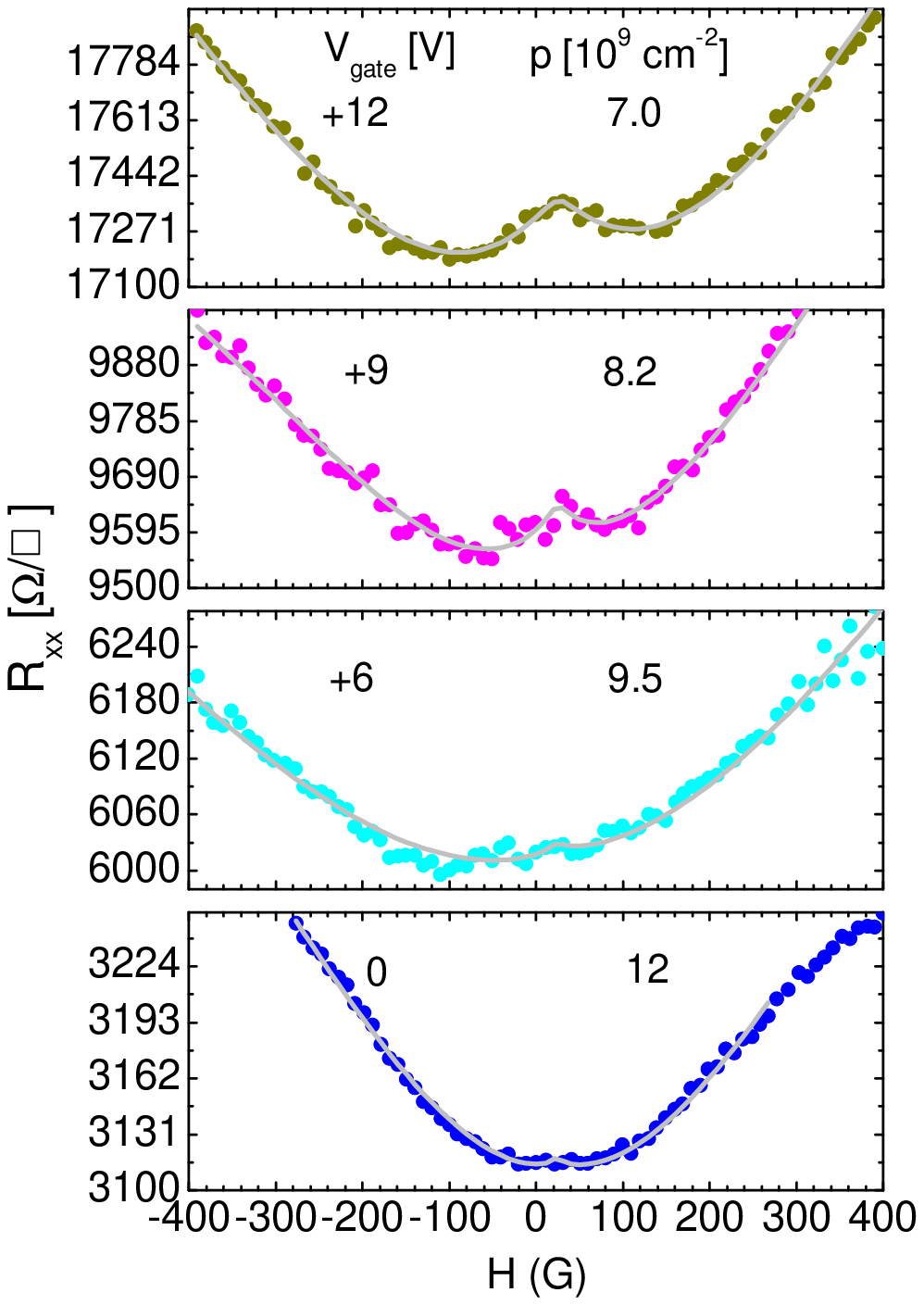}}
\caption{\label{mills_R(T)} The left hand panels show resistance per square as a function of temperature for 2D holes in an ultra-clean GaAs/AlGaAs heterostructure for various values of the gate bias.  The right hand panels show the variation of longitudinal resistance with perpendicular magnetic field for the same sample at various gate biases.  The solid grey lines represent the fit described in the text. From Mills \etal (2001).}
\end{figure}

However, the agreement with non-interacting theory breaks down at higher interaction strengths.  Mills~\etal (2001) studied $p$-type GaAs/AlGaAs heterostructures of much higher quality which remained metallic down to $p\approx3\cdot10^9$~cm$^{-2}$ (corresponding to $r_s\sim60$ provided the effective mass does not change), and found the coherent backscattering to be almost completely suppressed at these ultra-low hole densities.  In the right hand panel of Fig.~\ref{mills_R(T)}, the magnetoresistance traces are shown at $T\approx$9~mK for four different carrier densities.  The width of the characteristic peak at $B_\perp=0$, visible in the top two curves, is approximately as expected from the theory, but its magnitude is about a factor of 30 smaller than expected.  At slightly higher $p$ (the two bottom curves), the peak is not seen at all.

In principle, the strong disagreement between the expected and measured peak magnitudes might be due to the fact that the theory for coherent backscattering is not applicable to a system with resistivity of the order of $h/e^2$.  However, this is not the source of the disagreement, as the resistivity in the experiments of Mills~\etal (2001) is in the same range as in the experiments of Simmons~\etal (2000) ({\it cf} Figs.~\ref{simmons_R(T)} and \ref{mills_R(T)}). Therefore the suppression of the peak is apparently related to stronger interactions (higher $r_s$) in the Mills~\etal samples rather than to high values of $\rho$.

The left hand panel of Fig.~\ref{mills_R(T)} shows the temperature dependence of the $B=0$ resistance in the ultra low density sample of Mills~\etal over the temperature range 5 to 100~mK.  The dashed lines show the expected corrections to the resistivity caused by weak localization calculated using the equation $\Delta\sigma(T)=b\,e^2/h\,\mbox{ln}(\tau_\phi/\tau)$, where $b$ is a constant expected to be universal and equal to $1/\pi$.  The calculated dependence is clearly at variance with the measurements; rather, at very low temperatures, the resistance becomes nearly constant.  Fitting the theoretical expression to the experimental data, Mills~\etal found that the upper limits for $b$ are from one to nearly two orders of magnitude smaller than the $b=1/\pi$ expected from the theory.

\begin{figure}\vspace{-8mm}\hspace{-1mm}
\scalebox{.38}{\includegraphics{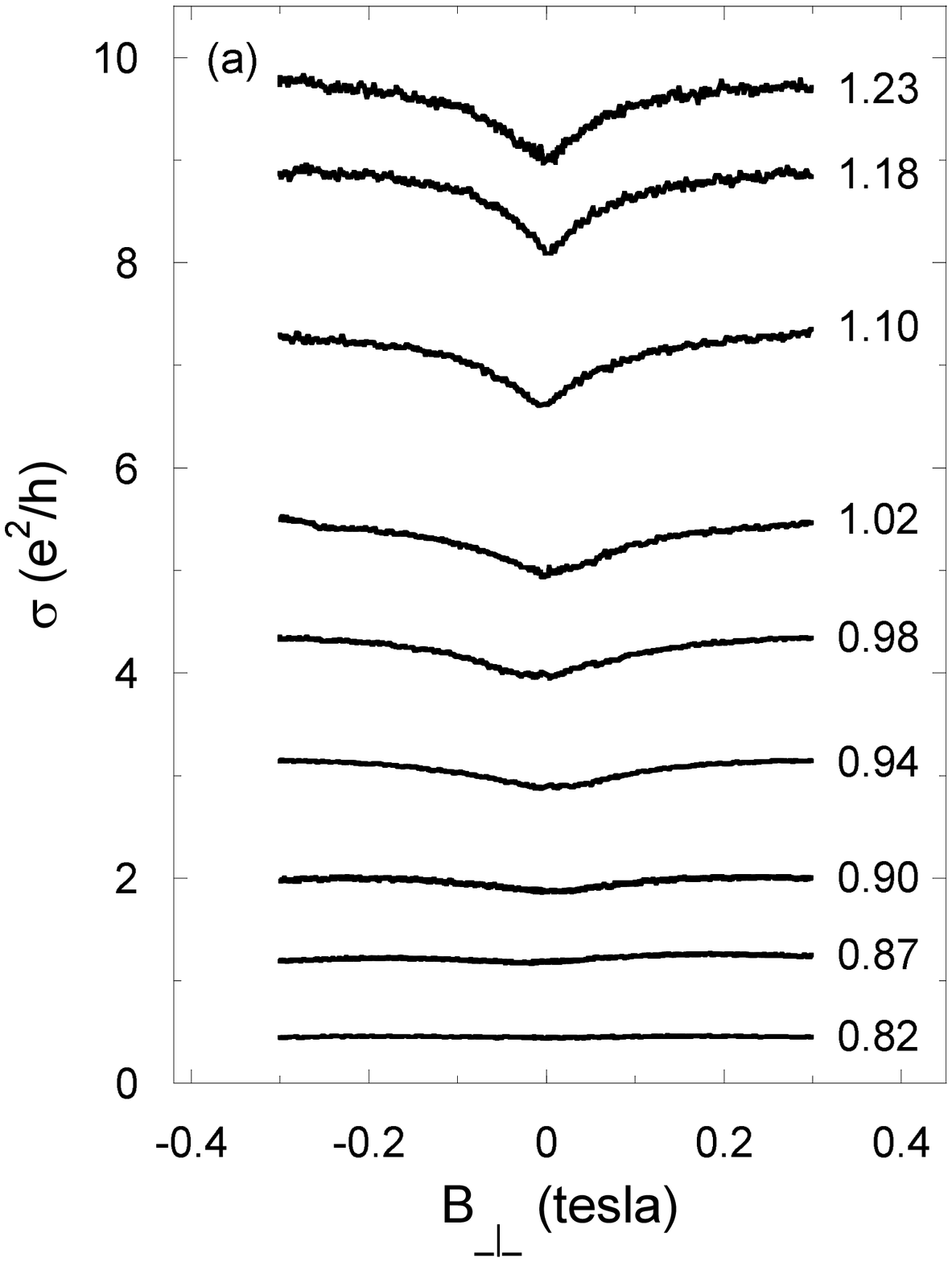}}\hspace{-2cm}
\scalebox{.38}{\includegraphics{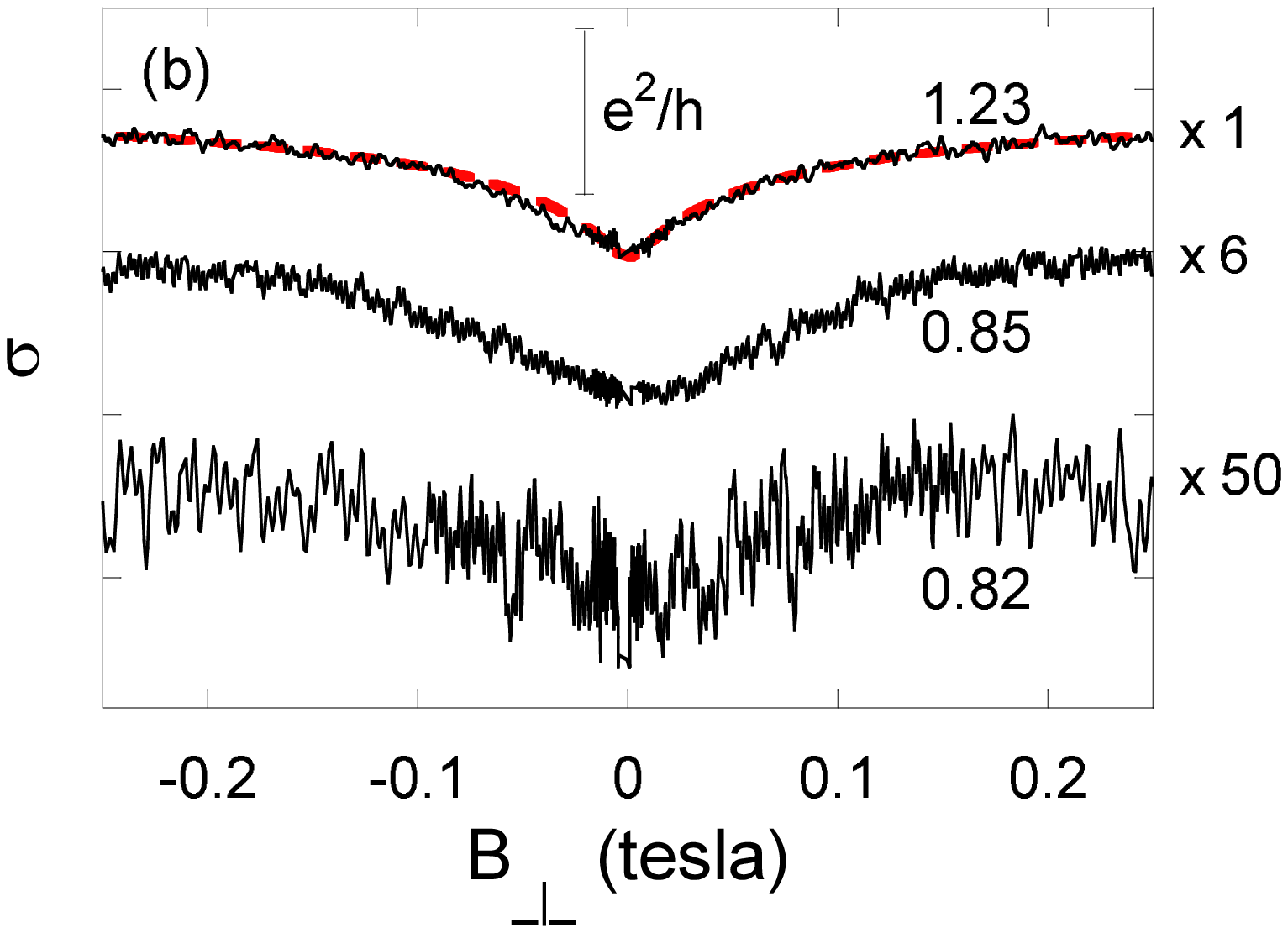}}\vspace{-1.5cm}
\caption{\label{anti_weak_loc} Longitudinal magnetoconductivity of a low-disordered silicon MOSFET in a weak perpendicular magnetic field at $T=42$~mK for a range of electron densities indicated near each curve in units of $10^{11}$~cm$^{-2}$. The curves in (b) are vertically shifted and the two lowest curves are multiplied by 6 (the middle one) and 50 (the lower one).  The thick dashed line in (b) is a fit by Eq.~\ref{HL} with $b=0.6$ and $\tau_\phi=30$~ps.  From Rahimi~\etal (2003).}
\end{figure}

The disappearance of weak localization corrections near the MIT has also been observed by Rahimi~\etal (2003) in low-disordered silicon MOSFETs. The results are shown in Fig.~\ref{anti_weak_loc}.  At higher $n_s$ (the upper curves in Fig.~\ref{anti_weak_loc}~(a)), the characteristic dip is observed in the magnetoconductance at zero magnetic field. As follows from Eq.~\ref{HL}, the magnitude of the dip is expected to be equal to $(b\, e^2g_v/h)$~ln$(\tau_\phi/\tau)$, and should therefore exhibit a weak (double-logarithmic) increase as the average conductivity decreases provided the variations in electron density are small, as they are in this case.  This is not what is observed in the experiment: as one approaches the transition, the magnitude of the dip decreases sharply, and at the critical electron density (the lowest curve in Fig.~\ref{anti_weak_loc}~(a)), the dip is no longer seen on the scale of this figure.  However, the shape of the magnetoconductivity does not change significantly with decreasing $n_s$ as illustrated by the middle curve in Fig.~\ref{anti_weak_loc}~(b), which shows $\sigma(B_\perp)$ multiplied by six ($n_s=0.85 \cdot 10^{11}$~cm$^{-2}$) to make it quantitatively similar to the upper curve.  This similarity demonstrates that the functional form of the $\sigma(B_\perp)$ dependence, described by the expression in square brackets in Eq.~\ref{HL}, does not change noticeably as the density is reduced from $1.23 \cdot 10^{11}$ to $0.85 \cdot 10^{11}$~cm$^{-2}$; instead, it is the magnitude of the effect that rapidly decreases upon approaching the MIT.  At yet lower density, $n_s=0.82 \cdot 10^{11}$~cm$^{-2}$, the magnitude of the dip does not exceed 2\% of that for $n_s=1.23 \cdot 10^{11}$~cm$^{-2}$ (compare the upper and the lower curves in Fig.~\ref{anti_weak_loc}~(b)).

It may seem surprising that a change in $n_s$ by only a factor of 1.5 (from $n_s=1.23 \cdot 10^{11}$ to $n_s=0.82 \cdot 10^{11}$~cm$^{-2}$) results in such a dramatic suppression of the quantum localization.  It is interesting to note in this connection that the experimental data on silicon MOSFETs described in sections \ref{sec:gm} and \ref{sec:r_s} reveal a sharp increase of the effective mass in the same region of electron densities where the suppression of the weak localization is observed.  Due to the strong renormalization of the effective mass, the ratio between the Coulomb and Fermi energies, $r^*=2\,r_s(m^*/m_b)$, grows much faster than $n_s^{-1/2}$ reaching values greater than 50 near the critical density.

\begin{figure}\hspace{3cm}
\scalebox{.7}{\includegraphics{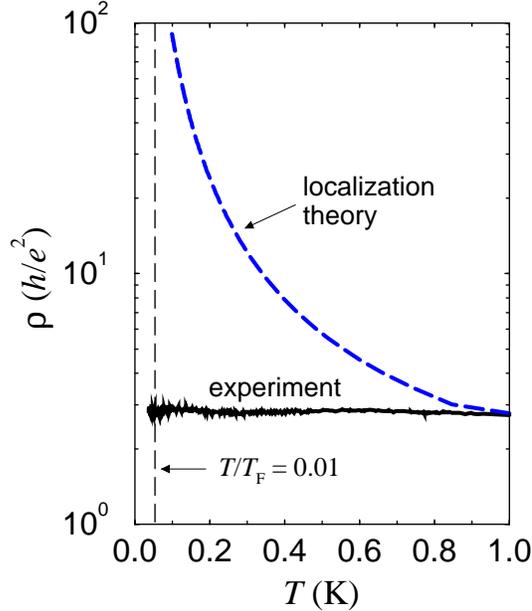}}
\caption{\label{gang_comparison} Resistivity at the separatrix in a low-disordered silicon MOSFET as a function of temperature (the solid curve) compared to that calculated from the one-parameter scaling theory using $\frac{d\,\mbox{ln}\,\rho(L_\phi)}{d\,\mbox{ln}\,L_\phi}=-\beta(\rho)$ (here $L_\phi\propto\rho^{-\gamma}T^{-p/2}$ is the phase-breaking length, $\beta(\rho)$ is the scaling function approximated by $\beta(\rho)=-\mbox{ln}(1+a\rho)$ following Altshuler~\etal (2000), $a=2/\pi$, $\rho$ is measured in units of $h/e^2$, and $p$ and $\gamma$ are constants equal to 3 and 0.5 respectively).  As the dashed line shows, the resistivity of a ``conventional'' (noninteracting) 2D system should have increased by a factor of more than 30 when the temperature has been reduced to 100~mK.  From Kravchenko and Klapwijk (2000a).}
\end{figure}

The apparent absence of localization at and just above the critical density may account for the existence of a flat separatrix at $n_s=n_c$ (see Fig.~\ref{r(t)universal}).  If the localization were present, the temperature-independent curve would require that the temperature dependence of $\rho$ due to localization be cancelled exactly over a wide temperature range by a temperature dependence of opposite sign due to interactions, a coincidence which seems very improbable for two unrelated mechanisms.  Note that at resistivity levels of order or greater than $h/e^2$, the quantum corrections to the resistivity are expected to be very strong and cannot be easily overlooked.  The calculated temperature dependence of the resistivity, expected for non-interacting electrons (Abrahams~\etal 1979), is shown in Fig.\ref{gang_comparison} by the dashed curve: at 100~mK, quantum localization is expected to cause a factor of more than 30 increase in resistivity, in strong contradiction with the experiment which shows it to be constant within $\pm5$\%.

\section{THE EFFECT OF A MAGNETIC FIELD}
\subsection{Resistance in a parallel magnetic field}

\begin{figure}\hspace{.5cm}
\scalebox{.44}{\includegraphics{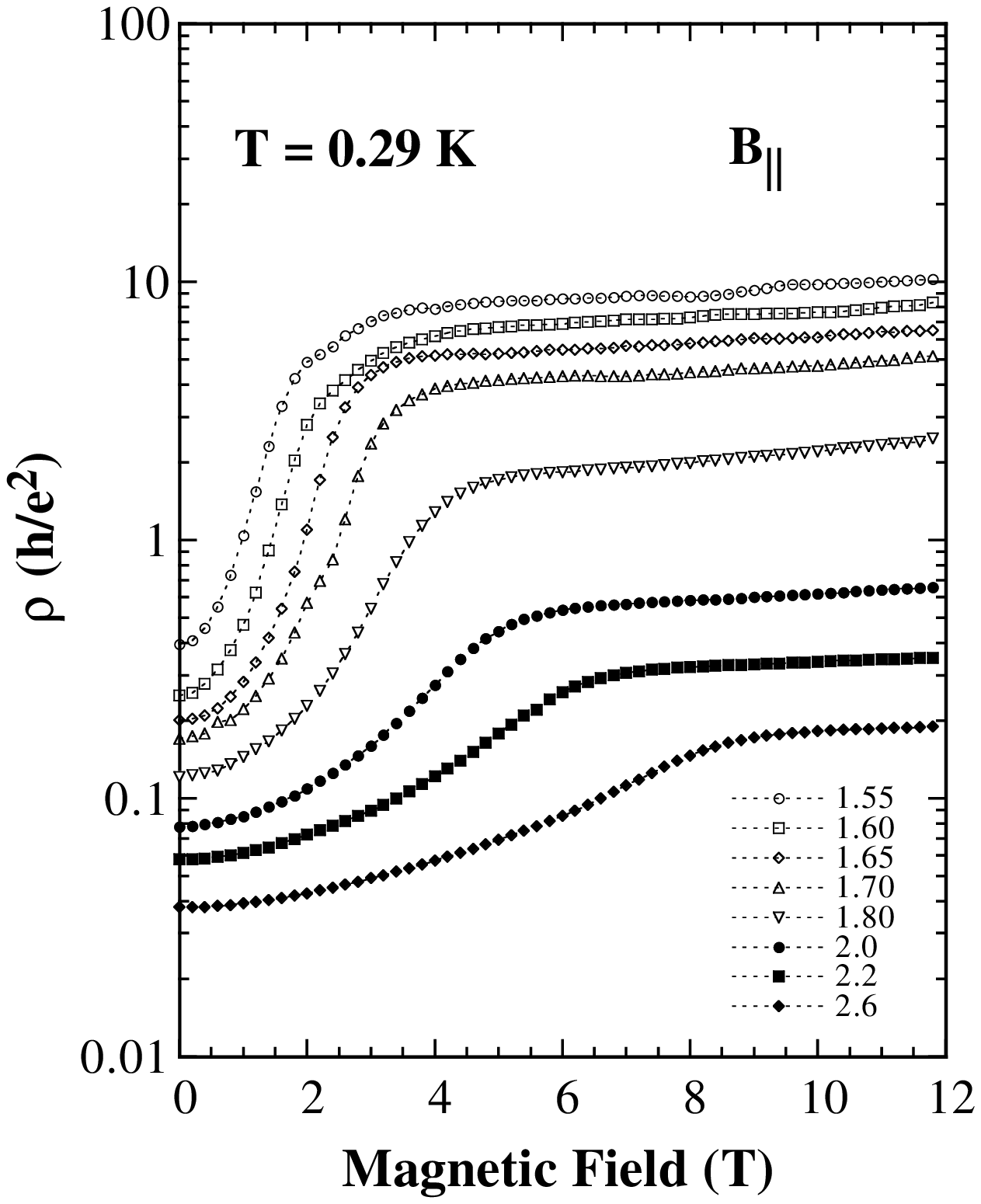}}\hspace{.5cm}
\scalebox{.407}{\includegraphics{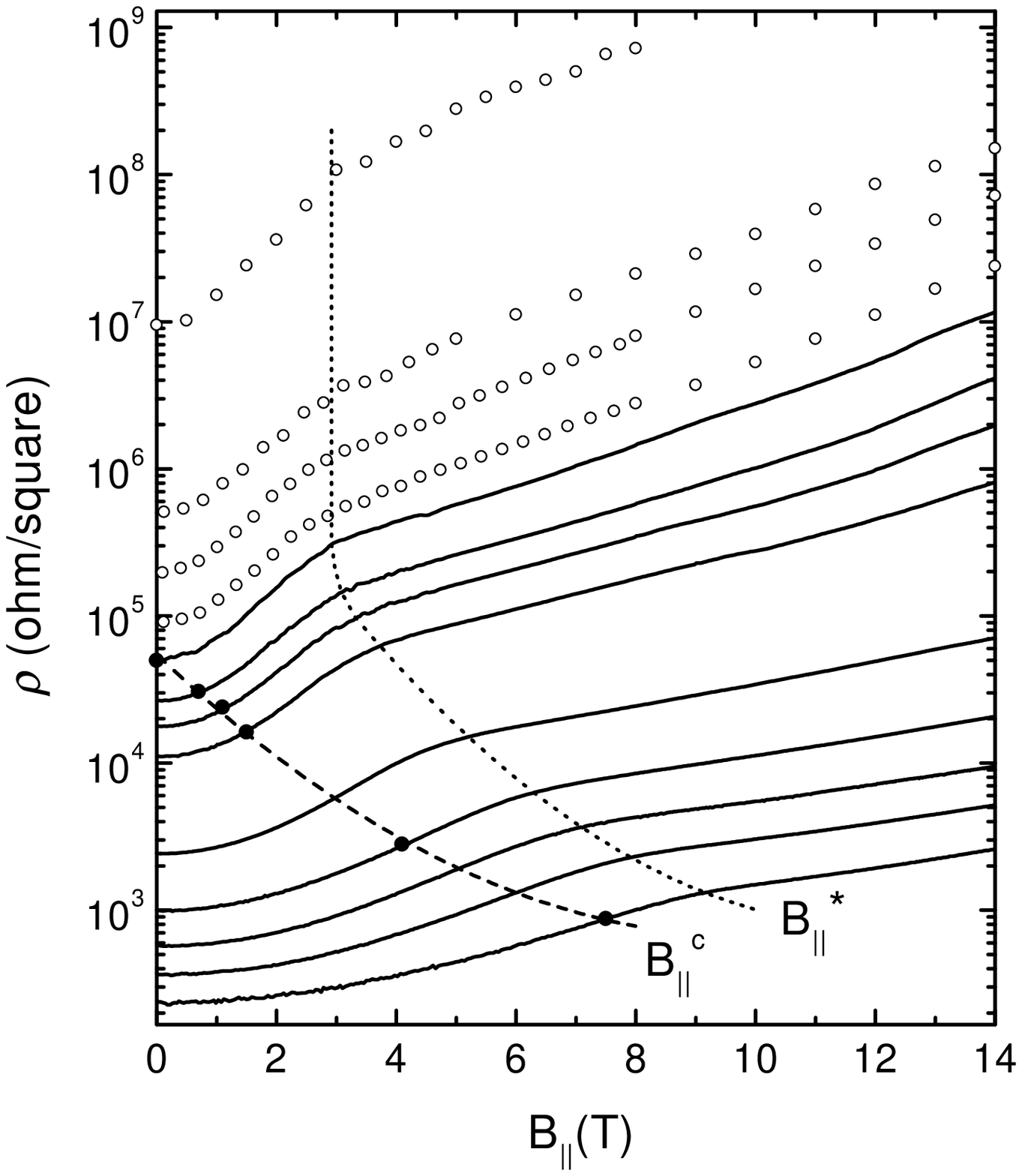}}
\caption{\label{RB_pudalov} Left hand panel: resistivity versus parallel magnetic field measured at $T=0.29$~K on low-disordered silicon sample.  Different symbols correspond to densities from 1.01 to $2.17\cdot10^{11}$~cm$^{-2}$ (adopted from Pudalov~\etal 1997).  Right hand panel: resistivity as a function of $B_\parallel$ in a p-GaAs/AlGaAs heterostructure at 50~mK at the following hole densities, from the bottom: 4.11, 3.23, 2.67, 2.12, 1.63, 1.10, 0.98, 0.89, 0.83, 0.79, 0.75, 0.67$\cdot10^{10}$~cm$^{-2}$.  The solid lines are for hole densities above $p_c$ and the open circles are for densities below $p_c$.  The solid circles denote the experimentally determined critical magnetic fields, and the dashed line is a guide to the eye. $B^*_\parallel$, the boundary separating the high and the low field regions, is shown by the dotted line.  Adopted from Yoon~\etal (2000).}
\end{figure}

In ordinary metals, the application of a parallel magnetic field
($B_\parallel$) does not lead to any dramatic changes in the transport
properties: if the thickness of the 2D electron system is small compared
to the magnetic length, the parallel field couples largely to the
electrons' spins while the orbital effects are suppressed.  Only weak
corrections to the conductivity are expected due to electron-electron
interactions (Lee and Ramakrishnan 1982, 1985).  It therefore came as a
surprise when Dolgopolov~\etal (1992) observed a dramatic suppression of
the conductivity in dilute Si MOSFETs by a parallel in-plane magnetic
field $B_\parallel$.  The magnetoresistance in a parallel field was
studied in detail by Simonian~\etal (1997b) and Pudalov~\etal (1997),
also in Si MOSFETs.  In the left hand part of Fig.~\ref{RB_pudalov}, the
resistivity is shown as a function of parallel magnetic field at a fixed
temperature of 0.3~K for several electron densities.  The resistivity
increases sharply as the magnetic field is raised, changing by a factor
of about 4 at the highest density shown and by more than an order of
magnitude at the lowest density, and then saturates and remains
approximately constant up to the highest measuring field,
$B_\parallel=12$~tesla.  The magnetic field where the saturation occurs,
$B_{\rm sat}$, depends on $n_s$,
varying from about 2~tesla at the lowest measured density to about 9
tesla at the highest.  The metallic conductivity is suppressed in a
similar way by magnetic fields applied at any angle relative to the 2D
plane (Kravchenko~\etal 1998) independently of the relative directions
of the measuring current and magnetic field (Simonian~\etal 1997b;
Pudalov~\etal 2002a).  All these observations suggest that the giant
magnetoresistance is due to coupling of the magnetic field to the
electrons' spins.  Indeed, from an analysis of the positions of
Shubnikov-de~Haas oscillations in tilted magnetic fields, Okamoto~\etal
(1999) and Vitkalov~\etal (2000, 2001a) have concluded that in MOSFETs at
relatively high densities, the magnetic field $B_{\rm sat}$ is equal to
that required to fully polarize the electrons' spins.

\begin{figure}\hspace{3.1cm}
\scalebox{.35}{\includegraphics{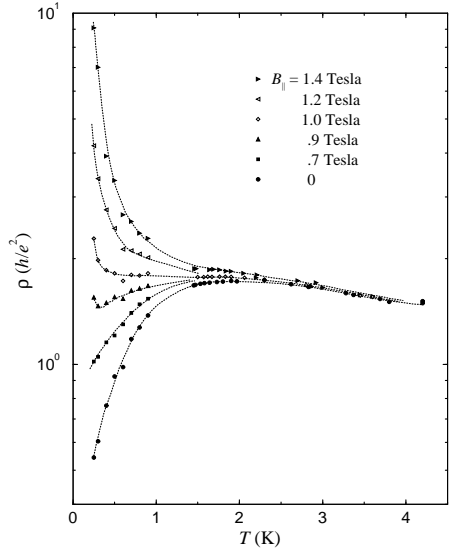}}
\caption{\label{simonian97} Resistivity versus temperature for five different fixed magnetic fields applied parallel to the plane of a low-disordered silicon MOSFET.  The electron density is $8.83\cdot10^{10}$~cm$^{-2}$. From Simonian~\etal (1997b).}
\end{figure}

In $p$-type GaAs/AlGaAs heterostructures, the effect of a parallel
magnetic field is qualitatively similar, as shown in the right hand part
of Fig.~\ref{RB_pudalov}.  The dependence of $\rho$ on
$B_\parallel$ does not saturate to a constant value as in Si MOSFETs,
but continues to increase with increasing field, albeit at a
considerably slower rate.  This is presumably due to strong coupling of
the parallel field to the orbital motion arising from the finite layer
thickness (see Das~Sarma and Hwang 2000), an effect that is more important
in GaAs/AlGaAs heterostructures than in silicon MOSFETs because of a much
thicker layer.  As in the case of Si MOSFETs, there is a distinct
knee that serves as a demarcation between the behaviour in low and high
fields.  For high hole densities, Shubnikov-de~Haas measurements
(Tutuc~\etal 2001) have shown that this knee is associated with full
polarization of the spins by the in-plane magnetic field.  However,
unlike Si MOSFETs, the magnetoresistance in p-GaAs/AlGaAs
heterostructures has been found to depend on the relative directions of
the measuring current, magnetic field, and crystal orientation
(Papadakis~\etal 2000); one should note that the crystal anisotropy of
this material introduces added complications.  In p-SiGe
heterostructures, the parallel field was found to induce negligible
magnetoresistance (Senz~\etal 1999) because in this system the parallel
field cannot couple to the spins due to very strong spin-orbit
interactions.

Over and above the very large magnetoresistance induced by an in-plane
magnetic fields, an even more important effect of a parallel field is
that it causes the zero-field 2D metal to become an insulator
(Simonian~\etal 1997b;  Mertes~\etal 2001; Shashkin~\etal 2001b;
Gao~\etal 2002).  Figure~\ref{simonian97} shows how the temperature
dependence of the resistance changes as the magnetic field is increased.
  Here, the resistivity of a Si MOSFET with fixed density on the
metallic side of the transition is plotted as a function of temperature
in several fixed parallel magnetic fields between 0 and 1.4~tesla.  The
zero-field curve exhibits behaviour typical for ``just-metallic''
electron densities: the resistivity is weakly-insulating at
$T>T_{\max}\approx2$~K and drops substantially as the temperature is
decreased below $T_{\max}$.  In a parallel magnetic field of only 1.4~tesla
(the upper curve), the metallic drop of the resistivity is completely
suppressed, so that the system is now strongly insulating in the entire
temperature range.  The effect of the field is negligible at
temperatures above $T_{\max}$, {\em i.e.}, above the temperature below
which the metallic behaviour in $B=0$ sets in.

\begin{figure}
\scalebox{.9}{\includegraphics{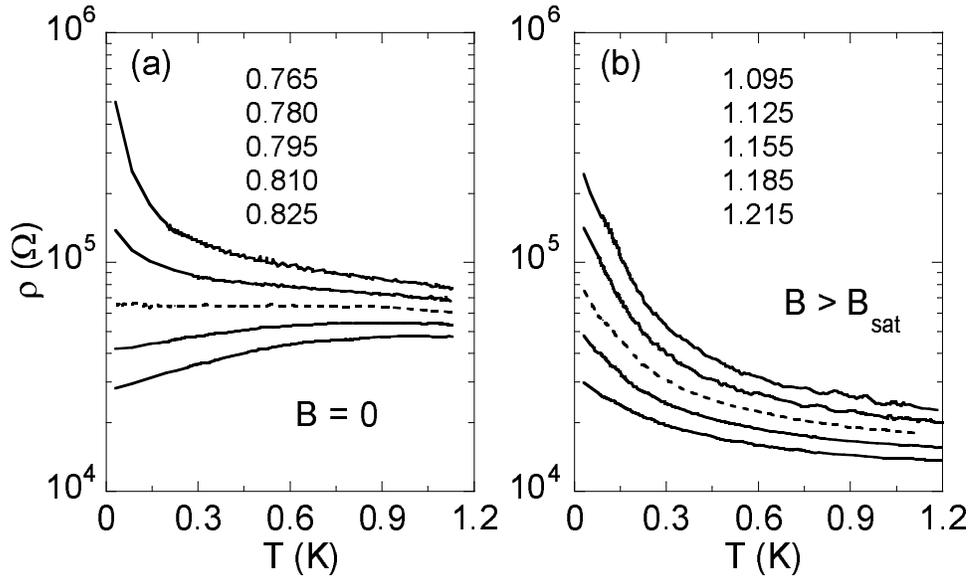}}
\caption{\label{R(T)inB_ours} Temperature dependence of the resistivity of a low-disordered silicon MOSFET at different electron densities near the MIT in zero magnetic field (a), and in a parallel magnetic field of 4~tesla (b). The electron densities are indicated in units of $10^{11}$~cm$^{-2}$. Dashed curves correspond to $n_s=n_{c1}$ which is equal to $0.795\cdot10^{11}$~cm$^{-2}$ in zero field and to $1.155\cdot10^{11}$~cm$^{-2}$ in $B_\parallel=4$~tesla. From Shashkin~\etal (2001b).}
\end{figure}

The extreme sensitivity to parallel field is also illustrated in
Fig.~\ref{R(T)inB_ours} where the temperature dependence of the
resistivity is compared in the absence (a) and in the presence (b) of a
parallel magnetic field.  For $B_\parallel=0$, the resistivity displays
the familiar, nearly symmetric (at temperatures above 0.2~K) critical
behaviour about the separatrix (the dashed line).  However, in a parallel
magnetic field of $B_\parallel=4$~tesla, which is high enough to cause
full spin polarization at this electron density, all the $\rho(T)$
curves display ``insulating-like'' behaviour, including those which start
below $h/e^2$ at high temperatures.  There is no temperature-independent
separatrix at any electron density in a spin-polarized electron system
(Simonian~\etal 1997b; Shashkin~\etal 2001b).

This qualitative difference in behaviour demonstrates convincingly that the spin-polarized and unpolarized states behave very differently.  This rules out explanations which predict qualitatively similar behaviour of the resistance regardless of the degree of spin polarization.  In particular, the explanation of the metallic behaviour suggested by Das~Sarma and Hwang (1999) (see also Lilly~\etal 2003), based on the temperature-dependent screening, predicts metallic-like $\rho(T)$ for both spin-polarized and unpolarized states, which is in disagreement with experiment (for more on this discrepancy, see Mertes~\etal 2001).

\subsection{Scaling of the magnetoresistance; evidence for a phase
transition}\label{scaling of magnetoresistance}

\begin{figure}\hspace{2.27cm}
\scalebox{.52}{\includegraphics{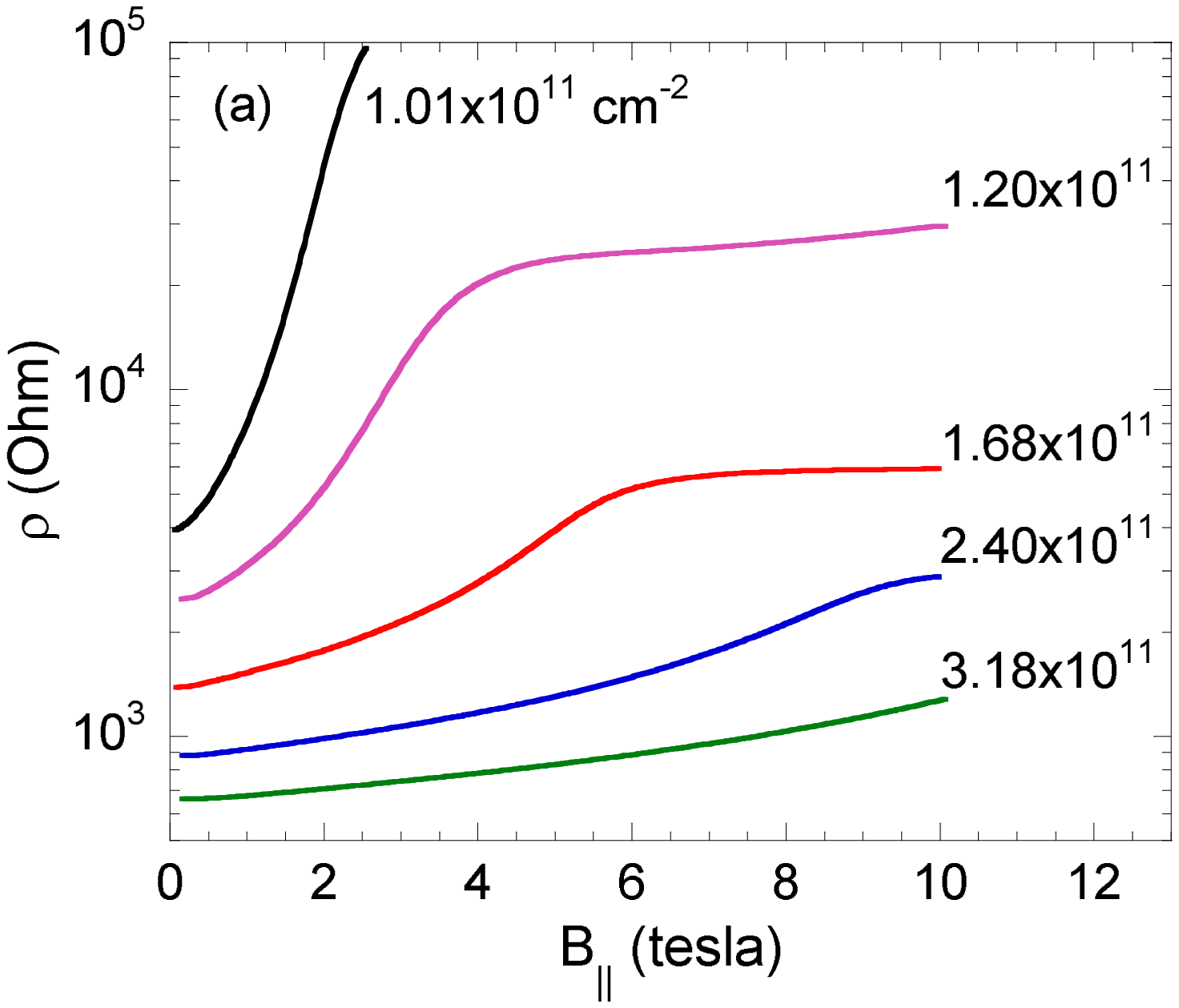}}
\begin{center}
\scalebox{.52}{\includegraphics{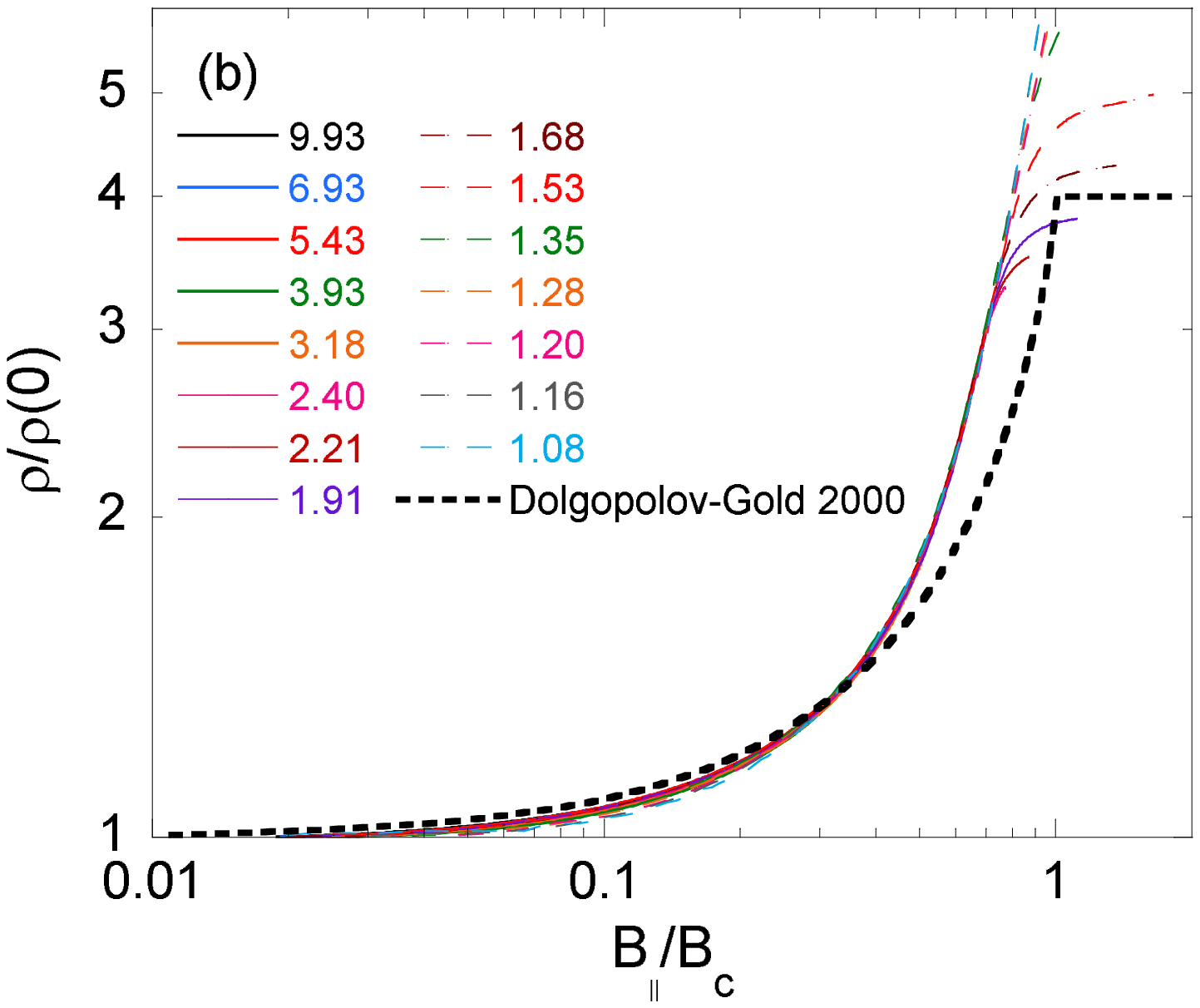}}
\end{center}\vspace{-3mm}
\caption{\label{ferro} (a)~Low-temperature magnetoresistance of a clean silicon MOSFET in parallel magnetic field at different electron densities above $n_c$.  (b)~Scaled curves of the normalized magnetoresistance versus $B_\parallel/B_c$.  The electron densities are indicated in units of $10^{11}$~cm$^{-2}$.  Also shown by a thick dashed line is the normalized magnetoresistance calculated by Dolgopolov and Gold (2000). Adopted from Shashkin~\etal (2001a).}
\end{figure}

There have been many attempts to obtain a quantitative description of
the magnetoresistance as a function of the carrier density and
temperature over the entire field range, including the saturation
region. Attempts to obtain a collapse of the magnetoresistance onto a
single scaled curve have yielded scaling at either low or high magnetic
field; over a wide range of temperatures but only at the metal-insulator
transition; or in a wide range of carrier densities, but only in the
limit of very low temperatures.  Simonian~\etal (1998) found that at the
transition, the deviation of the magnetoconductivity from its zero-field
value, $\Delta\sigma\equiv\sigma(B_\parallel)-\sigma(0)$, is a universal
function of $B_\parallel/T$.  Although the quality of the scaling is
good, it breaks down rather quickly as one moves into the metallic
phase.

Two scaling procedures have been recently proposed; although they differ in
procedure and yield results that differ somewhat in detail, the major
conclusions are essentially the same, as described below, and imply that
there is an approach to a quantum phase transition at a density near $n_c$.

\begin{figure}\hspace{3.2cm}
\scalebox{.45}{\includegraphics{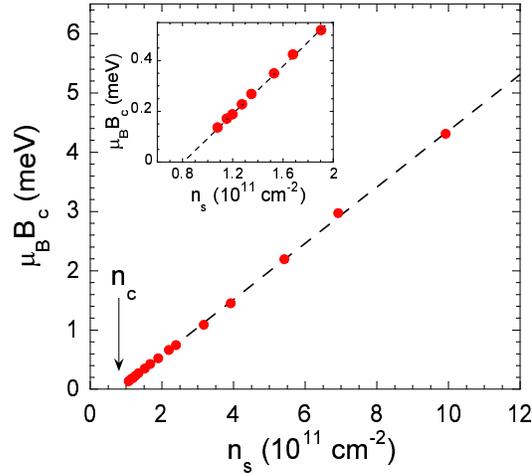}}
\caption{\label{polarization} Scaling parameter $B_c$ (corresponding to
the field required for full spin polarization) as a function of the
electron density.  An expanded view of the region near $n_c$ is
displayed in the inset.  From Shashkin~\etal (2001a).}
\end{figure}

Shashkin~\etal (2001a) scaled the magnetoresistivity in the spirit of the
theory of Dolgopolov and Gold (2000), who predicted that at $T=0$, the
normalized magnetoresistance is a universal function of the degree of spin
polarization, $P\equiv g^*\mu_BB_\parallel/2E_F=g^*m^*\mu_BB_\parallel/\pi\hbar^2n_s$ (here $m^*$
is the effective mass and $g^*$ is the g-factor). Shashkin~\etal scaled the
data obtained in the limit of very low temperatures where the
magnetoresistance becomes temperature-independent and, therefore, can be
considered to be at its $T=0$ value.  In this regime, the normalized
magnetoresistance, $\rho(B_\parallel)/\rho(0)$, measured at different
electron densities, collapses onto a single curve when plotted as a
function of $B_\parallel/B_c$ (here $B_c$ is the scaling parameter,
normalized to correspond to the magnetic field $B_{\rm sat}$ at which the
magnetoresistance saturates).  An example of how $\rho(B_\parallel)$,
plotted in Fig.~\ref{ferro}~(a), can be scaled onto a universal curve is
shown in Fig.~\ref{ferro}~(b).  The resulting function is described
reasonably well by the theoretical dependence predicted by Dolgopolov and
Gold.  The quality of the scaling is remarkably good for
$B_\parallel/B_c\le 0.7$ in the electron density range $1.08\cdot10^{11}$
to $10\cdot10^{11}$~cm$^{-2}$, but it breaks down as one approaches the
metal-insulator transition where the magnetoresistance becomes strongly
temperature-dependent even at the lowest experimentally achievable
temperatures.  As shown in Fig.~\ref{polarization}, the scaling parameter
is proportional over a wide range of electron densities to the deviation
of the electron density from its critical value: $B_c\propto(n_s-n_c)$.

\begin{figure}\vspace{-0.3cm}\hspace{2.5cm}
\scalebox{.43}{\includegraphics{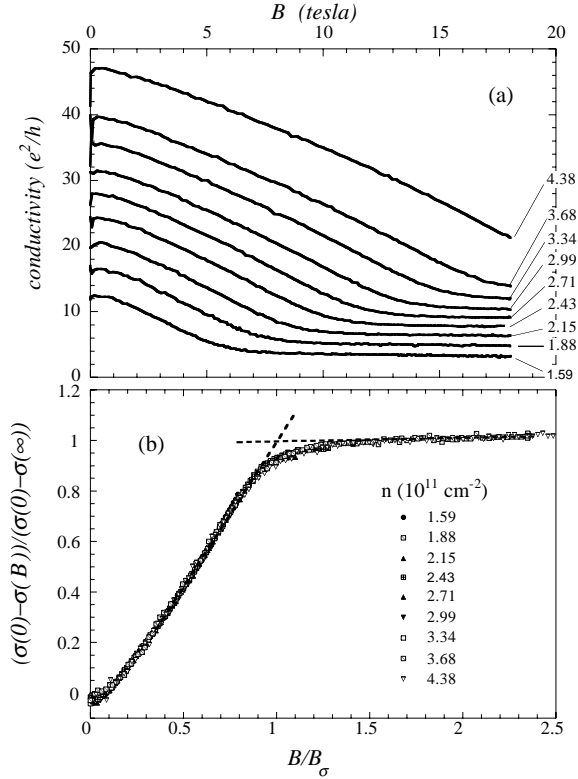}}
\caption{\label{vitkalov00}
(a)~Conductivity of a low-disordered silicon sample versus in-plane magnetic field at different electron densities in units of $10^{11}$~cm$^{-2}$, as labelled; $T=100$ mK. (b)~Data collapse obtained by applying the scaling procedure, Eq.\protect\ref{vitkalov}, to the curves shown in (a). Adopted from Vitkalov~\etal (2001b).}
\end{figure}

Vitkalov~\etal (2001b) succeeded in obtaining an excellent collapse of
magnetoconductivity data over a broad range of electron densities {\em
and temperatures} using a single scaling  parameter.  They separated the
conductivity into a field-dependent contribution,
$\sigma(B_\parallel)-\sigma(\infty)$,
and a contribution that is independent of magnetic field,
$\sigma(\infty)$. The field-dependent contribution to the conductivity,
$\sigma(0)-\sigma(B_\parallel)$, normalized to its full value,
$\sigma(0)-\sigma(\infty)$, was shown to be a universal function of
$B/B_\sigma$:
\begin{equation}
\frac{\sigma(0)-\sigma(B_\parallel)}{\sigma(0)-\sigma(\infty)}=
F(B_\parallel/B_{\sigma})\label{vitkalov}
\end{equation}
where $B_\sigma(n_s,T)$ is the scaling parameter.  Applied to the
magnetoconductance curves shown in Fig.~\ref{vitkalov00}~(a) for
different electron densities, the above scaling procedure yields the
data collapse shown in Fig.~\ref{vitkalov00}~(b).

\begin{figure}\vspace{.5cm}\hspace{2.7cm}
\scalebox{.46}{\includegraphics{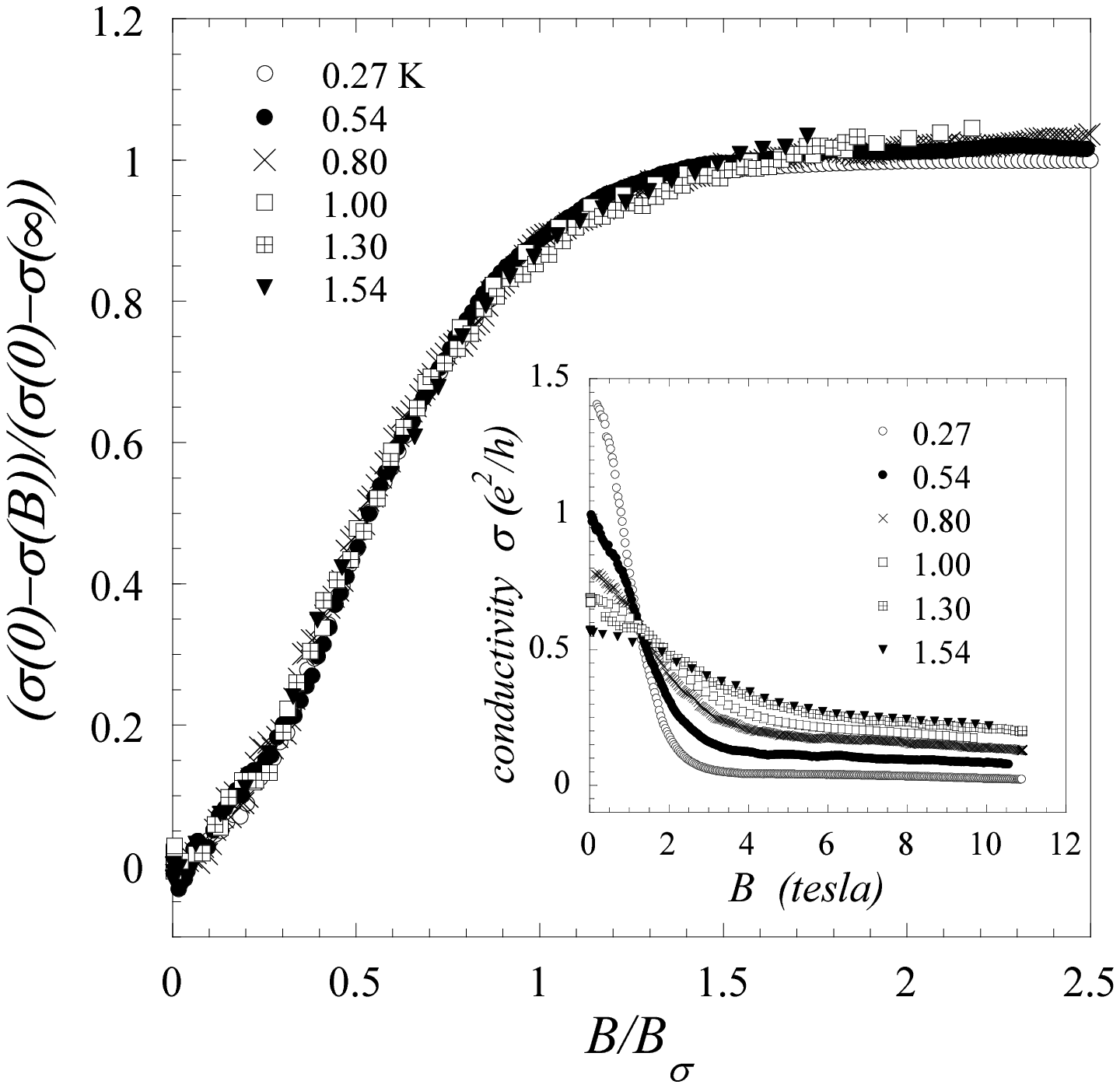}}
\caption{\label{vitkalov00_2} Data collapse obtained by applying the scaling procedure, Eq.\protect\ref{vitkalov}, to the magnetoconductivity at different temperatures for electron density $n_s=9.4\cdot10^{10}$~cm$^{-2}$.  The inset shows the conductivity at different temperatures as a function of magnetic field.  Adopted from Vitkalov~\etal (2001b).}
\end{figure}

Remarkably, similar scaling holds for curves obtained at different
temperatures.  This is demonstrated in Fig.~\ref{vitkalov00_2}, which
shows the scaled magnetoconductance measured at a fixed density and
different temperatures.  Altogether, the scaling holds for temperatures
up to $1.6$~K over a broad range of electron densities up to $4\, n_c$.

\begin{figure}\hspace{2cm}
\scalebox{.43}{\includegraphics{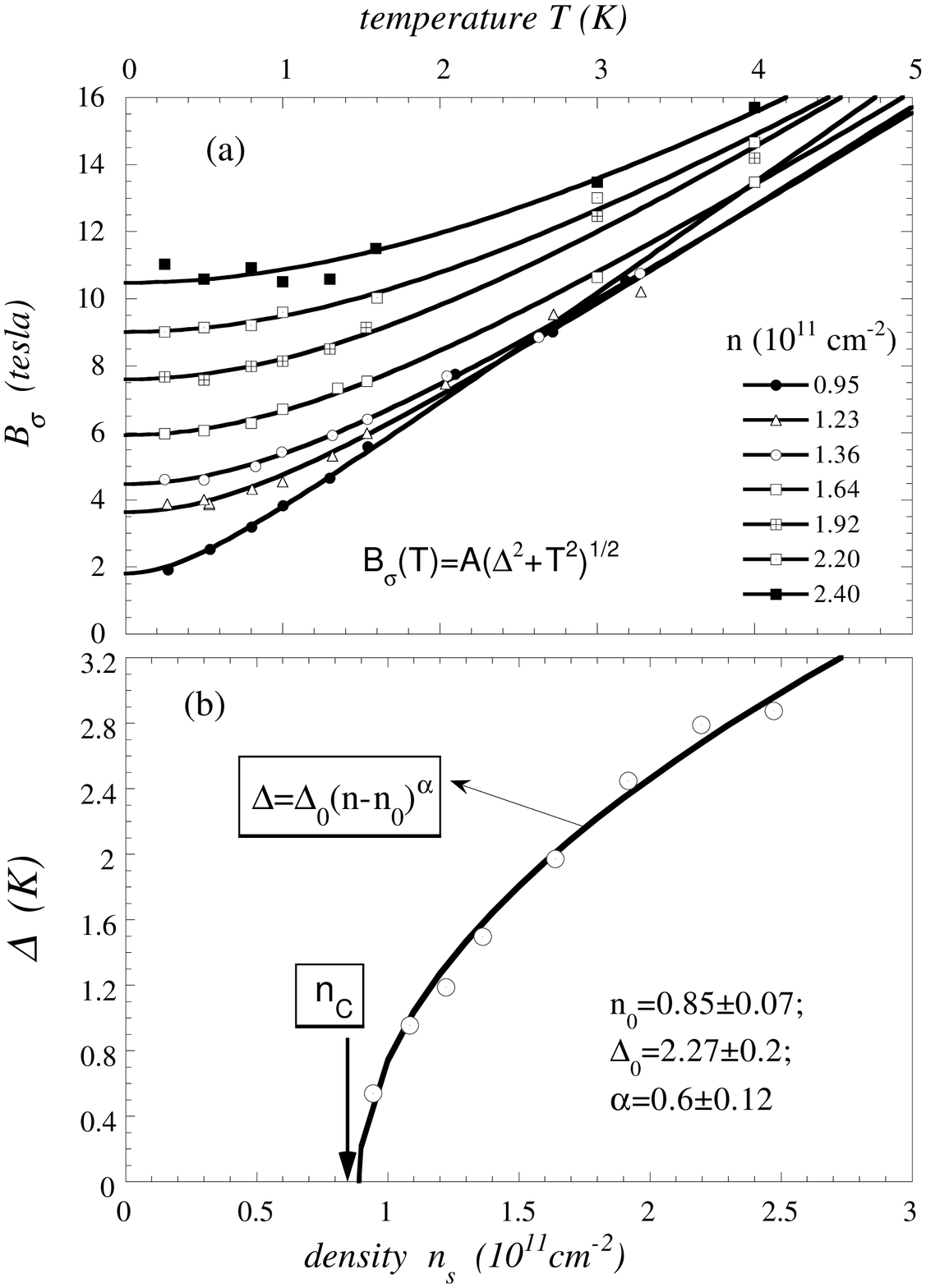}}
\caption{\label{myriam1} (a)~$B_{\sigma}$ versus temperature for
different electron densities; the solid lines are fits to the empirical
form $B_\sigma(n_s,T)=A(n_s) [[\Delta(n_s)]^2+T^2]^{1/2}$. (b)~The
parameter $\Delta$ as a function of electron density; the solid line is
a fit to the critical form $\Delta=\Delta_0(n_s-n_0)^\alpha$. From
Vitkalov~\etal (2001b).}
\end{figure}

Fig.~\ref{myriam1} shows the scaling parameter $B_{\sigma}$ plotted as a
function of temperature for different electron densities $n_s>n_c$.  The
scaling parameter becomes smaller as the electron density is reduced;
for a given density, $B_{\sigma}$ decreases as the temperature decreases
and approaches a value that is independent of temperature,
$B_\sigma(T=0)$. As the density is reduced toward $n_c$, the temperature
dependence of $B_{\sigma}$ dominates over a broader range and becomes
stronger, and the low-temperature asymptotic value becomes smaller.  Note
that for electron densities below $1.36\cdot10^{11}$~cm$^{-2}$,
$B_{\sigma}$ is approximately linear with temperature at high $T$.  The
behaviour of the scaling parameter $B_{\sigma}(T)$ can be approximated by
an empirical fitting function:
$$
B_\sigma(n_s,T)=A(n_s)[\Delta(n_s)^{2}+T^{2}]^{1/2}.
$$
The solid lines in Fig.~\ref{myriam1}~(a) are fits to this expression using $A(n_s)$ and $\Delta(n_s)$ as fitting parameters.  As can be inferred from the slopes of the curves of Fig.~\ref{myriam1}~(a), the parameter $A(n_s)$ is constant over most of the range and then exhibits a small increase (less than $20$\%) at lower densities.  As shown in Fig.~\ref{myriam1}~(b), the parameter $\Delta$ decreases sharply with decreasing density and extrapolates to zero at a density $n_0$ which is within $10$\% of the critical density $n_c\approx0.85\cdot10^{11}$~cm$^{-2}$ for the metal-insulator transition.

The scaling parameters $B_c$ and $\Delta$ in the analysis by
Shashkin~\etal (2001a) and Vitkalov~\etal (2001b) represent energy scales
$\mu_BB_c$ and $k_B\Delta$, respectively, which vanish at or near the
critical electron density for the metal-insulator transition.   At high
electron densities and low temperatures $T<\mu_BB_c/k_B$ (corresponding
to $T<\Delta$), the system is in the zero temperature limit.  As one
approaches $n_c$, progressively lower temperatures are required to reach
the zero temperature limit.  At $n_s=n_0$, the energies $\mu_BB_c$ and
$k_B\Delta$ vanish; the parameter $B_\sigma\rightarrow0$ as
$T\rightarrow0$; the system thus exhibits critical behaviour
(Sondhi~\etal 1997; Vojta 2003), signalling the approach to a new phase in the limit
$T=0$ at a critical density $n_0\approx n_c$.

\section{SPIN SUSCEPTIBILITY NEAR THE METAL-INSULATOR TRANSITION}
\subsection{Experimental measurements of the spin susceptibility}\label{spin}

In the Fermi-liquid theory, the electron effective mass and the $g$-factor (and, therefore, the spin susceptibility $\chi\propto g^*m^*$) are renormalized due to electron-electron interactions (Landau 1957). Earlier experiments (Fang and Stiles 1968; Smith and Stiles 1972), performed at relatively small values of $r_s \sim2$ to 5, confirmed the expected increase of the spin susceptibility. More recently, Okamoto~\etal (1999) observed a renormalization of $\chi$ by a factor of $\sim2.5$ at $r_s$ up to about 6 (see Fig.~\ref{pudalov2002}~(a)).  At yet lower electron densities, in the vicinity of the metal-insulator transition, Kravchenko~\etal (2000b) have observed a disappearance of the energy gaps at ``cyclotron'' filling factors which they interpreted as evidence for an increase of the spin susceptibility by a factor of at least 5.

It was noted many years ago by Stoner that strong interactions can drive an electron system toward a ferromagnetic instability (Stoner 1946). Within some theories of strongly interacting 2D systems (Finkelstein 1983, 1984; Castellani~\etal 1984), a tendency towards ferromagnetism is expected to accompany metallic behaviour.  The experiments discussed in Section~\ref{scaling of magnetoresistance} which indicate that $B_c$ and $\Delta$ vanish at a finite density prompted Shashkin~\etal (2001a) and Vitkalov~\etal (2001b) to suggest that spontaneous spin polarization may indeed occur at or near the critical electron density in the limit $T=0$. The data obtained in these experiments provide information from which the spin susceptibility, $\chi$, can be calculated in a wide range of densities. In the clean limit, the band tails are small (Vitkalov~\etal 2002; Dolgopolov and Gold 2002; Gold and Dolgopolov 2002) and can be neglected, and the magnetic field required to fully polarize the spins is related to the $g$-factor and the effective mass by the equation $g^*\mu_BB_c=2E_F=\pi\hbar^2n_s/m^*$ (here, the two-fold valley degeneracy in silicon has been taken into account). Therefore, the spin susceptibility, normalized by its ``non-interacting'' value, can be calculated as 
$$\frac{\chi}{\chi_0}=\frac{g^*m^*}{g_0m_b}= \frac{\pi\hbar^2n_s}{2\mu_BB_cm_b}.$$

\begin{figure}\hspace{2.5cm}
\scalebox{.475}{\includegraphics{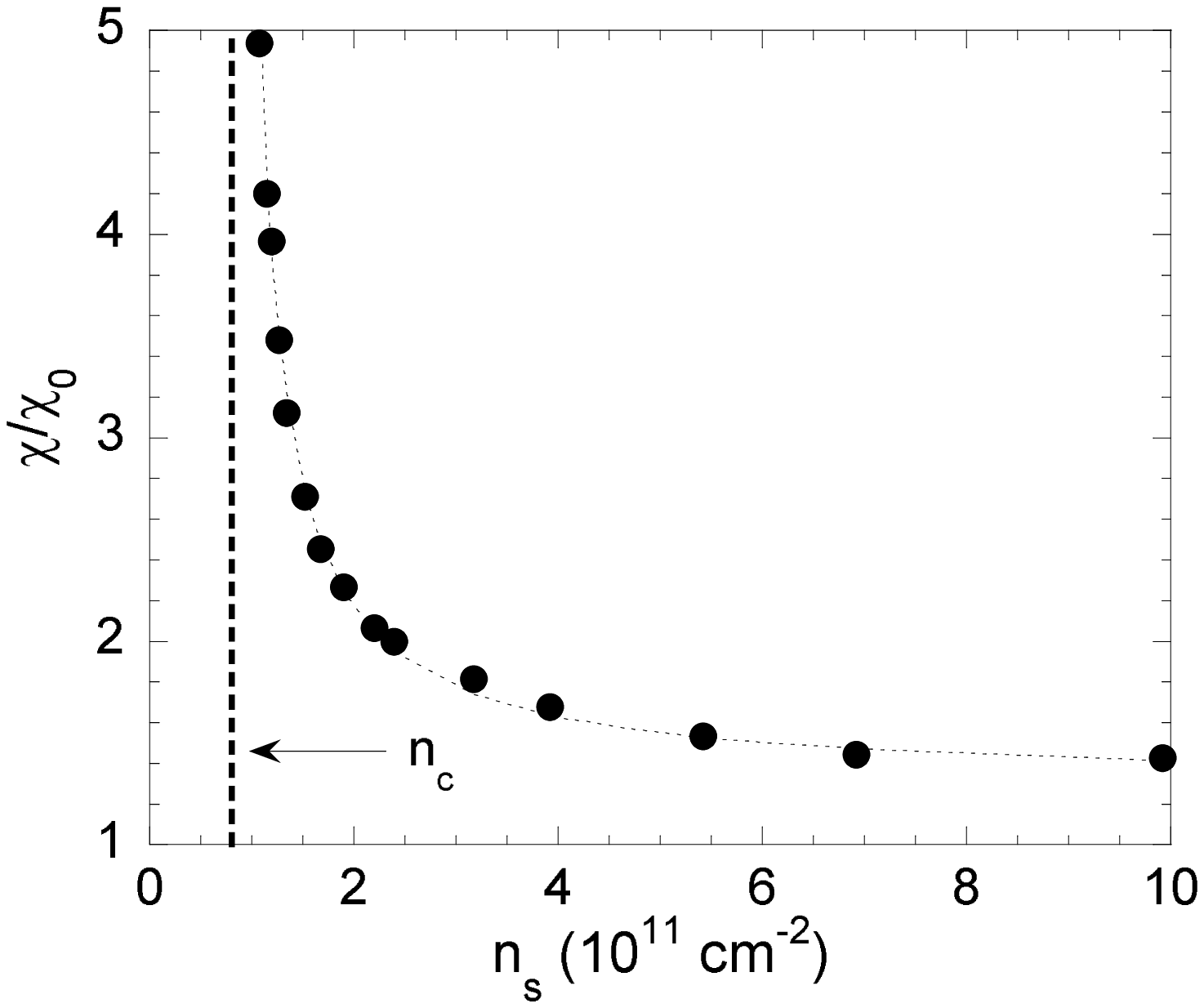}}
\begin{center}
\scalebox{.43}{\includegraphics{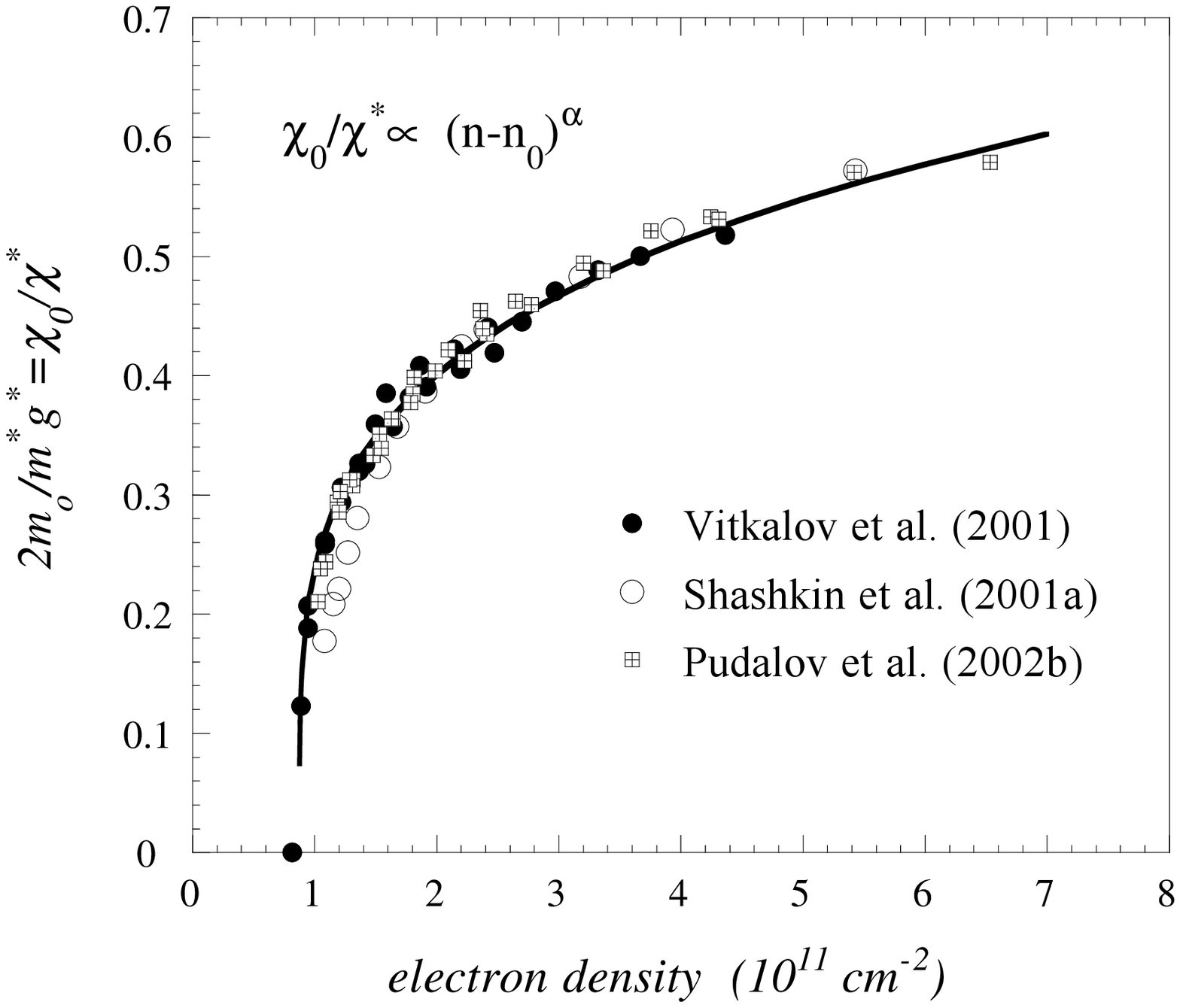}}
\end{center}
\caption{\label{vitkalov_gm} Upper graph: normalized spin
susceptibility {\it vs} $n_s$ obtained from the data in
Fig.~\ref{polarization}. The dashed line is a guide to the eye. The
vertical dashed line denotes the position of the critical density for
the metal-insulator transition.  Lower graph: the inverse of the
normalized spin susceptibility $\chi_0/\chi^*$ versus electron density
obtained by Vitkalov~\etal~(2001b), plotted with data of
Shashkin~\etal~(2001a) and Pudalov~\etal~(2002b).  The solid curve is a
fit to the critical form $\chi_0/\chi^*=A(n_s-n_0)^\alpha$ for the data of
Vitkalov~\etal~(2001b) (excluding the point shown at $\chi_0/\chi^*=0$).
Adopted from Vitkalov~\etal~(2002).}
\end{figure}

Fig.~\ref{vitkalov_gm} shows the normalized spin susceptibility (Shashkin~\etal 2001a) and its inverse (Vitkalov~\etal 2002) obtained using the above expression. The values deduced by both groups indicate that $g^*m^*$ diverges ($(g^*m^*)^{-1}$ extrapolates to zero) in silicon MOSFETs at a finite density close or equal to $n_c$.  Also shown on the same figure are the data of Pudalov~\etal obtained from an analysis of Shubnikov-de~Haas (SdH) measurements in crossed magnetic fields (see section \ref{sec:gm}).  The susceptibilities obtained by all three groups on different samples, by different methods and in different ranges of magnetic field, are remarkably similar (on the mutual consistency of the data obtained on different samples by different groups, see also Kravchenko~\etal (2002) and Sarachik and Vitkalov (2003)).

A novel and very promising method has recently been used by Prus~\etal
(2003) to obtain direct measurements of the thermodynamic spin
susceptibility.  The method entails modulating the (parallel) magnetic
field by an auxiliary coil and measuring the AC current induced between
the gate and the 2D electron gas, which is proportional to
$\partial\mu/\partial B$ (where $\mu$ is the chemical potential).  Using
Maxwell's relation,
$$\frac{\partial\mu}{\partial B}=-\frac{\partial M}{\partial n_s},$$
one can obtain the magnetization $M$ by integrating the induced current
over $n_s$.  The magnetic susceptibility can then be determined from the
slope of the $M(B)$ versus $B$ dependence at small fields.  To date,
Prus~\etal have reported data for one sample which becomes insulating at
a relatively high electron density ($1.3 \cdot 10^{11}$~cm$^{-2}$), and
the data obtained below this density are irrelevant for the metallic
regime we are interested in.  For this reason, the data collected so far
do not provide information about the most interesting regime just above
$n_\chi$.

\begin{figure}\hspace{2.7cm}
\scalebox{.95}{\includegraphics{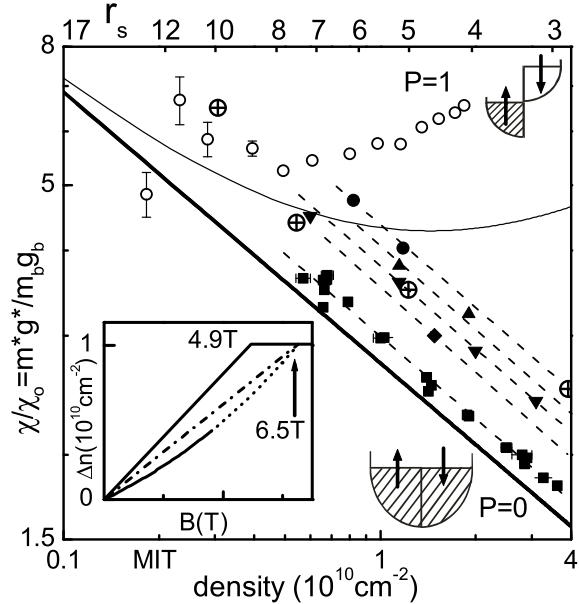}}
\caption{\label{zhu2003} Density-dependence of $m^*g^*$ in an ultra-clean 2D electron system in GaAs/AlGaAs determined by two different methods.  The solid data points are obtained from
tilted-field Shubnikov-de~Haas measurements with different configurations of Landau levels.  The parallel dashed lines indicate a power law dependence of $m^*g^*$ with a single exponent for all level configurations.  The open circles show nonmonotonic density-dependence of $m^*g^*$ derived from the full polarization field, $B_{c}$, using the parallel field method. The inset shows the net spin for $n_s=1 \cdot 10^{10}$~cm$^{-2}$ with interpolated regime (solid line) and extrapolated regime (dotted line).  $B_{c}$=4.9T from the in-plane field method and $B_{ext}$=6.5~T from extrapolation of the tilted-field method. The thick solid line represents extrapolation of $m^*g^*$ to the P=0 limit. Calculations from Attaccalite~\etal (2002) are shown as crossed circles. Adopted from Zhu~\etal (2003).}
\end{figure}

In GaAs/AlGaAs heterostructures, a similar strong increase of the spin
susceptibility at ultra-low carrier densities has now been established
based on an analysis of the Shubnikov-de~Haas oscillations (Zhu~\etal
2003).  The results are shown in Fig.~\ref{zhu2003} by solid symbols; $\chi
$ increases by more than a factor of two as the density decreases.
Zhu~\etal suggested an empirical equation $\chi\propto n_s^{-0.4}$ to
describe their experimental data, which works well in the entire range of
electron densities spanned. Although $\chi$ tends toward a divergence, it
is not clear from these experiments whether it does so at a finite density.
We note that due to the lower effective mass, higher dielectric constant
and the absence of valley degeneracy, the ratio $r^*$ between Coulomb
and Fermi energies in GaAs/AlGaAs is an order of magnitude smaller
than in silicon MOSFETs at the same electron density;
therefore, to reach the same relative interaction strength, samples are
required which remain metallic at densities about two orders of
magnitude lower than in silicon, {\it i.e.}, $<10^9$~cm$^{-2}$.  The
currently accessible electron densities in GaAs/AlGaAs heterostructures
are still too high to reliably establish the limiting behaviour of $\chi$.

\begin{figure}\vspace{3mm}
\begin{center}
\scalebox{1.3}{\includegraphics{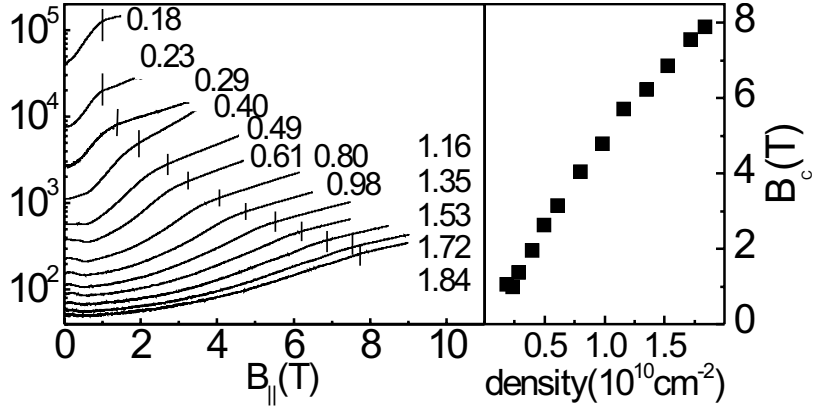}}
\end{center}
\caption{\label{23} Left hand panel:  magnetoresistance of GaAs/AlGaAs as a function of in parallel-field for different electron densities indicated in unit of $10^{10}$~cm$^{-2}$; the positions of the magnetic fields, $B_c$, required for full polarization are indicated following Tutuc~\etal (2002).  Right hand panel: $B_c$ as a function of density. Adopted from Zhu~\etal (2003).}
\end{figure}

The open circles in Fig.~\ref{zhu2003} denote $\chi(n_s)$ derived from
an alternative method for measuring $B_c$ based on a determination of the
parallel magnetic field corresponding to the ``knee'' of the
magnetoresistance curves, as shown in the left hand panel of Fig.~\ref{23}.  Earlier, this method yielded an anomalous and quite puzzling decrease of the susceptibility with decreasing density in both hole (Tutuc~\etal~2002) 
and electron (Noh~\etal~2002) systems in GaAs/AlGaAs.  This was in
disagreement with findings in Si MOSFETs, and argued against any
spontaneous spin polarization.  These results are now believed to
be in error for a number of possible reasons.  The field for full spin
polarization, marked by vertical bars in Fig.~\ref{23}~(left hand panel) and plotted as
a function of $n_s$ in Fig.~\ref{23}~(right hand panel), decreases with decreasing
carrier density and exhibits an apparent saturation below
$n_s\approx0.23\cdot10^{10}$~cm$^{-2}$.  Calculation based on this
saturation field would yield a spin susceptibility that
decreases below this density.  However, the saturation field may be
constant below this density due to the fact that the experiments are
performed at a temperature which is too high; if the temperature were
decreased further, the saturation field would probably continue to
decrease, consistent with the decrease of $B_\sigma$ shown in
Fig.~\ref{myriam1}~(a).  Another possible source of error is the finite
thickness of the electron layer in GaAs/AlGaAs heterostructures, which
leads to an increase in the effective mass with increasing parallel
magnetic field (Batke and Tu 1986).  Indeed, it has recently been shown
by Zhu~\etal~(2003) that the conclusion that the spin susceptibility
decreases with decreasing carrier density is erroneous and originates
from the fact that the effective mass depends on magnetic field.  The
effect of the finite thickness on the spin susceptibility was studied in
detail by Tutuc~\etal (2003).

\subsection{Effective mass and $g$-factor}\label{sec:gm}

\begin{figure}\hspace{2.3cm}
\scalebox{.4}{\includegraphics{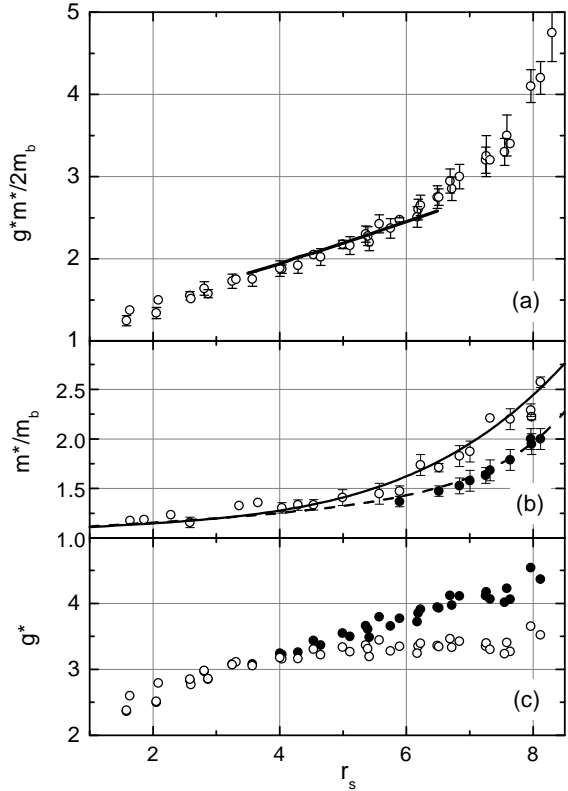}}
\caption{\label{pudalov2002} The parameters $g^*m^*$, $m^*$, and $g^*$ for different silicon samples as a function of $r_s$ (dots). The solid line in (a) shows the data of Okamoto~\etal (1999). The solid and open dots (b) and (c) correspond to two different methods of finding $m^*$ (see the text). The solid and dashed lines in (b) are polynomial fits for the two functions for $m^*(r_s)$.  Adopted from Pudalov~\etal (2002).}
\end{figure}

In principle, the increase of the spin susceptibility could be due to an
enhancement of either $g^*$ or $m^*$ (or both).  To obtain $g^*$ and $m^*$
separately, Pudalov~\etal (2002b) performed measurements of SdH
oscillations in superimposed and independently controlled parallel and
perpendicular magnetic fields.  Their results are shown in
Fig.~\ref{pudalov2002}.  The data were taken at relatively high
temperatures (above 300~mK), where the low-$n_s$ resistance depends
strongly on temperature.  Therefore, the conventional procedure of
calculating the effective mass from the temperature dependence of the
amplitude of the SdH oscillations is unreliable, as it assumes a
temperature-independent Dingle temperature.  Pudalov~\etal considered
two limiting cases: a temperature-independent $T_D$, and a $T_D$ that linearly increases with temperature; two sets of data based on these
assumptions are plotted in Fig.~\ref{pudalov2002}~(b) and (c).  Within
the uncertainty associated with the use of these two methods, both $g^*$
and $m^*$ were found to increase with decreasing $n_s$.

Shashkin~\etal (2002) used an alternative method to obtain $g^*$ and $m^*$ separately.  They analyzed the data for the temperature dependence of the conductivity in zero magnetic field using the recent theory of Zala~\etal (2001).  According to this theory, $\sigma$ is a linear function of temperature:
\begin{equation}
\frac{\sigma(T)}{\sigma_0}=1-A^*k_BT,\label{sigma}
\end{equation}
where the slope, $A^*$, is determined by the interaction-related parameters: the Fermi liquid constants $F_0^a$ and $F_1^s$:

\begin{equation}
A^*=-\frac{(1+8F_0^a)g^*m^*}{\pi\hbar^2n_s},\\
\frac{g^*}{g_0}=\frac{1}{1+F_0^a},\qquad\frac{m^*}{m_b}=1+F_1^s\label{A^*}
\end{equation}
(here $g_0=2$ is the ``bare'' g-factor.)  These relations allow a determination of the many-body enhanced $g^*$ factor
and mass $m^*$ separately using the data for the slope $A^*$ and the product $g^*m^*$.

\begin{figure}
\begin{center}
\scalebox{0.55}{\includegraphics{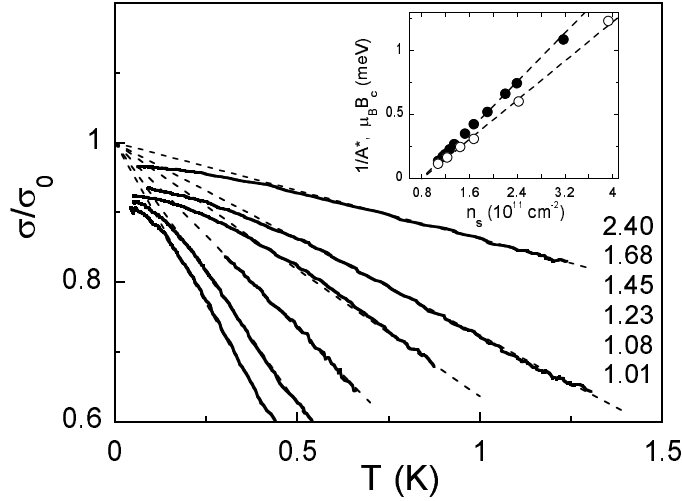}}
\end{center}
\caption{\label{sigma_gm} For a low-disordered silicon MOSFET, the temperature dependence of the normalized conductivity at different electron densities (indicated in units of $10^{11}$~cm$^{-2}$) above the critical electron density for the metal-insulator transition. The dashed lines are fits to the linear portions of the curves. The inset shows $1/A^*(n_s)$ (see Eq.~\protect\ref{sigma}) and $B_c(n_s)$ (open and solid circles, respectively). The dashed lines are continuations of the linear fits, which extrapolate to the critical electron density for the metal-insulator transition.  From Shashkin~\etal (2002).}
\end{figure}

\begin{figure}[b]
\begin{center}
\scalebox{.5}{\includegraphics{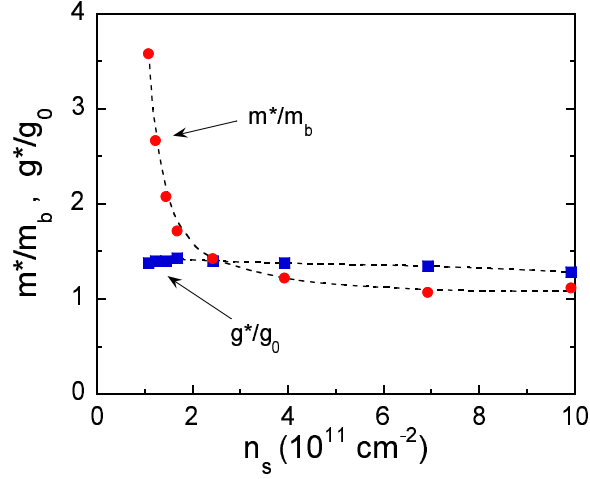}}
\end{center}
\caption{\label{gm} The effective mass (circles) and $g$ factor (squares) in a silicon MOSFET, determined from the analysis of the parallel field magnetoresistance and temperature-dependent conductivity, versus electron density.  The dashed lines are guides to the eye. From Shashkin~\etal (2002).}
\end{figure}

\begin{figure}
\begin{center}
\scalebox{.5}{\includegraphics{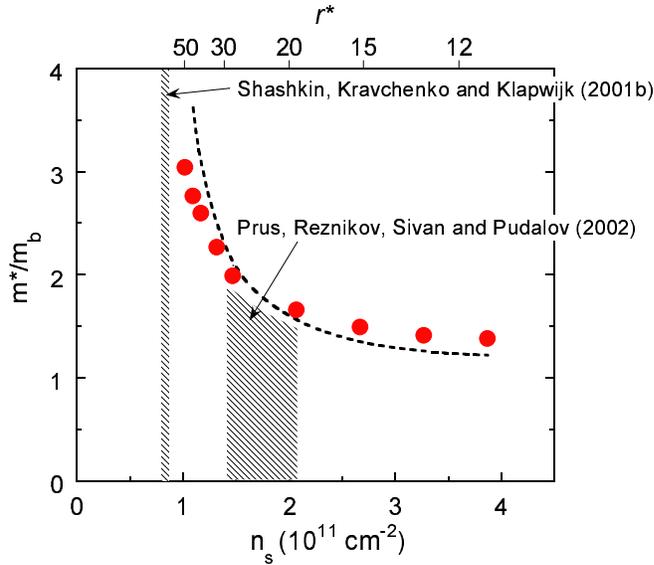}}
\end{center}
\caption{\label{m} For a silicon MOSFET, the effective mass obtained from an analysis of $\rho(B_\parallel)$ (dotted line) and from an analysis of SdH oscillations (circles).  Data of Shashkin~\etal (2003a).  The upper x-axis shows $r^*$ calculated using the latter value for the effective mass.  The shaded areas correspond to metallic regimes studied in papers by Shashkin~\etal (2001b) and Prus~\etal (2002), as labelled.}
\end{figure}

The dependence of the normalized conductivity on temperature,
$\sigma(T)/\sigma_0$, is displayed in Fig.~\ref{sigma_gm} at different
electron densities above $n_c$; the value of $\sigma_0$ used to normalize
$\sigma$ was obtained by extrapolating the linear interval of the
$\sigma(T)$ curve to $T=0$.  The inverse slope $1/A^*$ (open circles)
and $B_c(n_s)$ (solid circles) are plotted as a function of $n_s$ in the
inset to Fig.~\ref{sigma_gm}.  Over a wide range of electron densities,
$1/A^*$ and $\mu_BB_c$ are close to each other; the
low-density data for $1/A^*$ are approximated well by a linear
dependence which extrapolates to the critical electron density $n_c$ in
a way similar to $B_c$.

Values of $g^*/g_0$ and $m^*/m_b$ determined from this analysis are shown as a function of the electron density in Fig.~\ref{gm}.  In the high $n_s$ region (relatively weak interactions), the enhancement of both $g$ and $m$ is relatively small, both values increasing slightly with decreasing electron density in agreement with earlier data (Ando~\etal 1982).  Also, the renormalization of the $g$-factor is dominant compared to that of the effective mass, consistent with theoretical studies (Iwamoto 1991; Kwon~\etal 1994; Chen and Raikh 1999).  In contrast, the renormalization at low $n_s$ (near the critical region), where $r_s\gg1$, is much more striking.  As the electron density is decreased, the renormalization of the effective mass increases markedly with decreasing density while the $g$ factor remains relatively constant. Hence, this analysis indicates that it is the effective mass, rather than the $g$-factor, that is responsible for the strongly enhanced spin susceptibility near the metal-insulator transition.  To verify this conclusion, Shashkin~\etal (2003a, 2003b) used an independent method to determine the effective mass through an analysis of the temperature dependence of the SdH oscillations similar to the analysis done by Smith and Stiles (1972) and Pudalov~\etal (2002b), but extended to much lower temperatures where the Dingle temperature is expected to be constant and the analysis reliable.  In Fig.~\ref{m}, the effective mass obtained by this method is compared with the results obtained by the procedure described above (the dotted line).  The quantitative agreement between the results obtained by two independent methods supports the validity of both.  The data for the effective mass are also consistent with data for spin and cyclotron gaps obtained by magnetocapacitance spectroscopy (Khrapai~\etal 2003).

\begin{figure}
\begin{center}
\scalebox{.5}{\includegraphics{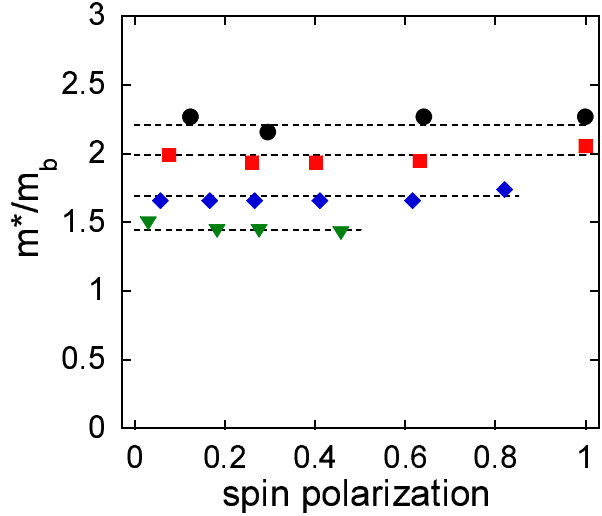}}
\end{center}
\caption{\label{mass}  For silicon MOSFETs, the effective mass versus the
degree of spin polarization for the following electron densities in units
of $10^{11}$~cm$^{-2}$: 1.32 (dots), 1.47 (squares), 2.07 (diamonds), and
2.67 (triangles). The dashed lines are guides to the eye. From
Shashkin~\etal (2003a).}
\end{figure}

To probe a possible connection between the effective mass enhancement
and spin and exchange effects, Shashkin~\etal (2003a, 2003b) introduced a
parallel magnetic field component to align the electrons' spins. In
Fig.~\ref{mass}, the effective mass is shown as a function of the spin
polarization, $P=(B_\perp^2+B_\parallel^2)^{1/2}/B_c$.  Within the
experimental accuracy, the effective mass does not depend on $P$.
Therefore, the enhancement of $m^*$ near the MIT is robust, and the
origin of the mass enhancement has no relation to the electrons' spins
and exchange effects.

\subsection{Electron-electron interactions near the transition}
\label{sec:r_s}

The interaction parameter, $r^*\equiv E_C/E_F$, is generally assumed to
be equal to the dimensionless Wigner-Seitz radius, $r_s=1/(\pi
n_s)^{1/2}a_B$, and, hence, proportional to $n_s^{-1/2}$.  However, this
is true only if the effective mass does not depend on $n_s$; close to
the transition, where the effective mass is strongly enhanced, the
deviations from the square root law become important, and the
interaction parameter is larger than the Wigner-Seitz radius by a factor
of $m^*/m_b$ (or $2m^*/m_b$ in silicon MOSFETs, where an additional factor
of 2 derives from the valley degeneracy).

\begin{figure}
\begin{center}
\scalebox{.45}{\includegraphics{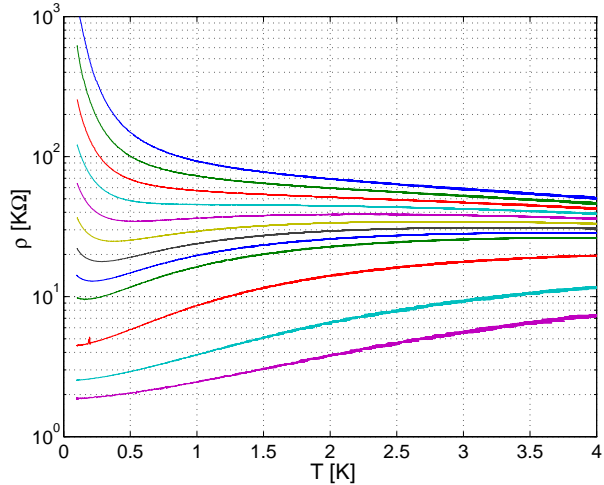}}
\end{center}
\caption{\label{reznikov_R(T)} $\rho(T)$ in a moderately disordered silicon MOSFET. Densities from top to bottom are $n_s=1.20$ to $1.44$ (in 0.03 steps), $1.56$, $1.8$ and $2.04\cdot10^{11}$~cm$^{-2}$. From Prus~\etal (2002).}
\end{figure}

Near the metal-insulator transition, $r^*$ grows rapidly due to the
sharp increase of the effective mass, as shown on the upper $x$-axis in
figure~\ref{m}; here $r^*$ is calculated using the effective mass
obtained from the analysis of the SdH oscillations.  We note that since
at low $n_s$ this method yields a somewhat smaller effective
mass than the method based on the analysis of $\rho(B_\parallel)$, the
plotted values represent the most conservative estimate for $r^*$.  Due
to the rapid increase of $m^*$ near the transition, even small changes
in the electron density may lead to large changes in $r^*$.  For
example, the transition to a localized state in a low-disordered sample
used by Shashkin~\etal (2001b) occurs at
$n_s=0.795 \cdot 10^{11}$~cm$^{-2}$ (see Fig.~\ref{R(T)inB_ours}), while
in a sample of lesser quality used by Prus~\etal (2002), it occurs at
$1.44 \cdot 10^{11}$~cm$^{-2}$ (see Fig.~\ref{reznikov_R(T)}). The areas
corresponding to the metallic regimes studied in these papers are shaded
in Fig.~\ref{m}; the corresponding values of $r^*$ differ dramatically
(the actual values of $r^*$ in the study by Shashkin~\etal are not known
because the effective mass has not been determined near the transition;
the lower boundary for $r^*$ is 50).  The difference in $r^*$ may
account for the qualitative change in the behaviour of the resistance in
the two samples: in the more disordered sample, there is no
temperature-independent curve (the separatrix), and some of the curves
which look ``metallic'' at higher temperature are insulating below a few
hundred mK, suggesting that localization becomes dominant in this sample
as $T\rightarrow0$.

\section{HOW DOES ALL THIS FIT THEORY?}

The possibility that a $B=0$ metallic state in 2D can be stabilized by interactions was first suggested by Finkelstein (1983, 1984) and Castellani~\etal (1984). In this theory, the combined effects of disorder and interactions were studied by perturbative renormalization group methods. It was found that as the temperature is decreased, the resistivity increases and then decreases at lower temperatures, suggesting that the system is approaching a low-temperature metallic state.  An external magnetic field, via Zeeman splitting, drives the system back to the insulating state.  These predictions of the theory are in qualitative agreement with experiments.

However, an interaction parameter scales to infinitely large values before zero temperature is reached, and the theory thus becomes uncontrolled; this scenario has consequently not received general acceptance.  It should also be noted that the theory may not be applicable to the current experiments since it was developed for the diffusive regime: $T\ll\hbar/\tau$, where $\tau$ is the elastic mean-free time extracted from the Drude conductivity (Boltzmann constant is assumed to be equal to 1 throughout this section).  This condition corresponds to the low-temperature limit $T\ll T_F\rho/(h/e^2)$.  Since the Fermi temperature, $T_F$, is rather low at the small carrier densities considered here, the above condition is satisfied only close to the transition, where $\rho$ becomes of the order of $h/e^2$.  In the experiments, however, the characteristic decrease of the resistance with decreasing temperature often persists into the relatively high-temperature {\em ballistic} regime $T>\hbar/\tau$ (or $T>T_F\rho/(h/e^2)$).  Altshuler~\etal (2001) and Brunthaler~\etal (2001) have interpreted this observation as evidence that the mechanism responsible for the strong temperature dependence cannot originate from quantum interference.

For the ballistic region, calculations in the random phase approximation were carried out two decades ago by Stern (1980), Gold and Dolgopolov (1986), Das~Sarma (1986) and were recently refined by Das~Sarma and Hwang (1999).  These theories predict a linear temperature dependence for the conductivity with metallic-like slope (${\rm d}\sigma/{\rm d}T<0$) regardless of the strength of the interactions. The spin degree of freedom is not important in this theory and enters only through the Fermi energy, which is a factor of two larger for the spin-polarized than for the unpolarized system.  Therefore, the application of a (parallel) magnetic field does not eliminate the metallic temperature dependence.

The two limits --- diffusive and ballistic --- had until recently been
assumed to be governed by different physical processes.  However,
Zala~\etal (2001) have shown that the temperature dependence of the
conductivity in the ballistic regime originates from the same physical
process as the Altshuler-Aronov-Lee corrections: coherent scattering of
electrons by Friedel oscillations. In this regime the correction is
linear in temperature, as is the case for the results mentioned in the
previous paragraph.  However, the value and even the {\em sign} of the
slope depends on the strength of electron-electron interaction, the
slope being directly related to the renormalization of the spin
susceptibility (see also Gold 2001, 2003).  By aligning the spins, a
magnetic field causes a positive magnetoresistance and changes the
temperature dependence of the conductivity from metallic-like to
insulating-like (Herbut 2001; Zala~\etal 2002; Gold 2003), in agreement 
with experiments.

\subsection{The diffusive regime: Renormalization group analysis}
\label{sec:rg}

It is well known that in the diffusive limit ($T\tau\ll\hbar$), electron-electron interactions in two dimensional disordered systems lead to logarithmically divergent corrections to the conductivity given by:
\begin{equation}
\delta\sigma=-\frac{e^2}{2\pi^2\hbar}\ln\left(\frac{\hbar}{T\tau}\right)
\Big[1 + 3\left(1-\frac{\ln(1+F_0^\sigma)}{F_0^\sigma}
\right) \Big],
\end{equation}
where $F_0^\sigma$ is the interaction constant in the triplet channel
which depends on the interaction strength.  The sign of this
logarithmically divergent correction may be positive (metallic-like) or
negative (insulating-like), depending on the value of $F_0^\sigma$.

Experimentally, the diffusive regime is realized in a relatively narrow range of
electron densities near the metal-insulator transition, where the
resistivity is of the order of $h/e^2$. Punnoose and Finkelstein (2002)
have convincingly demonstrated that in this region, the temperature
dependence of the resistivity can be understood within the
renormalization group theory describing the effect of the
electron-electron interactions on the propagation of diffusive
collective modes, with the {\em delocalizing} component becoming
dominant in dilute 2D systems.

In 2D, the renormalization group equation describing the evolution of
the resistivity has been derived previously by Finkelstein:
\begin{equation}
\frac{dg}{d\xi }=g^{2}\left[1+1-3\left(\frac{1+\gamma_{2}}{\gamma_{2}}\ln
(1+\gamma_{2})-1\right)\right]~.
\label{eqn:1-val-g}
\end{equation}
Here $\xi=-\ln (T\tau/\hbar )$, the dimensionless parameter $g$ is the
conductance in units of $e^2/\pi h$ and $\gamma_2$ is the coupling
constant.  The first term in the square brackets of
Eq.~\ref{eqn:1-val-g} corresponds to the weak localization correction
(quantum interference; Abrahams~\etal 1979), while the second term is
the contribution of the singlet density mode which is due to the long
range nature of the Coulomb interaction (Altshuler, Aronov and Lee 1980).
The last term describes the contribution of the three triplet modes.
Note that the last two terms have opposite signs, favouring localization
and delocalization, respectively.  The resulting dependence $g(\xi )$
becomes delocalizing when $\gamma_{2}>\gamma_{2}^{\ast}=2.04$.  This
requires the presence of rather strong electron correlations.

In the case of two distinct valleys (as in (100) silicon MOSFETs),
Eq.~\ref{eqn:1-val-g} can be easily generalized to:
\begin{equation}
\frac{dg}{d\xi }=g^{2}\left[2+1-15\left(\frac{1+\gamma_{2}}{\gamma_{2}}\ln
(1+\gamma_{2})-1\right)\right]~.
\label{eqn:2-val-g}
\end{equation}
The difference between the numerical factors in Eq.~\ref{eqn:1-val-g}
and Eq.~\ref{eqn:2-val-g} results from the number of degrees of freedom
in each case. The weak localization term becomes twice as large. The
difference in the number of the multiplet modes increases the
coefficient of the $\gamma_{2}$ term from $3$ to $15$. As a result of
these modifications, the value of $\gamma_{2}$ required for the
dependence $g(\xi)$ to become delocalizing is reduced to
$\gamma_{2}^{\ast}=0.45.$ which makes it easier in the case of two
valleys to reach the condition where the resistivity decreases with
decreasing temperature.

In conventional conductors the initial values of $\gamma_{2}$ are small,
and the net effect is in favour of localization. In dilute systems,
however, this value is enhanced due to electron-electron correlations.
In addition, $\gamma_{2}$ also experiences logarithmic corrections due
to the disorder (Finkelstein 1983, 1984; Castellani~\etal 1984, 1998).
The equation describing the renormalization group evolution of
$\gamma_{2}$ is the same for the case of one and two valleys:
\begin{equation}
\frac{d\gamma_{2}}{d\xi }=g~\frac{(1+\gamma_{2})^{2}}{2}~.
\label{eqn:val-r}
\end{equation}
It follows from this equation that $\gamma_{2}$ increases monotonically
as the temperature is decreased.  Once it exceeds $\gamma_{2}^{\ast}$,
the resistivity passes through a maximum. Even though the initial values
of $g$ and $\gamma_{2}$ are not universal, the flow of $g$ can be
described by a universal function $R(\eta)$ (Finkelstein, 1983):
\begin{equation}
g=g_{\max}R(\eta)\mbox{ and }\eta=g_{\max}\ln(T_{\max }/T)~,
\label{eqn:rgsol}
\end{equation}
where $T_{\max}$ is the temperature at which $g$ reaches its maximum
value $g_{\max }$.

\begin{figure}
\begin{center}
\scalebox{.65}{\includegraphics{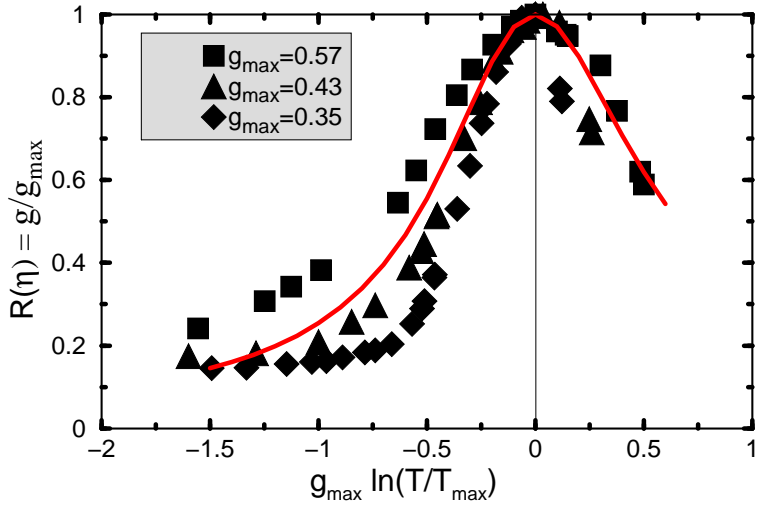}}
\end{center}
\caption{\label{punnoose} Data for silicon MOSFETs at $n_s=0.83$, $0.88$ and $0.94\cdot 10^{11}$~cm$^{-2}$ from Pudalov~\etal (1998) are scaled according to Eq.~\protect\ref{eqn:rgsol}. The solid line corresponds to the solution of the renormalization group, Eqs.~\protect\ref{eqn:2-val-g} and \protect\ref{eqn:val-r}; no adjustable parameters were used in this fit. From Punnoose and Finkelstein (2002).}
\end{figure}

In Fig.~\ref{punnoose}, the theoretical calculations are compared with the experimental data obtained by Pudalov~\etal (1998).  Since the renormalization group equations were derived in the lowest order in $g$ and cannot be applied in the critical region where $g>1$, only curves with maximum $g$ ranging from $g_{\max}\approx0.3$ to $g_{\max}\approx0.6$ are shown in the figure.  The decrease in the resistivity by up to a factor of five, as well as its saturation, are both captured in the correct temperature interval by this analysis without any adjustable parameters.

In agreement with experiment, this universal behaviour can be observed only
in ultra clean samples (with negligible inter-valley scattering) and will
not be found in samples that are only moderately clean.  In disordered
silicon MOSFETs, a description in terms of an effective {\em single} valley
is relevant, and the large value of $\gamma_2^\star=2.04$ makes it
difficult for the non-monotonic $\rho(T)$ to be observed as the initial
values of $\gamma_2$ are usually much less than $2.04$.  As a result, for
$g$ near the critical region, the resistivity becomes insulating-like
instead of going through the maximum.

Therefore, in ultra clean silicon MOSFETs, the behaviour of the
resistivity not far from the transition is quantitatively well described
by the renormalization group analysis that considers the interplay of
the electron-electron interactions and disorder when the electron band
has two distinct valleys.  The theory in this case remains in control
down to exponentially low temperatures, unlike the single valley case,
where $\gamma_{2}$ diverges at $\eta\approx1$, or at temperatures just
below $T_{\max}$. A parallel magnetic field provides a Zeeman splitting
and gives rise to positive magnetoresistance (Lee and Ramakrishnan 1982;
Finkelstein 1984; Castellani~\etal 1998). In a very strong magnetic
field, when the electrons are completely spin-polarized, the system
becomes ``spinless''. In this case, the system always scales to an
insulator, and in the weak disorder limit, a universal logarithmic
temperature dependence is expected: $$\sigma(T)=\sigma_0+(e^2/\pi
h)(2-2\ln2)\ln(T\tau/\hbar).$$

\subsection{Farther from the transition (the ballistic regime)}

Far from the transition ($n_s\gg n_c$), the inverse scattering time
$\tau^{-1}$ is often smaller than the temperature at which the experiments are
performed, and one is in the ballistic regime: $T\tau\gg\hbar$.  The
interaction theory by Zala~\etal (2001) considers coherent electron
scattering off the Friedel oscillations.  The phase of the Friedel
oscillation is such that the wave scattered from the impurity interferes
constructively with the wave scattered from the oscillation, leading to
a quantum correction to the Drude conductivity, $\sigma_0$.  As has
already been mentioned, this correction ($\Delta\sigma$), which is
logarithmic in the diffusive regime, becomes linear in the ballistic
regime:
\begin{equation}
\Delta\sigma\left( T\right)=\frac{e^2}{\pi\hbar}\frac{T\tau}\hbar\left(
1+\frac{3F_0^\sigma}{1+F_0^\sigma}\right)=\sigma_0\left(
1+\frac{3F_0^\sigma}{1+F_0^\sigma }\right)\frac{T}{T_F}\label{eq1}
\end{equation}
where $F_0^\sigma $ is the Fermi liquid interaction parameter in the
triplet channel.  As in the diffusing regime, the sign of ${\rm d}\rho/{\rm
d}T$ depends on the constant $F_0^\sigma $.

\begin{figure}
\begin{center}
\scalebox{1}{\includegraphics{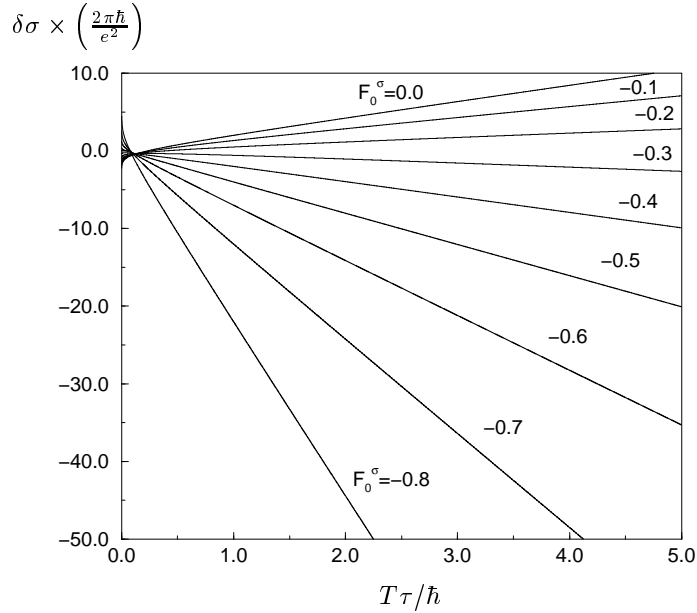}}
\end{center}
\caption{\label{aleiner} Total interaction correction to the
conductivity. The curve for $F_0^\sigma=0$ corresponds to the universal
behaviour of a completely spin polarized electron gas.  From Zala~\etal
(2001).}
\end{figure}

The total correction to the conductivity is shown in Fig.~\ref{aleiner}
for different values of $F_0^\sigma$. The divergence at low temperature
is due to the usual logarithmic correction (Altshuler, Aronov and Lee 1980;
Finkelstein 1983, 1984; Castellani~\etal 1984). Although the exact value
of $F_0^\sigma$ cannot be calculated theoretically (in particular, its
relation to the conventional measure of the interaction strength, $r_s$,
is unknown for $r_s>1$), it can in principle be found from a measurement
of the Pauli spin susceptibility.  The correction to the conductivity is
almost always monotonic, except for a narrow region
$-0.45<F_0^\sigma<-0.25$.

As in the diffusive limit, magnetic field in the ballistic regime
suppresses the correction in the triplet channel in Eq.~(\ref{eq1}),
resulting in a universal, positive correction to the Drude conductivity
in magnetic field, $\sigma_0^B$, and hence in the insulating-like
behaviour of $\sigma(T)$:
\begin{equation}
\Delta \sigma =\sigma _0^B\frac T{T_F}\mbox{ at }B\geq B_s, \label{eq2}
\end{equation}
where $B_s$ is the field corresponding to full spin polarization of the
2D system.

\begin{figure}\vspace{5mm}
\begin{center}
\scalebox{.55}{\includegraphics{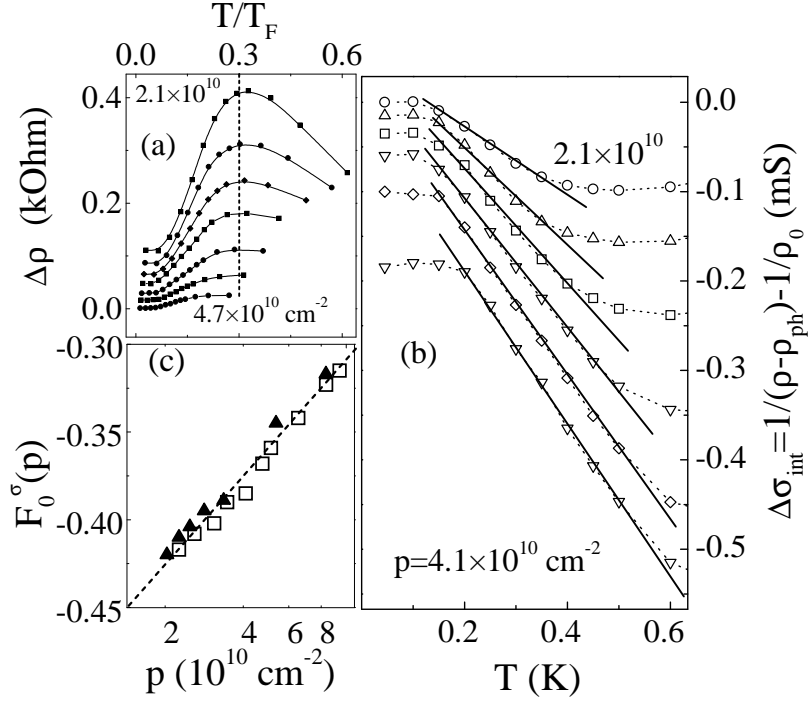}}
\end{center}
\caption{\label{proskuryakov} (a)~$\Delta\rho$ as a function of the dimensionless temperature ($T/T_F$) at different $p$ in a $p$-type GaAs/AlGaAs heterostructure.  (For clarity, curves in (a) and (b) are offset vertically.) (b)~The same data as in (a) plotted as conductivity, with linear fits. (c)~Fermi liquid parameter versus hole density.  Open symbols show the result obtained from the analysis of $\rho(T)$ at zero magnetic field; closed symbols show results obtained from the parallel-field magnetoresistance.  From Proskuryakov~\etal (2002).}
\end{figure}

Proskuryakov~\etal (2002) were the first to perform a quantitative comparison of experimental data (obtained on $p$-type GaAs/AlGaAs) with the above theory.  Their data for the temperature-induced corrections to the conductivity are plotted in Fig.~\ref{proskuryakov}~(b).  (To extract corrections to the conductivity, Proskuryakov~\etal used the $\Delta\rho(T)$ dependence shown in Fig.~\ref{proskuryakov}~(a) obtained from raw resistivity data by subtracting the contribution of phonon scattering.)  A linear fit of $\Delta\sigma(T)$ yields the parameter $F_0^\sigma $ shown in Fig.~\ref{proskuryakov}~(c) for different $p$.  The interaction constant obtained is negative and provides the metallic slope of $\sigma(T)$.  As expected, the slope increases with decreasing density. When extrapolated to the crossover point from metallic to insulating behaviour ($p\approx1.5\cdot10^{10}$~cm$^{-2}$), the interaction constant is much higher than the value of $F_0^\sigma=-1$ for the Stoner instability.  The transition from metallic-like to insulating-like $\sigma(T)$ in their sample is unlikely to be related to any dramatic change in magnetic properties and is probably caused by Anderson localization.

\begin{figure}
\begin{center}
\scalebox{.82}{\includegraphics{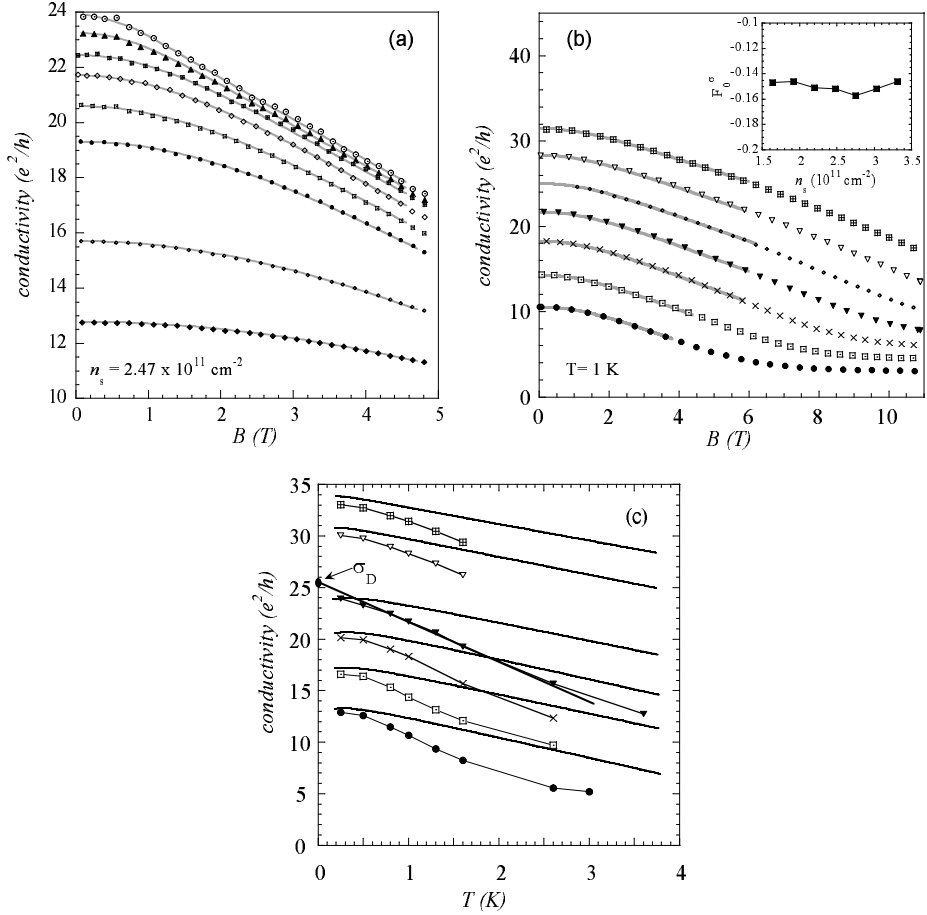}}
\end{center}
\caption{\label{james} For parameters $\Delta=1$~K and $F_0^\sigma=-0.15$, the magnetoconductivity of a silicon MOSFET: (a)~For different temperatures at a fixed density $n_s=2.47\cdot 10^{11}$~cm$^{-2}$; $T=0.25$, $0.5$, $0.8$, $1.0$, $1.3$, $1.6$, $2.6$ and $3.6$~K (from top). (b)~For different densities at a fixed temperature of $1$~K; $n_s=3.3$, $3.0$, $2.75$, $2.47$, $2.19$, $1.92$ and $1.64\cdot10^{11}$~cm$^{-2}$ (from top).  (c)~The conductivity as a function of temperature for the densities $n_s=3.3$, $3.0$, $2.47$, $2.19$, $1.92$ and $1.64\cdot10^{11}$~cm$^{-2}$ (from top).  Symbols and solid lines denote data and theory, respectively. From Vitkalov~\etal (2003).}
\end{figure}

Comparison between the experimentally measured $\sigma(T)$ in the ballistic regime with the predictions of Zala~\etal was also carried out in silicon MOSFETs by Shashkin~\etal (2002; see sec.~\ref{sec:gm}), Vitkalov~\etal (2003) and Pudalov~\etal (2003) (see also Das~Sarma and Hwang 2003 and Shashkin~\etal 2003c) and in $p$-type SiGe heterostructures by Coleridge~\etal (2002).  As shown in Fig.~\ref{james}~(a) and (b), Vitkalov~\etal reported quantitative agreement between the theory and their magnetoresistance data in silicon MOSFETs.  However, the values of the valley splitting $\Delta$ and Fermi liquid parameter $F_0^\sigma$ required to obtain fits to the field dependence yield curves that are inconsistent with the observed temperature dependence in zero field, as shown in Fig.~\ref{james}~(c). This was attributed to the neglect of additional scattering terms that affect the temperature dependence more strongly than the field dependence.  Despite this quantitative discrepancy, the theory of Zala~\etal provides a reasonable description of the conductivity of strongly interacting 2D systems in the ballistic regime.

\subsection{Approaches not based on Fermi liquid}

The calculations of Finkelstein (1983, 1984), Castellani~\etal (1984) and Zala~\etal (2001) use the Fermi liquid as a starting point.  As discussed earlier, $r_s$ becomes so large that the theory's applicability is in question near the transition (Chamon~\etal 2001; for a review, see Varma~\etal 2002). Moreover, the behaviour of the effective mass reported by Shashkin~\etal (2003a, 2003b) in the vicinity of the transition is unlikely to be consistent with a Fermi liquid model.

\begin{figure}
\begin{center}
\scalebox{.55}{\includegraphics{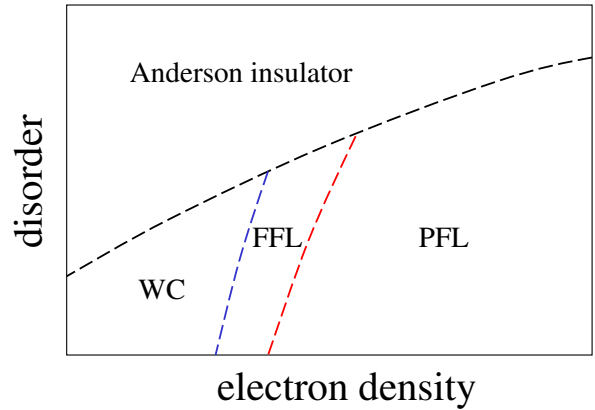}}
\end{center}
\caption{\label{diagram} Schematic phase diagram corresponding to the numerical results of Attaccalite~\etal 2002. The symbols WC, FFL and PFL correspond to the Wigner crystal, ferromagnetic Fermi liquid and paramagnetic Fermi liquid phases, respectively.}
\end{figure}

Very little theory has been developed for strongly interacting systems for which $r_s$ is large but below the expected Wigner crystallization. Several candidates have been suggested for the ground state of the strongly interacting 2D system, among them (i)~a Wigner crystal characterized by spatial and spin ordering (Wigner 1934), (ii)~an itinerant ferromagnet with spontaneous spin ordering (Stoner 1946), and (iii)~a paramagnetic Fermi liquid (Landau 1957). According to numerical simulations (Tanatar and Ceperley 1989), Wigner crystallization is expected in a very dilute regime, when $r_s$ reaches approximately 35.  This value has already been exceeded in the best $p$-type GaAs/AlGaAs heterostructures; however, no dramatic change in transport properties has occurred at the corresponding density. Recent detailed numerical simulations (Attaccalite~\etal 2002) have predicted that in the range of the interaction parameter $25<r_s<35$ prior to the crystallization, the ground state of the system becomes an itinerant ferromagnet.  The corresponding schematic phase diagram is shown in Fig.~\ref{diagram}. As discussed earlier, there are experimental indications that a spontaneous spin polarization may occur at a finite electron density in silicon MOSFETs (and possibly in GaAs/AlGaAs heterostructures). Chakravarty~\etal (1999) have proposed a more complicated phase diagram where the low-density insulating state is a Wigner glass, a phase that has quasi-long-range translational order and competing ferromagnetic and antiferromagnetic spin-exchange interactions.  The transition between insulating and metallic states within this theory is the melting of the Wigner glass, where the transition is second order due to the disorder.

\begin{figure}
\scalebox{.8}{\includegraphics{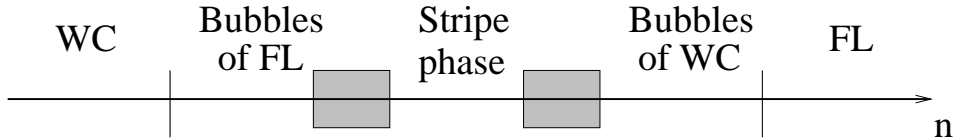}}
\caption{\label{spivak} Phase diagram of the 2D electron system at $T=0$ suggested by Spivak (2002). The symbols WC and FL correspond to the Wigner crystal and the Fermi liquid phases, respectively. The shaded regions correspond to phases which are more complicated than the bubble and the stripe phases.}
\end{figure}

Spivak (2002) predicted the existence of an intermediate phase between the Fermi liquid and the Wigner crystal with a first order transition in a clean electron system.  The suggested phase diagram is shown in Fig.~\ref{spivak}.  In analogy with He$^3$, where $m^*$ increases approaching the crystallization point, Spivak (2001) suggested in an earlier paper that the renormalization of $m^*$ is dominant compared to that of the $g$-factor as the transition is approached, and that $m^*$ should increase with magnetic field.  Although the increase of the mass is in agreement with the experimental results of Shashkin~\etal (2002, 2003a, 2003b), the suggested increase of $m^*$ with the degree of spin polarization is not confirmed by their data.  The strong increase of the effective mass near crystallization also follows from the approach used by Dolgopolov (2002), who has applied Gutzwiller's variational method (Brinkman and Rice 1970) to silicon MOSFETs, and from the dynamical mean-field theory (Tanaskovi\'{c}~\etal 2003).  However, the predicted dependence of $m^*$ on the degree of spin polarization is again at odds with the experiment.  Dharma-wardana (2003) has argued that in two-valley systems, correlation effects outweigh exchange, and a coupled-valley state is formed at $r_s\approx5.4$ leading to strong enhancement of the effective mass, which in this case is only weakly dependent on the degree of spin polarization.

\section{CONCLUSIONS}

Significant progress has been achieved during the past few years in
understanding the metallic state found a decade ago in strongly
interacting 2D systems.  There is now considerable evidence that the
strong metallic temperature dependence of the resistance in these systems
is due to the delocalizing effects of strong electron-electron
interactions.  In ultra-clean silicon systems, the temperature dependence
of the resistivity is universal near the metal-insulator transition and
is quantitatively well described by the renormalization group
theory; deep in the metallic state, in the ballistic regime, the
temperature dependence of the resistance can be explained by coherent
scattering of electrons by Friedel oscillations.  In both cases, an
external magnetic field quenches the delocalizing effect of interactions
by aligning the spins, causing a giant positive magnetoresistance.

The metal-insulator transition is not yet understood theoretically.  In
silicon MOSFETs, various experimental methods provide evidence for a sharp
increase and possible divergence of the spin susceptibility at some
finite sample-independent electron density, $n_\chi$, which is at or very
near the critical density for the MIT in high mobility samples.  Unlike
the Stoner instability which entails an increase in the $g$-factor, the
increase in the susceptibility in these systems is due to an increase of
the effective mass.  The effective mass is, in turn, found to be
independent of the degree of spin polarization, implying that the
increase is not due to spin exchange, in disagreement with the Fermi
liquid model.  A similar increase of the spin susceptibility is observed
in GaAs/AlGaAs heterostructures, but it is not yet clear whether or not
it points to a spontaneous spin polarization at a finite carrier density.

The fact that the $B=0$ metal-insulator transition in the least disordered
silicon samples occurs at or very close to $n_\chi$ indicates that the
transition in such samples is a property of a clean electron system and is
not driven by disorder.  Quantum localization appears to be suppressed near
the transition in these systems.  In lower mobility samples, the
localization transition occurs at electron densities much higher than
$n_\chi$ and may be driven by disorder.

In closing, we note that most of the work done in these dilute two-dimensional systems concerns the transport behaviour, as transport measurements are relatively straightforward (despite problems associated with high impedance electrical contacts at low densities, the need to ensure proper cooling of the electrons system and so on).  Studies done to date, many of which are reviewed here, include measurements of the resistivity as a function of temperature and magnetic field, the Hall coefficient, Shubnikov-de~Haas oscillations and measurements of noise. Results have also been reported for the compressibility (Dultz and Jiang 2000; Ilani~\etal 2000, 2001; see also Si and Varma 1998 and Fogler 2003) and capacitive measurements (Khrapai~\etal 2003) from which one can obtain information about the chemical potential, the density of states, and which provide a measure of the magnetization, as discussed in sec.~\ref{spin}.

Many properties that would yield crucial information have not been investigated.  Thermodynamic measurements such as specific heat and direct measurements of magnetization would be particularly illuminating; however, these are very difficult, if not impossible, to perform at this time due to the very small number of electrons available in a dilute, two-dimensional layer.  Other experiments which could provide valuable information include tunnelling and different resonance techniques.  Measurements of one, several, or perhaps all of these may be required for a full understanding of the enigmatic and very interesting behaviour of strongly interacting electrons (or holes) in two dimensions.

\ack

We are grateful to G~B Bachelet, V~T Dolgopolov, A~M Finkelstein, T~M Klapwijk, S Moroni, B~N Narozhny, D Neilson, A~A Shashkin, C Senatore, B Spivak and S~A Vitkalov for useful discussions.  SVK is supported by the National Science Foundation grant DMR-0129652 and the Sloan Foundation; MPS is supported by Department of Energy grant no.\ DE-FG02-84ER45153 and National Science Foundation grant DMR-0129581.

\References

\item[] Abrahams E, Anderson P W,  Licciardello D C and Ramakrishnan T V 1979 Scaling theory of localization: Absence of quantum diffusion in two dimensions \PRL {\bf 42} 673-676

\item[] Abrahams E, Kravchenko S~V and Sarachik M~P 2001 Metallic behavior and related phenomena in two dimensions \RMP {\bf 73} 251-266

\item[] Altshuler B~L, Aronov A~G and Lee P~A 1980 Interaction effects in disordered Fermi systems in two dimensions \PRL {\bf 44} 1288-1291

\item[] Altshuler B~L and Maslov D~L 1999 Theory of metal-insulator transitions in gated semiconductors \PRL {\bf 82} 145-148

\item[] Altshuler B~L, Maslov D~L and Pudalov V~M 2000 Metal-insulator transition in 2D: Anderson localization in temperature-dependent disorder? \PSS b {\bf 218} 193-200

\item[] Altshuler B~L, Maslov D~L and Pudalov V~M 2001 Metal-insulator transition in 2D: resistance in the critical region {\it Physica} E {\bf 9} 209-225

\item[] Ando T, Fowler A~B and Stern F 1982 Electronic-properties of two-dimensional systems \RMP {\bf 54} 437-672

\item[] Attaccalite C, Moroni S, Gori-Giorgi P and Bachelet G~B 2002 Correlation energy and spin polarization in the 2D electron gas \PRL {\bf 88} 256601

\item[] Batke E and Tu C~W 1986 Effective mass of a space-charge layer on GaAs in a parallel magnetic-field \PR B {\bf 34} 3027-3029

\item[] Bishop D~J, Tsui D~C and Dynes R~C 1980 Nonmetallic conduction in electron inversion layers at low temperatures \PRL {\bf 44} 1153-1156

\item[] Bishop D~J, Dynes R~C and Tsui D~C 1982 Magnetoresistance in Si metal-oxide-semiconductor field-effect transistors -- evidence of weak localization and correlation \PR B {\bf 26} 773-779

\item[] Bogdanovich S and Popovi\'{c} D 2002 Onset of glassy dynamics in a two-dimensional electron system in silicon \PRL {\bf 88} 236401

\item[] Brinkman W~F and Rice T~M 1970 Application of Gutzwiller's variational method to the metal-insulator transition \PR B\ {\bf 2} 4302-4304

\item[] Brunthaler G, Prinz A, Bauer G and Pudalov V~M 2001 Exclusion of quantum coherence as the origin of the 2D metallic state in high-mobility silicon inversion layers \PRL {\bf 87} 096802

\item[] Castellani C, Di~Castro C, Lee P~A and Ma M 1984 Interaction-driven metal-insulator transitions in disordered fermion systems \PR B {\bf 30} 527-543

\item[] Castellani C, Di~Castro C and Lee P~A 1998 Metallic phase and metal-insulator transition in two-dimensional electronic systems \PR B {\bf 57} R9381-R9384

\item[] Chakravarty S, Kivelson S, Nayak C and Voelker K 1999 Wigner glass, spin liquids and the metal-insulator transition {\it Phil.\ Mag.\ B} {\bf 79} 859-868

\item[] Chamon C, Mucciolo E~R and Castro~Neto A~H 2001 {\it P}-wave pairing and ferromagnetism in the metal-insulator transition in two dimensions \PR B {\bf 64} 245115

\item[] Chen G-H and Raikh M~E 1999 Exchange-induced enhancement of spin-orbit coupling in two-dimensional electronic systems \PR B {\bf 60} 4826-4833

\item[] Coleridge P~T, Williams R~L, Feng Y and Zawadzki P 1997 Metal-insulator transition at $B=0$ in $p$-type SiGe \PR B {\bf 56} R12764-R12767

\item[] Coleridge P~T, Sachrajda AS and Zawadzki P 2002 Weak localization, interaction effects, and the metallic phase in $p$-SiGe \PR B {\bf 65} 125328

\item[] Das~Sarma S 1986 Theory of finite-temperature screening in a disordered two-dimensional electron-gas \PR B {\bf 33} 5401-5405

\item[] Das~Sarma S and Hwang E~H 1999 Charged impurity-scattering-limited low-temperature resistivity of low-density silicon inversion layers \PRL {\bf 83} 164-167

\item[] Das~Sarma S and Hwang E~H 2000 Parallel magnetic field induced giant magnetoresistance in low density quasi-two-dimensional layers \PRL {\bf 84} 5596-5599

\item[] Das~Sarma S and Hwang E~H 2003 On the temperature dependence of 2D ``metallic'' conductivity in Si inversion layers at intermediate temperatures {\it Preprint} cond-mat/0310260

\item[] Dharma-wardana M~W~C 2003 The effective mass and the g-factor of the strongly-correlated 2-D electron fluid: Evidence of a coupled-valley state in the Si system {\it Preprint} cond-mat/0307153

\item[] Dobrosavljevi\'{c} V, Abrahams E, Miranda E and Chakravarty S 1997 Scaling theory of two-dimensional metal-insulator transitions \PRL {\bf 79} 455-458

\item[] Dolan G~J and Osheroff D~D 1979 Nonmetallic conduction in thin metal films at low temperatures \PRL {\bf 43} 721-724

\item[] Dolgopolov V~T, Kravchenko G~V, Shashkin A~A and Kravchenko S~V 1992 Properties of electron insulating phase in Si inversion-layers at low-temperatures {\it JETP Lett}.\ {\bf 55} 733-737

\item[] Dolgopolov V~T and Gold A 2000 Magnetoresistance of a two-dimensional electron gas in a parallel magnetic field {\it JETP Lett}.\ {\bf 71} 27-30

\item[] Dolgopolov V~T 2002 On effective electron mass of silicon field structures at low electron densities {\it JETP Lett}.\ {\bf 76} 377-379

\item[] Dolgopolov V~T and Gold A 2002 Comment on ``Weak anisotropy and disorder dependence of the in-plane magnetoresistance in high-mobility (100) Si-inversion layers'' \PRL {\bf 89} 129701

\item[] Dultz S~C and Jiang H~W 2000 Thermodynamic signature of a two-dimensional metal-insulator transition \PRL {\bf 84} 4689-4692

\item[] Efros A~L and Shklovskii B~I 1975 Coulomb gap and low temperature conductivity of disordered systems \JPC {\bf 8} L49-L51

\item[] Fang F~F and Stiles P~J 1968 Effects of a tilted magnetic field on a two-dimensional electron gas \PR {\bf 174} 823-828

\item[] Finkelstein A~M 1983 Influence of Coulomb interaction on the properties of disordered metals {\it Sov.\ Phys.\ --- JETP} {\bf 57} 97-108

\item[] Finkelstein A~M 1984 Weak localization and coulomb interaction in disordered-systems \ZP B {\bf 56} 189-196

\item[] Fogler M~M 2003 Nonlinear screening and percolative transition in a two-dimensional electron liquid {\it Preprint} cond-mat/0310010

\item[] Gao X~P~A, Mills A~P, Ramirez A~P, Pfeiffer L~N and West K~W 2002 Weak-localization-like temperature-dependent conductivity of a dilute two-dimensional hole gas in a parallel magnetic field \PRL {\bf 89} 016801


\item[] Gold A and Dolgopolov V~T 1986 Temperature dependence of the conductivity for the two-dimensional electron gas: Analytical results for low temperatures \PR B {\bf 33} 1076-1084

\item[] Gold A 2001 Linear temperature dependence of mobility in quantum wells and the effects of exchange and correlation \JPCM {\bf 13} 11641-11650

\item[] Gold A and Dolgopolov V~T 2002 On the role of disorder in transport and magnetic properties of the two-dimensional electron gas \JPCM {\bf 14} 7091-7096

\item[] Gold A 2003 Linear temperature dependence of the mobility in two-dimensional electron gases: many-body and spin-polarization effects \JPCM {\bf 15} 217-223

\item[] Hanein Y, Meirav U, Shahar D, Li C~C, Tsui D~C and Shtrikman H 1998a The metalliclike conductivity of a two-dimensional hole system \PRL {\bf 80} 1288-1291

\item[] Hanein Y, Shahar D, Yoon J, Li C~C, Tsui D~C and Shtrikman H 1998b Properties of the apparent metal-insulator transition in two-dimensional systems \PR B {\bf 58} R7520-R7523

\item[] Heemskerk R and Klapwijk T~M 1998 Nonlinear resistivity at the metal-insulator transition in a two-dimensional electron gas \PR B {\bf 58} R1754-R1757

\item[] Herbut I F 2001 The effect of a parallel magnetic field on the Boltzmann conductivity and the Hall coefficient of a disordered two dimensional Fermi liquid \PR B {\bf 63} 113102

\item[] Hikami S, Larkin A I and Nagaoka Y 1980 Spin-orbit interaction and magnetoresistance in the two dimensional random system {\it Prog.\ Theor.\ Phys.} {\bf 63} 707-710

\item[] Ilani S, Yacoby A, Mahalu D and Shtrikman H 2000 Unexpected behavior of the local compressibility near the $B=0$ metal-insulator transition \PRL {\bf 84} 3133-3136

\item[] Ilani S, Yacoby A, Mahalu D and Shtrikman H 2001 Microscopic structure of the metal-insulator transition in two dimensions {\it Science} {\bf 292} 1354-1357

\item[] Iwamoto N 1991 Static local-field corrections of 2-dimensional electron liquids \PR B {\bf 43} 2174-2182

\item[] Jaroszy\'nski J, Popovi\'c D and Klapwijk T~M 2002 Universal behavior of the resistance noise across the metal-insulator transition in silicon inversion layers \PRL {\bf 89} 276401

\item[] Khrapai V~S, Shashkin A~A and Dolgopolov V~T 2003 Direct measurements of the spin and the cyclotron gaps in a 2D electron system in silicon \PRL {\bf 91} 126404

\item[] Kravchenko S V, Kravchenko G V, Furneaux J E, Pudalov V M and D'Iorio M 1994 Possible metal-insulator-transition at $B=0$ in 2 dimensions \PR B {\bf 50} 8039-8042

\item[] Kravchenko S V, Mason W E, Bowker G E, Furneaux J E, Pudalov V M and D'Iorio M 1995 Scaling of an anomalous metal-insulator-transition in a 2-dimensional system in silicon at $B=0$ \PR B {\bf 51} 7038-7045

\item[] Kravchenko S~V, Simonian D, Sarachik M~P, Mason W and Furneaux J E 1996 Electric field scaling at a $B=0$ metal-insulator transition in two dimensions \PRL {\bf 77} 4938-4941

\item[] Kravchenko S~V, Simonian D, Sarachik M~P, Kent A~D and Pudalov V~M 1998 Effect of a tilted magnetic field on the anomalous $H=0$ conducting phase in high-mobility Si MOSFET's \PR B {\bf 58} 3553-3556

\item[] Kravchenko S~V and Klapwijk T~M 2000a Metallic low-temperature resistivity in a 2D electron system over an extended temperature range \PRL {\bf 84} 2909-2912

\item[] Kravchenko S~V, Shashkin A~A, Bloore D~A and Klapwijk T~M 2000b Shubnikov-de~Haas oscillations near the metal-insulator transition in a two-dimensional electron system in silicon \SSC {\bf 116} 495-499

\item[] Kravchenko S~V, Shashkin A~A and Dolgopolov V~T 2002 Comment on ``Low-density spin susceptibility and effective mass of mobile electrons in Si inversion layers'' \PRL {\bf 89} 219701

\item[] Kwon Y, Ceperley D~M and Martin R~M 1994 Quantum Monte-Carlo calculation of the Fermi-liquid parameters in the 2-dimensional electron-gas \PR B {\bf 50} 1684-1694

\item[] Landau L~D 1957 The theory of a Fermi liquid {\it Sov.\ Phys.\ --- JETP} {\bf 3} 920-925

\item[] Lee P~A and Ramakrishnan T~V 1982 Magnetoresistance of weakly disordered electrons \PR B {\bf 26} 4009-4012

\item[] Lee P~A and Ramakrishnan T~V 1985 Disordered electronic systems \RMP {\bf 57} 287-337

\item[] Lilly M~P, Reno J~L, Simmons J~A, Spielman I~B, Eisenstein J~P, Pfeiffer L~N, West K~W, Hwang E~H and Das~Sarma S 2003 Resistivity of dilute 2D electrons in an undoped GaAs heterostructure \PRL {\bf 90} 056806

\item[] Mason W, Kravchenko S~V, Bowker G~E and Furneaux J~E 1995 Experimental evidence for a Coulomb gap in 2 dimensions \PR B {\bf 52} 7857-7859

\item[] Mertes K~M, Zheng H, Vitkalov S~A, Sarachik ~P and Klapwijk T~M 2001 Temperature dependence of the resistivity of a dilute two-dimensional electron system in high parallel magnetic field \PR B {\bf 63} 041101(R)

\item[] Mills A~P, Ramirez A P, Pfeiffer L N and West K W 1999 Nonmonotonic temperature-dependent resistance in low density 2D hole gases \PRL {\bf 83} 2805-2808

\item[] Mills A P, Ramirez A P, Gao X~P~A, Pfeiffer L N, West K W and Simon S H 2001 Suppression of weak localization effects in low-density metallic 2D holes {\it Preprint} cond-mat/0101020

\item[] Noh H, Lilly M~P, Tsui D~C, Simmons J~A, Hwang E~H, Das~Sarma S, Pfeiffer L~N and West K~W 2002 Interaction corrections to two-dimensional hole transport in the large $r_{s}$ limit \PR B {\bf 68} 165308

\item[] Okamoto T, Hosoya K, Kawaji S and Yagi A 1999 Spin degree of freedom in a two-dimensional electron liquid \PRL {\bf 82} 3875-3878

\item[] Papadakis S~J and Shayegan M 1998 Apparent metallic behavior at $B=0$ of a two-dimensional electron system in AlAs \PR B {\bf 57} R15068-R15071

\item[] Papadakis S~J, De~Poortere E~P, Shayegan M and Winkler R 2000 Anisotropic magnetoresistance of two-dimensional holes in GaAs \PRL {\bf 84} 5592-5595

\item[] Pastor A~A and Dobrosavljevi\'{c} V 1999 Melting of the electron glass \PRL {\bf 83} 4642-4645

\item[] Pepper M, Pollitt S and Adkins C~J 1974 The spatial extent of localized state wavefunctions in silicon inversion layers \JPC {\bf 7} L273-L277

\item[] Phillips P, Wan Y, Martin I, Knysh S and Dalidovich D 1998 Superconductivity in a two-dimensional electron gas {\it Nature} (London) {\bf 395} 253-257

\item[] Polyakov D~G and Shklovskii B~I 1993 Conductivity-peak broadening in the quantum Hall regime \PR B {\bf 48} 11167-11175

\item[] Popovi\'{c} D, Fowler A~B and Washburn S 1997 Metal-insulator transition in two dimensions: Effects of disorder and magnetic field \PRL {\bf 79} 1543-1546

\item[] Proskuryakov Y~Y, Savchenko A~K, Safonov S~S, Pepper M, Simmons M~Y and Ritchie D~A 2002 Hole-hole interaction effect in the conductance of the two-dimensional hole gas in the ballistic regime \PRL {\bf 89} 076406

\item[] Prus O, Reznikov M, Sivan U and Pudalov V~M 2002 Cooling of electrons in a silicon inversion layer \PRL {\bf 88} 016801

\item[] Prus O, Yaish Y, Reznikov M, Sivan U and Pudalov V~M 2003 Thermodynamic spin magnetization of strongly correlated two-dimensional electrons in a silicon inversion layer \PR B {\bf 67} 205407

\item[] Pudalov V~M, D'Iorio M, Kravchenko S~V and Campbell J~W 1993 Zero magnetic field collective insulator phase in a dilute 2D electron system \PRL {\bf 70} 1866-1869

\item[] Pudalov V~M, Brunthaler G, Prinz A and Bauer G 1997 Instability of the two-dimensional metallic phase to a parallel magnetic field {\it JETP Lett}.\ {\bf 65} 932-937

\item[] Pudalov V~M, Brunthaler G, Prinz A and Bauer G 1998 Metal-insulator transition in two dimensions {\it Physica} E {\bf 3} 79-88

\item[] Pudalov V~M, Brunthaler G, Prinz A and Bauer G 1999 Weak-field Hall resistance and effective carrier density measurements across the metal-insulator transition in Si-MOS structures {\it JETP Lett}.\ {\bf 70} 48-53

\item[] Pudalov V~M, Brunthaler G, Prinz A and Bauer G 2001 Effect of the in-plane magnetic field on conduction of the Si-inversion layer: magnetic field driven disorder {\it Preprint} cond-mat/0103087

\item[] Pudalov V~M, Brunthaler G, Prinz A and Bauer G 2002a Weak anisotropy and disorder dependence of the in-plane magnetoresistance in high-mobility (100) Si-inversion layers \PRL {\bf 88} 076401

\item[] Pudalov V~M, Gershenson M~E, Kojima H, Butch N, Dizhur E~M, Brunthaler G, Prinz A and Bauer G 2002b Low-density spin susceptibility and effective mass of mobile electrons in Si inversion layers \PRL {\bf 88} 196404

\item[] Pudalov V~M, Gershenson M~E, Kojima H, Brunthaler G, Prinz A and Bauer G 2003 Interaction effects in conductivity of Si inversion layers at intermediate temperatures \PRL {\bf 91} 126403

\item[] Punnoose A and Finkelstein A~M 2002 Dilute electron gas near the metal-insulator transition: Role of valleys in silicon inversion layers \PRL {\bf 88} 016802

\item[] Rahimi M, Anissimova S, Sakr M~R, Kravchenko S~V and Klapwijk T~M 2003 Coherent back-scattering near the two-dimensional metal-insulator transition \PRL {\bf 91} 116402

\item[] Sarachik M~P and Kravchenko S~V 1999 Novel phenomena in dilute electron systems in two dimensions {\it Proc.\ Natl.\ Acad.\ Sci.\ USA} {\bf 96} 5900-5902

\item[] Sarachik M~P 2002 Disorder-dependence of the critical density in two-dimensional systems: An empirical relation {\it Europhys.\ Lett}.\ {\bf 57} 546-549

\item[] Sarachik M~P and Vitkalov S~A 2003 Does $m^*g^*$ diverge at a finite electron density in silicon inversion layers? \JPSJ (Suppl.\ A) {\bf 72} 57-62

\item[] Senz V, D\"{o}tsch U, Gennser U, Ihn T, Heinzel T, Ensslin K, Hartmann R and Gr\"{u}tzmacher D 1999 Metal-insulator transition in a 2-dimensional system with an easy spin axis \AP {\bf 8} (special issue) 237-240

\item[] Shahar D, Tsui D~C, Shayegan M, Cunningham J~E, Shimshoni E and Sondhi S~L 1996 Evidence for charge-flux duality near the quantum Hall liquid-to-insulator transition {\it Science} {\bf 274} 589-592

\item[] Shashkin A~A, Dolgopolov V~T and Kravchenko G~V 1994 Insulating phases in a 2-dimensional electron-system of high-mobility Si MOSFETs \PR B {\bf 49} 14486-14495

\item[] Shashkin A~A, Kravchenko S~V, Dolgopolov V~T and Klapwijk T~M 2001a Indication of the ferromagnetic instability in a dilute two-dimensional electron system \PRL {\bf 87}, 086801

\item[] Shashkin A~A, Kravchenko S~V and Klapwijk T~M 2001b Metal-insulator transition in 2D: equivalence of two approaches for determining the critical point \PRL {\bf 87}, 266402

\item[] Shashkin A~A, Kravchenko S~V, Dolgopolov V~T and Klapwijk T~M 2002 Sharp increase of the effective mass near the critical density in a metallic two-dimensional electron system \PR B {\bf 66} 073303

\item[] Shashkin A~A, Rahimi M, Anissimova S, Kravchenko S~V, Dolgopolov V~T and Klapwijk T~M 2003a Spin-independent origin of the strongly enhanced effective mass in a dilute 2D electron system \PRL {\bf 91} 046403

\item[] Shashkin A~A, Kravchenko S~V, Dolgopolov V~T and Klapwijk T~M 2003b Sharply increasing effective mass: a precursor of a spontaneous spin polarization in a dilute two-dimensional electron system \JPA {\bf 36} 9237-9247

\item[] Shashkin A~A, Dolgopolov V~T and Kravchenko S~V 2003c Comment on ``Interaction effects in conductivity of Si inversion layers at intermediate temperatures'' {\it Preprint} cond-mat/0311174

\item[] Si Q~M and Varma C~M 1998 Metal-insulator transition of disordered interacting electrons \PRL {\bf 81} 4951-4954

\item[] Simmons M~Y, Hamilton A~R, Pepper M, Linfield E~H, Rose P~D, Ritchie D~A, Savchenko A~K and Griffiths T~G 1998 Metal-insulator transition at $B=0$ in a dilute two dimensional GaAs-AlGaAs hole gas \PRL {\bf 80} 1292-1295

\item[] Simmons M~Y, Hamilton A~R, Pepper M, Linfield E~H, Rose P~D and Ritchie D~A 2000 Weak localization, hole-hole interactions, and the ``metal''-insulator transition in two dimensions \PRL {\bf 84} 2489-2492

\item[] Simonian D, Kravchenko S~V and Sarachik M~P 1997a Reflection symmetry at a $B=0$ metal-insulator transition in two dimensions \PR B {\bf 55} R13421-R13423

\item[] Simonian D, Kravchenko S~V, Sarachik M~P and Pudalov V~M 1997b Magnetic field suppression of the conducting phase in two dimensions \PRL {\bf 79} 2304-2307

\item[] Simonian D, Kravchenko S~V, Sarachik M~P and Pudalov V~M 1998 $H/T $ scaling of the magnetoconductance near the conductor-insulator transition in two dimensions \PR B {\bf 57} R9420-R9422

\item[] Smith J~L and Stiles P~J 1972 Electron-electron interactions continuously variable in the range $2.1>r_s>0.9$ \PRL {\bf 29} 102-104

\item[] Sondhi S~L, Girvin S~M, Carini J~P and Shahar D 1997 Continuous quantum phase transitions \RMP {\bf 69} 315-333

\item[] Spivak B 2001 Properties of the strongly correlated two-dimensional electron gas in Si MOSFET's \PR B {\bf 64} 085317

\item[] Spivak B 2002 Phase separation in the two-dimensional electron liquid in MOSFET's \PR B {\bf 67} 125205

\item[] Stern F 1980 Calculated temperature dependence of mobility in silicon inversion layers \PRL {\bf 44} 1469-1472

\item[] Stoner E~C 1946 Ferromagnetism \RPP {\bf 11} 43-112

\item[] Tanaskovi\'{c} D, Dobrosavljevi\'{c} V, Abrahams E and Kotliar G 2003 Disorder screening in strongly correlated systems \PRL {\bf 91} 066603

\item[] Tanatar B and Ceperley D~M 1989 Ground-state of the two-dimensional electron-gas \PR B {\bf 39} 5005-5016

\item[] Tutuc E, De~Poortere E~P, Papadakis S~J and Shayegan M 2001 In-plane magnetic field-induced spin polarization and transition to insulating behavior in two-dimensional hole systems \PRL {\bf 86} 2858-2861

\item[] Tutuc E, Melinte S and Shayegan M 2002 Spin polarization and g factor of a dilute GaAs two-dimensional electron system \PRL {\bf 88} 36805

\item[] Tutuc E, Melinte S, De~Poortere E~P, Shayegan M and Winkler R 2003 Role of finite layer thickness in spin polarization of GaAs two-dimensional electrons in strong parallel magnetic fields \PR B {\bf 67} 241309(R)

\item[] Uren M~J, Davies R~A and Pepper M 1980 The observation of interaction and localisation effects in a two-dimensional electron gas at low temperatures \JPC {\bf13} L985-L993

\item[] Varma C~M, Nussinov Z and van Saarloos W 2002 Singular or non-Fermi liquids {\it Phys. Rep}.\ {\bf 361} 267-417

\item[] Vitkalov S~A, Zheng H, Mertes K~M, Sarachik M~P and Klapwijk Y~M 2000 Small-angle Shubnikov-de~Haas measurements in a 2D electron system: The effect of a strong in-plane magnetic field \PRL {\bf 85} 2164-2167

\item[] Vitkalov S~A, Sarachik M~P and Klapwijk T~M 2001a Spin polarization of two-dimensional electrons determined from Shubnikov-de~Haas oscillations as a function of angle \PR B {\bf 64} 073101

\item[] Vitkalov S~A, Zheng H, Mertes K~M, Sarachik M~P and Klapwijk T~M 2001b Scaling of the magnetoconductivity of silicon MOSFETs: Evidence for a quantum phase transition in two dimensions \PRL {\bf 87} 086401

\item[] Vitkalov S~A, Sarachik M~P and Klapwijk T~M 2002 Spin polarization of strongly interacting two-dimensional electrons: The role of disorder \PR B {\bf 65} 201106(R)

\item[] Vitkalov S~A, James K, Narozhny B~N, Sarachik M~P and Klapwijk T~M 2003 In-plane magnetoconductivity of Si MOSFETs: A quantitative comparison of theory and experiment \PR B {\bf 67} 113310

\item[] Vojta M 2003 Quantum phase transitions \RPP {\bf 66} 2069-2110

\item[] Wigner E 1934 On the interaction of electrons in metals \PR {\bf 46} 1002-1011

\item[] Yaish Y, Prus O, Buchstab E, Shapira S, Ben Yoseph G, Sivan U and Stern A 2000 Interband scattering and the ``metallic phase'' of two-dimensional holes in GaAs/AlGaAs \PRL {\bf 84} 4954-4957

\item[] Yoon J, Li C~C, Shahar D, Tsui D~C and Shayegan M 1999 Wigner crystallization and metal-insulator transition of two-dimensional holes in GaAs at $B=0$ \PRL {\bf 82} 1744-1747

\item[] Yoon J, Li C~C, Shahar D, Tsui D~C and Shayegan M 2000 Parallel magnetic field induced transition in transport in the dilute two-dimensional hole system in GaAs \PRL {\bf 84} 4421-4424

\item[] Zala G, Narozhny B~N and Aleiner I~L 2001 Interaction corrections at intermediate temperatures: Longitudinal conductivity and kinetic equation \PR B {\bf 64} 214204

\item[] Zala G, Narozhny B~N and Aleiner I~L 2002 Interaction corrections at intermediate temperatures: Magnetoresistance in a parallel field \PR B {\bf 65} 020201(R)

\item[] Zhu J, Stormer H~L, Pfeiffer L~N, Baldwin K~W and West K~W 2003 Spin susceptibility of an ultra-low-density two-dimensional electron system \PRL {\bf 90} 056805

\endrefs

\end{document}